\documentclass[twocolumn]{aastex631}
\usepackage{CJK}
\usepackage{units}
\usepackage{gensymb}
\usepackage{subfigure}
\usepackage[percent]{overpic}
\usepackage{appendix}

\graphicspath{{./}{figures/}}

\begin{document}
\begin{CJK*}{UTF8}{gbsn}

\title{Fast-rotating A- and F-type stars with $\mathrm{H\alpha}$ emissions in NGC 3532, candidate UV-dim stars?}

\correspondingauthor{Chenyu He}
\email{hechy39@mail.sysu.edu.cn}

\author[0000-0001-9131-6956]{Chenyu He (贺辰昱)}
\affiliation{School of Physics and Astronomy, Sun Yat-sen University, Zhuhai, 519082, China}
\affiliation{CSST Science Center for the Guangdong--Hong Kong--Macau Greater Bay Area, Zhuhai, 519082, China}

\author[0000-0002-3084-5157]{Chengyuan Li (李程远)}
\affiliation{School of Physics and Astronomy, Sun Yat-sen University, Zhuhai, 519082, China}
\affiliation{CSST Science Center for the Guangdong--Hong Kong--Macau Greater Bay Area, Zhuhai, 519082, China}

\author[0000-0001-9313-251X]{Gang Li (李刚)}
\affiliation{Institute for Astronomy, KU Leuven, Celestijnenlaan 200D bus 2401, 3001 Leuven, Belgium}

\date{December 2024}

\begin{abstract}
Extended main-sequence stars that are dim in the ultraviolet passbands of \textit{Hubble Space Telescope} (UV-dim stars) are found in several young and intermediate-age Magellanic Cloud star clusters. The obscuring of the dust in the discs of stars expelled due to fast rotation have been suggested to be responsible for the appearance of UV-dim stars, and play an important role in the formation of extended main-sequences. In this paper, we report a population of A- and F-type stars who show $\mathrm{H\alpha}$ emission features in their spectra in a young ($\sim \unit[340]{Myr}$-old) Galactic neighboring star cluster NGC 3532. By fitting the observed absorption profiles, we found that most $\mathrm{H\alpha}$ emitters are fast rotating stars, indicating that they form decretion discs by fast rotation like Be stars. As A- and F-type stars dominate the extended main-sequence turn-off regions of intermediate-age clusters, their appearance provides observational evidence to support the dust extinction scenario for these clusters, and might be the counterparts of UV-dim stars that are detected in remote Magellanic Cloud star clusters like NGC 1783.

\end{abstract}


\section{introduction\label{sec:intro}}

High- resolution and precision photometric observations with the Hubble Space Telescope (\textit{HST}) and \textit{Gaia} Telescope revealed that star clusters younger than 2 Gyr in the Magellanic Clouds (MCs) and the Milky ways show extended main-sequence turnoffs \citep[eMSTOs; e.g.,][]{2008ApJ...681L..17M, 2009A&A...497..755M,
2018MNRAS.477.2640M,2018ApJ...869..139C}. In star clusters younger than $\unit[600]{Myr}$, additional broadened features appear in their upper MSs \citep[][]{2015MNRAS.450.3750M,2016MNRAS.458.4368M,2017MNRAS.467.3628C,2018MNRAS.477.2640M,2017ApJ...844..119L, 2019ApJ...883..182S}. Extended star formation
histories \citep[eSFHs; e.g.,][]{2009A&A...497..755M,2009AJ....137.4988G,2011ApJ...737....4G}, stellar variability in the MSTOs \citep{2016ApJ...832L..14S} and different stellar interior mixing \citep{2019A&A...632A..74J} were proposed to account for the extending of eMSTOs. 

However, the most widely accepted explanation for the eMSTOs and broadened upper MSs \footnote{hereafter, we call both phenomenon as extended MSs (eMSs) in some content as they are believed to be caused by similar physical mechanisms \citep{LI2024}.} is the different stellar rotation of early-type stars in those regions \citep[e.g.,][]{2009MNRAS.398L..11B, 2014Natur.516..367L, 2015MNRAS.453.2637D, 2017NatAs...1E.186D, 2018MNRAS.477.2640M,2018ApJ...869..139C}. This scenario has been confirmed by spectroscopic studies in some MC and Galactic clusters \citep{2018ApJ...863L..33M, 2018AJ....156..116M,2019ApJ...876..113S,2019ApJ...883..182S,2020MNRAS.492.2177K, 2023MNRAS.518.1505K} where they all found that stars on the redder sides of the eMSs have larger mean projected rotation rates ($v\sin{i}$) than those of stars on the bluer sides. Rotation may contribute to the formation of the eMS through two ways: First, the centrifugal force from rotation distorts stars and causes temperature inhomogeneity \citep[known as the gravity-darkening effect;][]{1924MNRAS..84..665V, 2011A&A...533A..43E, 2013ApJS..208....4P, 2019ApJS..243...10P}. Therefore, the projected temperatures and luminosities are scattered with different stellar inclinations, leading to the phenomenon of the eMS. Second, internal rotation affects the equilibrium of stellar structure and induces different extents of element transport inside stars. As a result, it significantly modifies the lifetime and evolutionary tracks of stars \citep{2000ARA&A..38..143M, Heger2000ApJ.II, Heger2005ApJ, 2019ApJS..243...10P, Mombarg2022ApJ}, providing a possible explanation of the eMS formation. Early-type stars in both the field \citep{2012A&A...537A.120Z,2013A&A...550A.109D,2021ApJ...921..145S} and star clusters \citep[e.g.,][]{2018ApJ...863L..33M, 2018AJ....156..116M,2019ApJ...876..113S,2019ApJ...883..182S,2020MNRAS.492.2177K,2023MNRAS.518.1505K,2022ApJ...938...42H,2024ApJ...968...22B} show a wide range of rotation rates, spanning from several to hundreds of kilometers per second based on spectroscopic studies. Asteroseismic measurement also revealed their rotation periods to be from hundreds of days to as short as eight hours with a median rotation period of approximately one day \citep{Ligang2020_611, Ligang2024_NGC2516, Mombarg2024}. Bimodal distributions of rotation rates were observed in the stars of both the field \citep{2012A&A...537A.120Z,2013A&A...550A.109D,2021ApJ...921..145S} and star clusters \citep[e.g.,][]{2018ApJ...863L..33M,2018AJ....156..116M,2019ApJ...883..182S}. To explain the slow rotation of early-type field stars, tidal interactions and disc-locking during the pre-main-sequence phase have been suggested \citep{2012A&A...537A.120Z}. Tidal interactions \citep{2015MNRAS.453.2637D} and disc-locking \citep{2020MNRAS.495.1978B} were then proposed to account for the formation of slowly-rotating stars in star clusters. A kind of tidal interactions between stars and binary orbits, so-called `inverse tides', which transfer angular momentum from the stars to the binary orbitals, were found to also contribute to the different rotation rates of early-type stars \citep{Ligang_2020_35binaries, Fuller2021_inverse_tides}. Binary mergers were suggested as another possible path to form slowly rotating blue MS stars \citep{2022NatAs...6..480W}. Some works \citep[e.g.,][]{2021MNRAS.508.2302K,2021ApJ...912...27Y,2021MNRAS.502.4350S,Wang_2023,2023MNRAS.525.5880H,2024arXiv241102508M, 2024arXiv241200520B} have been conducted to test these scenarios. For a comprehensive description for the researches of eMSs, we suggested the review of \cite{LI2024}.

A population of eMSTO stars that was dim in the ultraviolet (UV) filter (F275W) of the \textit{HST} passbands was recently found in the intermediate-age \citep[$\sim$ 1.8 Gyr;][]{2008ApJ...681L..17M} Large MC (LMC) cluster NGC 1783 \citep{2023A&A...672A.161M}. Five MC clusters younger than $\unit[200]{Myr}$ were later found to harbor stars that are dim in the \textit{HST} F225W and F275W bands \citep{2023MNRAS.524.6149M}, indicating that UV-dim stars are common in star clusters. These stars are much redder than the bulk of stars when the color index includes UV bands, while well mixed with other stars for the colors made of optical bands. They are called `UV-dim' stars \citep{2023A&A...672A.161M}. To explain the UV-dim stars in NGC 1783, \cite{2023MNRAS.521.4462D} carefully studied the effect of dust extinction of the circumstellar discs on the morphology of the MSTO. They found that the model with dusty stars can exclusively reproduce the entire eMSTO stellar distribution including that of UV-dim stars. UV-dim stars were proposed to be fast-rotating stars obscured by dust in the discs along the line of sight, which is expelled due to fast stellar rotation \citep{2023MNRAS.521.4462D}. Then, they correspond to large $v\sin{i}$, which is consistent with the observations \citep[e.g.,][]{2020MNRAS.492.2177K} where stars with larger $v\sin{i}$ are located in the redder part \citep{2023MNRAS.521.4462D}. Their work indicates that dust self-extinction of stars may also play a significant role in the formation of the eMSTOs besides different stellar ages and rotation rates \citep{2023MNRAS.521.4462D}. If this scenario is on the right track, the influence of dust self-extinction would be not negligible when determining the parameters (e.g., the ages) of clusters using CMDs and stellar models, and studying the effect of rotation on stars for stellar models. 

Since A- and F- type stars dominate the MSTOs of intermediate-age clusters like NGC 1783, the dust self-extinction scenario indicates that star clusters should harbor a population of fast-rotating A- or F-type stars with excretion discs. Be stars, who are fast-rotating B-type stars exhibiting $H_{\alpha}$ emissions in their spectra, have been widely found in many young star clusters \citep{2017MNRAS.465.4795B,2018MNRAS.477.2640M}. Their $\mathrm{H\alpha}$ emissions are thought to be the result of the ionization of the gas in the decretion discs that are expelled due to very rapid rotation \citep{2013A&ARv..21...69R}. They provide the observational evidence for eMSs to have a population of fast rotating stars \citep{2017MNRAS.465.4795B}, and are found to be located in the red parts of eMSs, in favour of that different rotation of stars causes eMSs \citep{2018MNRAS.477.2640M}. Shell stars are stars obscured by discs along the light of sight whose spectra show broadened photospheric lines from the stars and narrow absorption lines from the discs \citep{2006A&A...459..137R}. The shell stars in the Be population of the LMC star cluster NGC 1850 ($\sim \unit[100]{Myr}$) are found to be redder than other normal Be stars \citep{2023MNRAS.518.1505K}. This indicates that self-extinction of the discs indeed introduces additional reddening other than stellar rotation \citep{2023MNRAS.518.1505K}. However, a large population of A- and F-type stars with $\mathrm{H\alpha}$ emissions has not been detected in star clusters yet. Since the weak ionization radiation of A- and F-type stars \citep{2013A&ARv..21...69R}, it is hard to detect both types of stars with discs using $\mathrm{H\alpha}$ emissions. The field A- and F-type with discs were generally detected as shell stars through the absorption profiles of discs in the near-infrared spectra \citep[e.g.,][]{1986PASP...98..867S}. Some of them show weak $\mathrm{H\alpha}$ emissions, however, they disappear quickly for later subtypes\citep{2013A&ARv..21...69R}.

In this paper, we report a population of A- and early F-type stars who show $\mathrm{H\alpha}$ emission features in a young ($\sim \unit[340]{Myr}$) Galactic neighboring ($\sim \unit[484]{pc}$) star cluster NGC 3532, using the high-resolution public spectra from the European Southern Observatory (ESO) archive. This cluster was one of the clusters in the Milky Way who were detected to show evident eMSs by \cite{2018ApJ...869..139C}. The $\mathrm{H\alpha}$ emitters account for $\sim 30\%$ of the spectroscopic samples, and have masses (1.4--$\unit[1.6]{M_\odot}$) close to that \citep[$\sim 1.5 \mathrm{M_\odot}$][]{2023MNRAS.521.4462D} of the eMSTO stars of NGC 1783. By fitting the observed absorption profiles in their spectra, we found that most $\mathrm{H\alpha}$ emitters are rapidly rotating stars. These results imply that a fraction of A- and F-type stars with discs expelled due to fast rotation appear in star clusters. They might provide vital observational support for the dust self-extinction scenario for the eMSs, and be the counterparts of the UV-dim stars found in remote MC clusters, in particular of those found in intermediate-age clusters like NGC 1783. 

This paper is organized as follows. In Section \ref{sec:Data R}, we describe our data reduction process. In Section \ref{sec:R}, we show the spectra of the $\mathrm{H\alpha}$ emitters and their $v\sin{i}$ properties. We
discuss our results in Section \ref{sec:discussion}, and conclude in Section \ref{sec:con}.

\section{Data Reduction\label{sec:Data R}}

\subsection{Isochrone fitting}

\begin{figure*}[ht]
    \centering
    \includegraphics[width=0.4\textwidth]{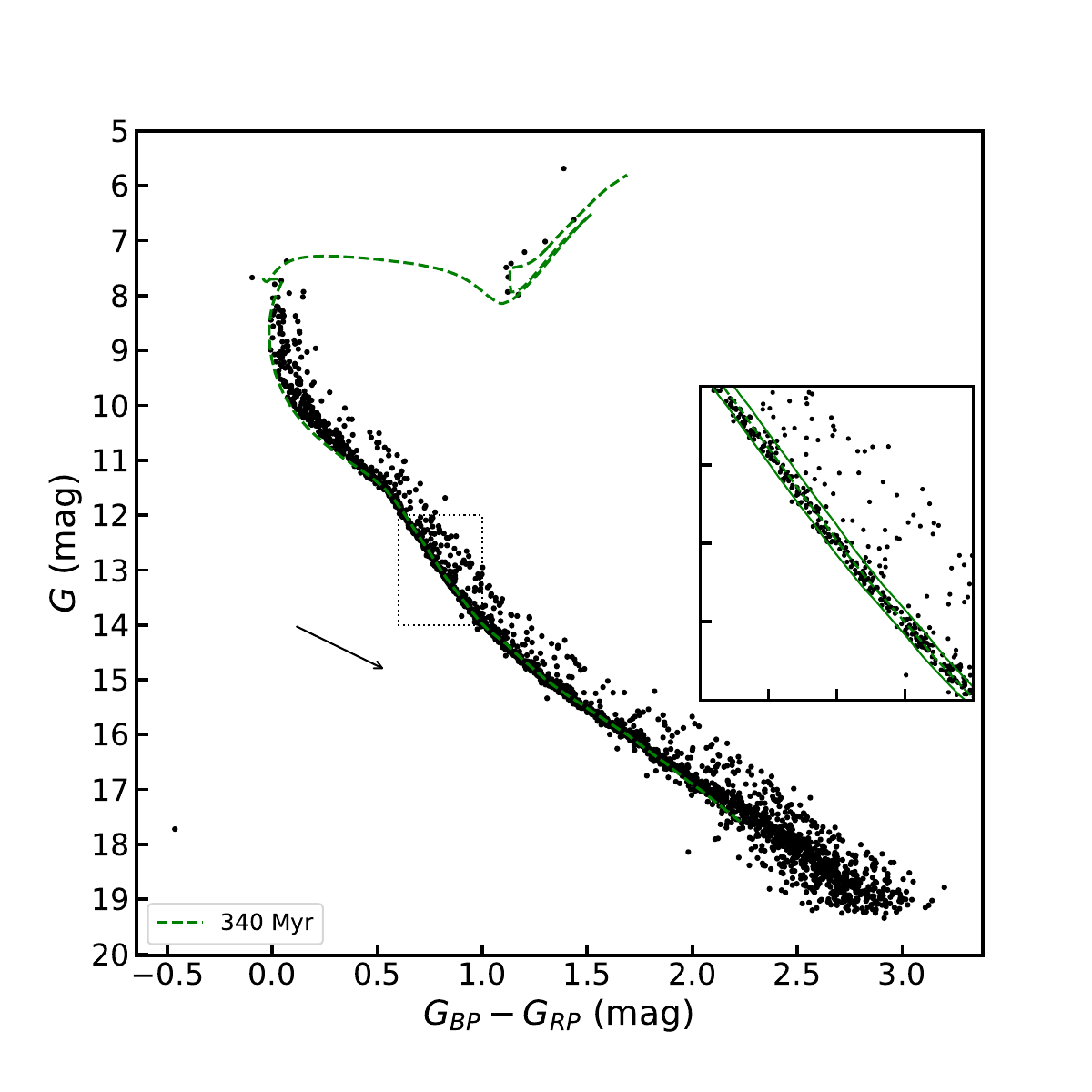}
	\caption{The CMD of the stars of NGC 3532 (black dots). The catalog of the member stars are from \cite{2022ApJ...931..156P}. The photometric data are from \textit{Gaia} EDR3 \citep{2016A&A...595A...1G,2021A&A...649A...1G}. The green dashed line represents the best-fitting isochrone for the blue edge of the eMSTO, who has an age of $\unit[340]{Myr}$, $Z=Z_\odot$, $(m-M)_{0}=\unit[8.38]{mag}$ and $A_\mathrm{V} = \unit[0.09]{mag}$. The isochrone is from the PARSEC model \citep[version 2.0S,][]{2022A&A...665A.126N}, corresponding to no rotation rates. The black arrow shows the reddening vector, corresponding to $\Delta A_\mathrm{V}=+1.0$ mag. The inset magnifies the CMD of stars with $\unit[12]{mag}<G<\unit[14]{mag}$，$\unit[0.6]{mag}<(G_{\mathrm{BP}}-G_{\mathrm{RP}})<\unit[1]{mag}$ (within the dotted rectangle). The green dashed line in the inset represents the best-fitting isochrone, while the green solid lines are the isochrones obtained by shifting the best-fitting isochrone by $\unit[\pm 0.016]{mag}$ in color.\label{fig:3532 CMD}}
\end{figure*}

We adopted the catalog of NGC 3532 member stars provided by \cite{2022ApJ...931..156P}. The CMD of the cluster is shown in Figure \ref{fig:3532 CMD}. In this Figure, the photometric data are from \textit{Gaia} Early Data Release 3 \citep[EDR3;][]{2016A&A...595A...1G,2021A&A...649A...1G}. We used the isochrones from the PARSEC model \citep[version 2.0S,][]{2022A&A...665A.126N} to fit the CMD of this cluster. The adopted isochrones correspond to stellar models with no rotation rates. The extinction coefficients for the \textit{Gaia} bands were derived with the $R_V = 3.1$ extinction curves from \citet{1989ApJ...345..245C} and \citet{1994ApJ...422..158O}. The best-fitting isochrone for the blue edge of the eMSTO is shown in Figure \ref{fig:3532 CMD}. It has an age of $\unit[340]{Myr}$, metallicity $Z=Z_\odot$ and a distance modulus $(m-M)_{0}=\unit[8.38]{mag}$. The distribution of the post-MS stars is well consistent with the position of the best-fitting isochrone. The distance modulus corresponds to a distance of $\unit[475]{pc}$, which is close to the distance ($\sim\unit[484]{pc}$) inferred from the average parallax of the cluster provided by \cite{2018A&A...616A..10G}. The age of the best-fitting isochrone is different from those provided by \cite{2018A&A...616A..10G}, \cite{2021A&A...647A..19T} and \cite{2022ApJ...931..156P} ($\unit[398]{Myr}$ from these references). This may be caused by the extending of the MSTO. The extinction of the cluster $A_\mathrm{V} = \unit[0.09]{mag}$, corresponding to $E(B-V)=0.029$, which is much smaller than the median $E(B-V)$ (0.13) of 324 neighboring open clusters within \unit[500]{pc} reported by \cite{2023ApJS..265...12Q}. To explore the effect of differential reddening on the cluster CMD, we inspected the distribution of stars with $\unit[12]{mag}<G<\unit[14]{mag}$ and $\unit[0.6]{mag}<(G_{\mathrm{BP}}-G_{\mathrm{RP}})<\unit[1]{mag}$ (within the dotted rectangle in Figure \ref{fig:3532 CMD}), which is shown in the inset. The solid green lines in the inset represent the shifted best-fitting isochrones by $\unit[\pm 0.016]{mag}$ in color, which magnitude corresponds to the average $3\sigma$ measurement error of $G_{\mathrm{BP}}-G_{\mathrm{RP}}$ of the stars with $\unit[12]{mag}<G<\unit[14]{mag}$. Most stars that reside on the MS ridgeline in the inset are located between the shifted best-fitting isochrones, indicating that the effect of differential reddening on the CMD of NGC 3532 is not significant. This implied that differential reddening, which was suggested to be responsible for the eMSTO in Trumpler 20 \citep{2012ApJ...751L...8P}, contribute little to the eMS in NGC 3532.

\begin{figure*}[ht!]
    \centering
    \begin{tabular}{cc}
    \includegraphics[width=0.4\textwidth]{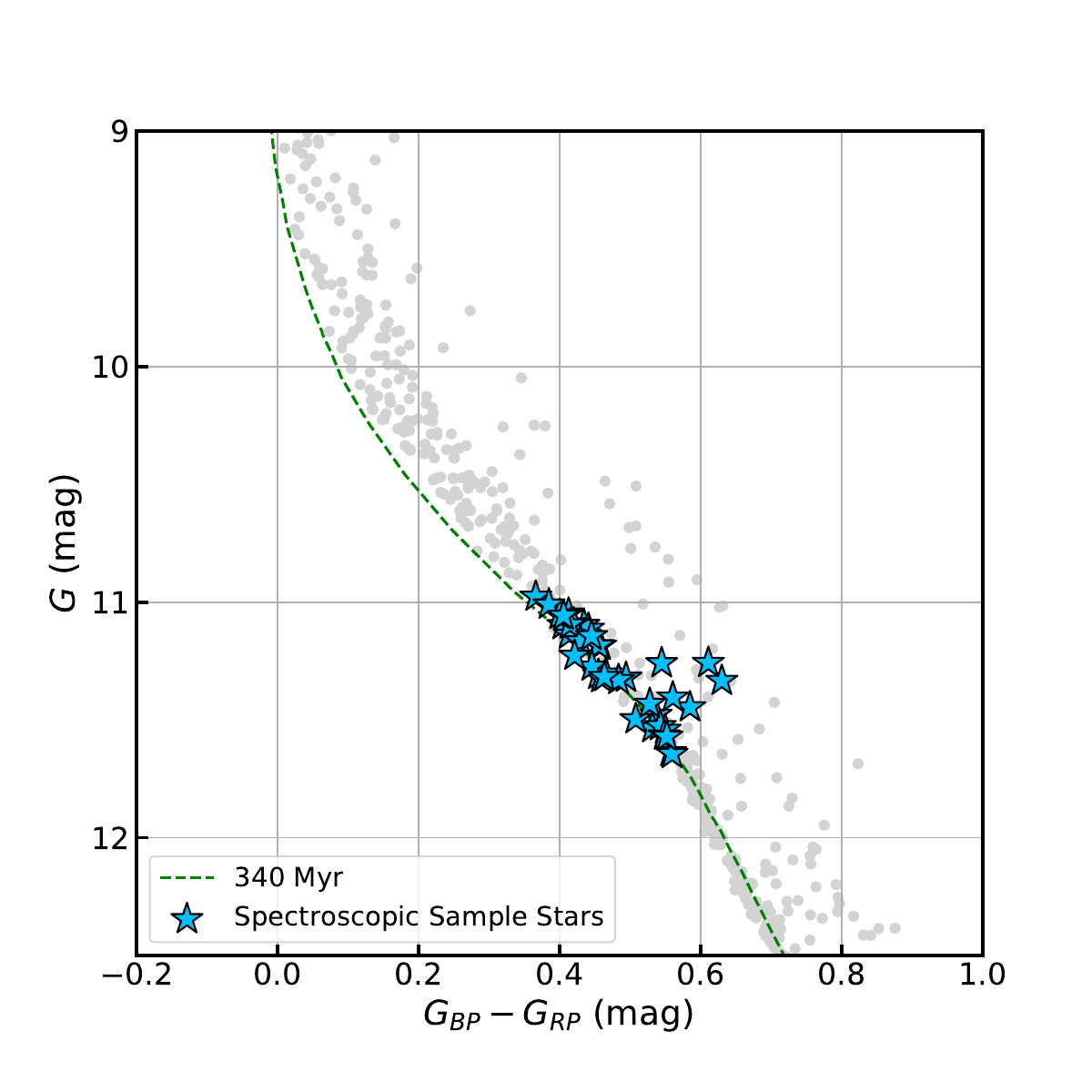}&
    \includegraphics[width=0.4\textwidth]{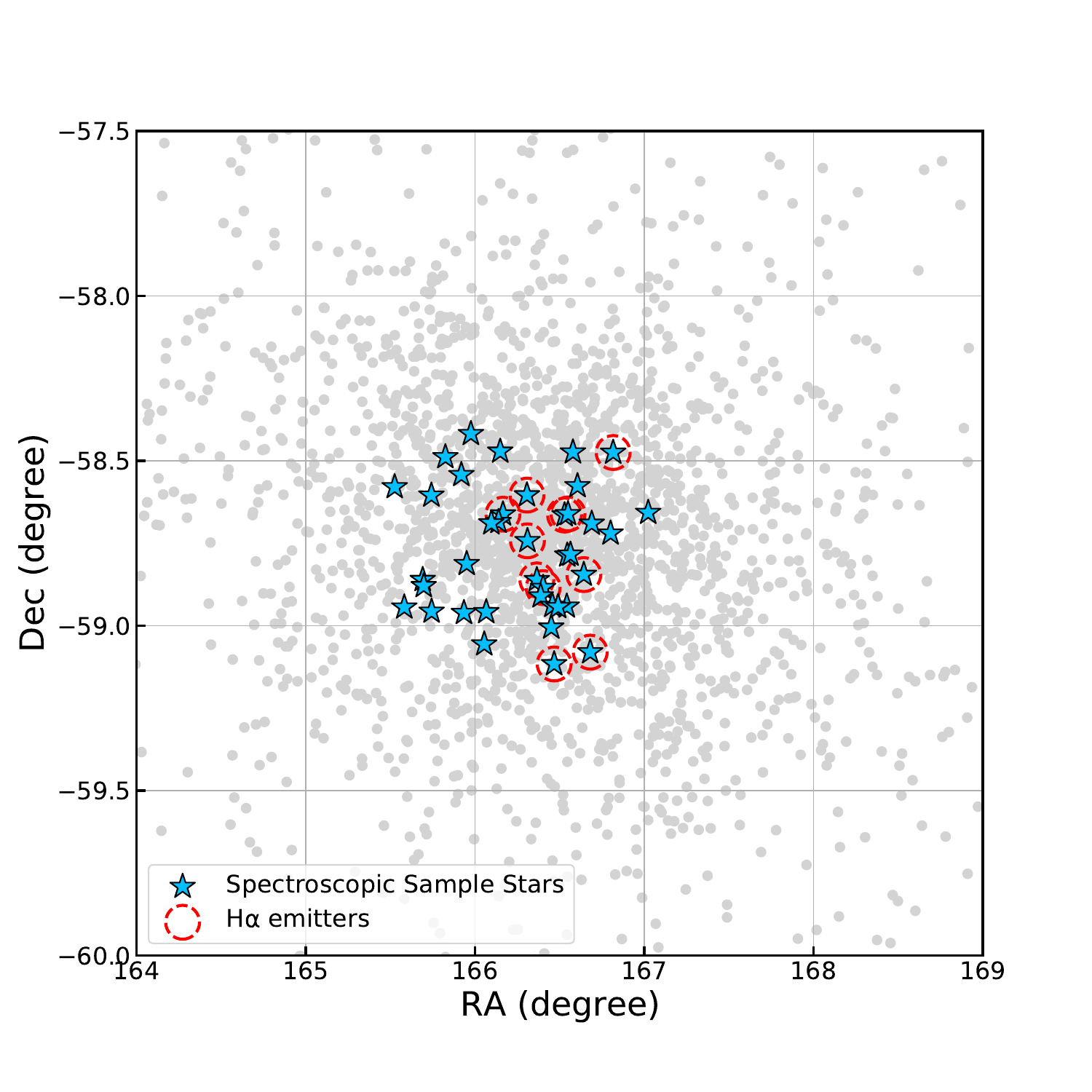}\\
    \end{tabular}
	\caption{The locations of the 39 spectroscopic sample stars (blue stars) in the CMD (left) and the cluster field (right). Other cluster member stars are shown with the light-grey dots in both panels. The green dashed line in the left panel represents the best-fitting isochrone. In the right panel, the stars with $\mathrm{H\alpha}$ emissions are highlighted with red open dashed circles.\label{fig:3532 Ha radec}}
 
\end{figure*}

\subsection{Spectrum Fitting\label{sec:SF}}
We obtained public spectra of 39 A- and early-F stars of NGC 3532 with $G<\unit[11.65]{mag}$ who have spectra in wavelength range of 644--$\unit[682]{nm}$ from the archive of European Southern Observatory (ESO). These spectra were observed by the program 193.B--0936 \citep[PI: Gilmore;][]{2012Msngr.147...25G}. The minimum $G$ magnitude of the spectroscopic sample stars is $\unit[10.97]{mag}$. Their mass range is 1.39--$\unit[1.59]{M_\odot}$ inferred from the best-fitting isocrhone. The spectra were observed with the GIRAFFE spectrograph mounted on the Very Large Telescope (VLT) with a resolution of $R=17000$ and signal-to-noise ratios (SNRs) of from 136 to 397. The loci of these spectroscopic sample stars in the CMD are shown in the left panel of Figure \ref{fig:3532 Ha radec}. 

To get the $v\sin{i}$ of these stars, we fitted the observed absorption profiles of the Fe {\sc I} 6633.7{\AA} and 6678.0{\AA} lines in the observed spectra. We obtained theoretical stellar spectra with different surface effective temperatures $T_\mathrm{eff}$, surface gravity $\log g$,
and metallicity [Fe/H] from the Pollux database \citep{2010A&A...516A..13P}. These spectra were derived based on the Plane-parallel ATLAS12 model atmospheres in local thermodynamic equilibrium \citep{2005MSAIS...8...14K}, using the SYNSPEC tool \citep{1992A&A...262..501H}. Their $T_\mathrm{eff}$ were set to be 5800--\unit[10,000]{K} in steps of \unit[100]{K}. The $\log g$ were set to be 3.5--\unit[5.0]{dex} in steps of \unit[0.1]{dex}. In all the synthetic spectra, a fixed microturblulent velocity of $\unit[2]{km\,s^{-1}}$ was introduced. The metallicity of the synthetic spectra was set to be solar based on the metallicity of the best-fitting isochrone. We used the tool PyAstronomy \citep{pya} to convolve the synthetic spectra with the effects of instrumental and rotating ($v\sin i$) broadening, and shift the wavelengths based on the input radial velocities (RVs). The range of $v\sin{i}$ was set to be 5--$\unit[400]{km\,s^{-1}}$ in steps of $\unit[5]{km\,s^{-1}}$. The RVs were set to be within $\unit[-25]{km\,s^{-1}}$ and $\unit[35]{km\,s^{-1}}$ in steps of $\unit[1]{km\,s^{-1}}$ based on the average RVs (5--$\unit[6]{km\,s^{-1}}$) of NGC 3532 members reported by \citet{2021A&A...647A..19T} and \citet{2022ApJ...931..156P} \footnote{We noted that the best-fitting RVs of some stars were close to the upper or lower limits of the RV fitting range. These stars might be in close binaries that show large RV differences from most cluster members. For their spectra, we extended their RV fitting ranges.}. Then we used the tool Astrolib PySynphot \citep{2013ascl.soft03023S} to derive the flux corresponding to each wavelength of the observed spectra. We compared the absorption profiles of the Fe {\sc I} 6633.7{\AA} and 6678.0{\AA} lines in the model spectra with those of the observed spectra, and determined the best-fitting model using a minimum-$\chi^2$ method. The uncertainty of the $v\sin i$ measurement was determined by combining the $\chi^2$ values corresponding to different input parameters \citep[][]{1976ApJ...210..642A,1996QJRAS..37..519W,2013ApJ...765....4D}. We fitted the correlation of the $\chi^2$ values with the $v\sin i$ values for the best-fitting $T_\mathrm{eff}$, $\log g$ and RVs. Then the 1$\sigma$ uncertainty of $v\sin{i}$ is the difference between the best-fitting $v\sin i$ value and the $v\sin i$ values corresponding to $\chi^2_{minimum}$+1. In the middle and right panels of Figure~\ref{fig:3532 star 2 and 6}, we show the observed spectra with the best-fitting models for two spectroscopic samples.

\section{Main Results\label{sec:R}}
\begin{figure*}
\centering
\begin{tabular}{cc}

    \includegraphics[width=0.4\textwidth]{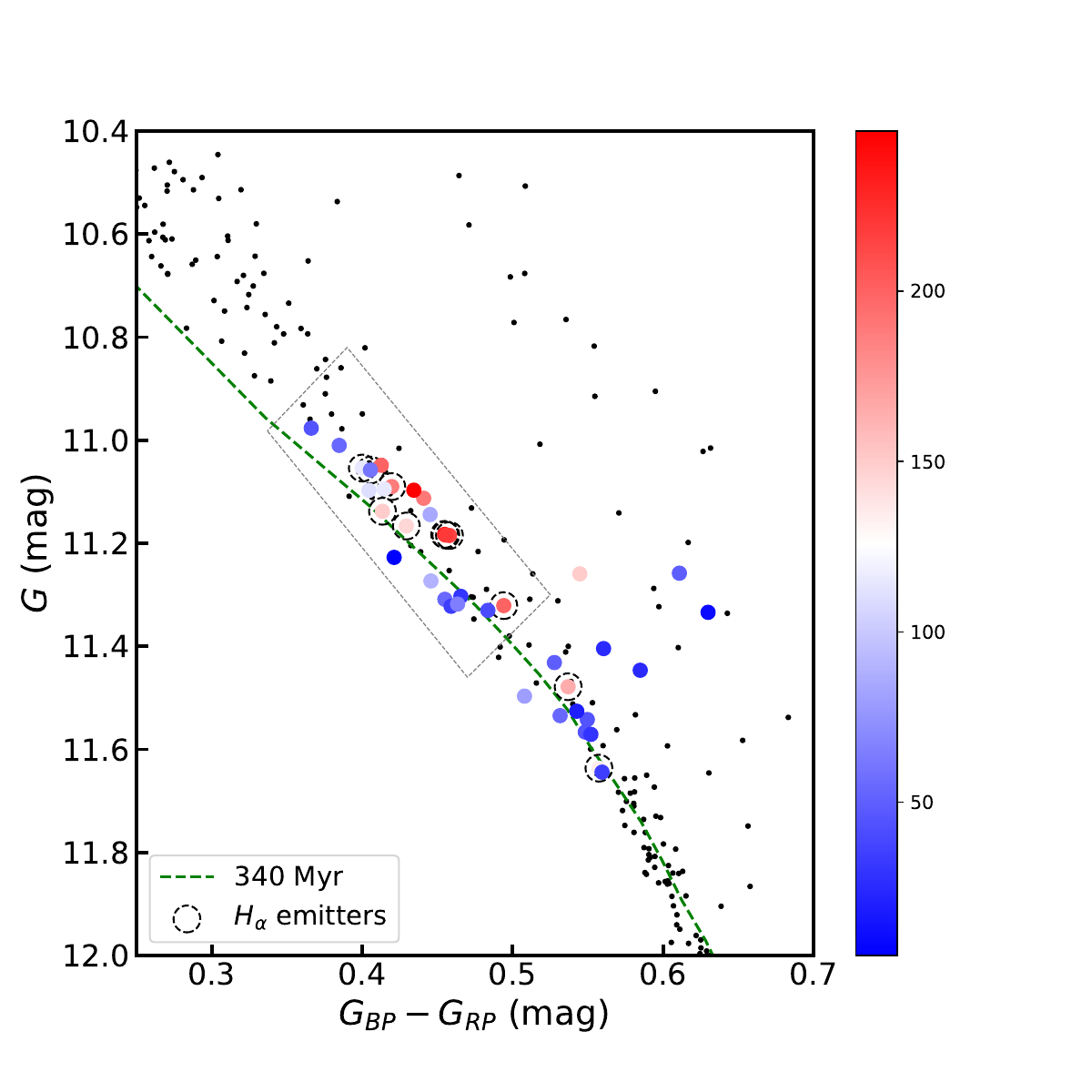}  & \includegraphics[width=0.4\textwidth]{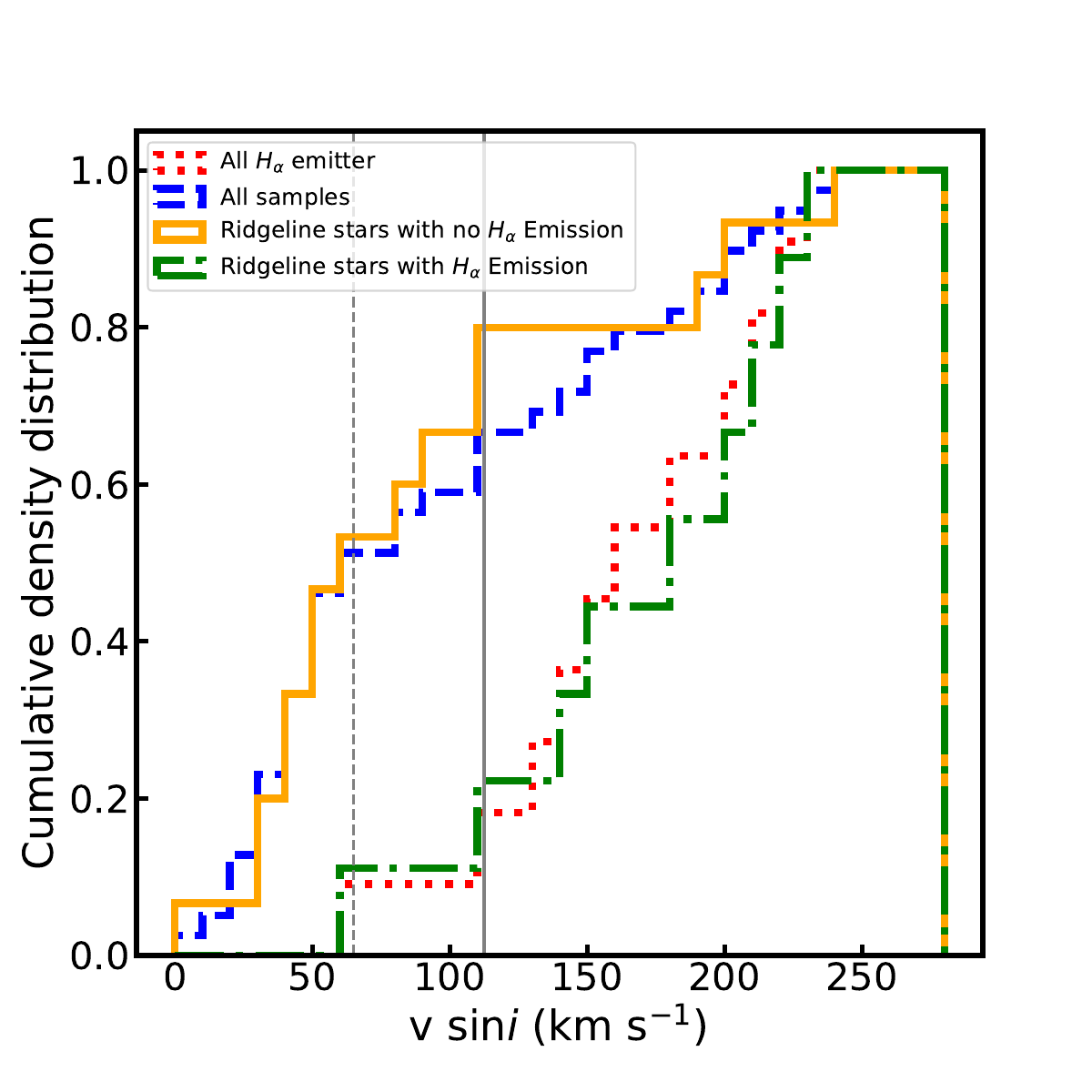}  \\
\end{tabular}
	\caption{The left panel shows the loci of stars that are spectroscopically explored (color dots), color-coded based on their $v\sin{i}$ values, and other cluster members (black dots). The stars who have $H_{\alpha}$ emission features are labeled with dashed open circles. The grey dashed parallelogram encircles the stars above the kink of eMS excluding possible unresolved binaries. The green dashed line is the best-fitting isochrone. In the right panel, the blue-dashed and red-dotted lines show the cumulative density distribution of the $v\sin{i}$ of all the 39 spectroscopic samples, and all the stars with $H_{\alpha}$ emission, respectively. The yellow-solid and the green-dash-dotted lines show the distributions for the stars without and with $H_{\alpha}$ emission within the dashed parallelogram of the left panel, respectively. The vertical dashed and solid black lines represent the median $v\sin{i}$ values of all the spectroscopic samples, and the spectroscopic samples within the dashed parallelogram of the left panel, respectively}\label{fig:spectrum samples and dust_vsini_Dist}
\end{figure*}

Table \ref{tab:3532 results} shows the measured $v\sin{i}$ of the 39 stars along with their \textit{Gaia} passband information. For each star who has spectra observed at multiple epochs, the $v\sin{i}$ listed in Table \ref{tab:3532 results} is the average value of the $v\sin{i}$ of the multiple-epoch spectra. Based on Table \ref{tab:3532 results}, the sample stars have a wide range of $v\sin{i}$ of 5--$\unit[247]{km\,s^{-1}}$. In the left panel of Figure \ref{fig:spectrum samples and dust_vsini_Dist}, we plot the loci of these spectroscopic samples color-coded based on their measured $v\sin{i}$. For stars near the MS ridgeline, stars with larger $v\sin{i}$ are located in the redder side. This is consistent with the positive correlation of $v\sin{i}$ and stellar colors observed in many young star clusters \citep[e.g.,][]{2018ApJ...863L..33M,2019ApJ...883..182S,2023MNRAS.518.1505K}.

We carefully inspected the spectra of all the spectroscopic sample stars, and found that 11 out of the 39 stars show $\mathrm{H\alpha}$ emission features. In Table \ref{tab:3532 results}, we denote these stars as Star 1--11. They account for $28.2\%$ of the whole spectroscopic sample stars. Their spectra for the $\mathrm{H\alpha}$, Fe I 6633.7{\AA} and 6678.0{\AA} lines, along with the best-fitting models for Fe I 6633.7{\AA} and Fe I 6678.0{\AA} lines are shown in Figure \ref{fig:3532 star 2 and 6}, and Figure \ref{fig:3532 star 1} to \ref{fig:3532 star 9} in the Appendix. Based on these figures, the emission features are evident compared with the best-fitting models, and are not the results of the observational errors of flux. The right panel of Figure \ref{fig:3532 Ha radec} shows their positions in the field of the cluster. In Figure \ref{fig:3532 Ha radec}, the stars with $\mathrm{H\alpha}$ emissions disperse in the cluster field. Therefore, their $\mathrm{H\alpha}$ emission features are not likely caused by the light contamination from a strong $\mathrm{H\alpha}$ emitter.

\begin{table*}
\scriptsize
\centering
\caption{Photometric data and $v\sin{i}$ of the 39 spectroscopic samples \label{tab:3532 results}}
\begin{tabular}{cccllcccll}
\hline
\hline
Gaia ID & G & $G_\mathrm{BP}-G_\mathrm{RP}$ & $v\sin i$  & labels & Gaia ID & G & $G_\mathrm{BP}-G_\mathrm{RP}$ & $v\sin i$  & labels \\
(1) & (mag) (2) & (mag) (3) & ($\mathrm{km\,s^{-1}}$) (4) & (5) & (1) & (mag) (2) & (mag) (3) & ($\mathrm{km\,s^{-1}}$) (4) & (5) \\
\hline
5338651637541468032 & 10.98 & 0.37 &  $45_{-1.1}^{+1.1}$ & -- & 5338676853241853312 & 11.45 & 0.58 &  $25_{-1.7}^{+0.2}$ & -- \\
5338659265404154880 & 11.01 & 0.38 &  $55_{-2.1}^{+0.3}$ & --& 5338657783587036032 & 11.50 & 0.51 &  $80_{-1.2}^{+1.8}$ & --\\
5340218097982590976 & 11.05 & 0.41 &  $200_{-2.1}^{+4}$ & --  & 5340160992121127168 & 11.53 & 0.54 &  $20_{-0.5}^{+0.5}$ & -- \\
5338679778169530880 & 11.09 & 0.41 &  $115_{-2.0}^{+2.4}$  & -- & 5338630162702473728 & 11.53 & 0.53 &  $55_{-1.4}^{+0.4}$ & --\\
5338674211891232640 & 11.10 & 0.43 &  $247_{-6}^{+8}$ & -- & 5338682011552542848 & 11.54 & 0.55 &  $45_{-0.4}^{+1.9}$ & --\\
5338650744188318848 & 11.10 & 0.40 &  $110_{-3}^{+1.7}$ & --& 5338709224455685888 & 11.57 & 0.55 &  $45_{-0.8}^{+1.0}$ & --\\
5338714653294510464 & 11.11 & 0.44 &  $190_{-1.9}^{+4}$ & -- & 5338627890614688384 & 11.57 & 0.55 &  $30_{-0.0}^{+2.1}$ & --\\
5340214112272818176 & 11.14 & 0.45 &  $85_{-0.4}^{+2.3}$ & --& 5340165149649780992 & 11.64 & 0.56 &  $35_{-0.2}^{+1.8}$ & --\\
5338651362663550720 & 11.23 & 0.42 &  $5_{-0.0}^{+4}$ & --& 5338652118577887616 & 11.32 & 0.49 &  $200_{-4}^{+4}$ & Star 1\\
5338709705490314624 & 11.26 & 0.54 &  $150_{-4}^{+2.6}$ & --& 5338624390266300416 & 11.18 & 0.46 &  $215_{-9}^{+6}$ & Star 2\\ 
5340164153217132928 & 11.26 & 0.61 &  $50_{-0.9}^{+0.4}$ & -- & 5338665067851545600 & 11.18 & 0.45 &  $237_{-11}^{+8}$ & Star 3$^{\mathrm{*}}$\\ 
5338715782825371008 & 11.27 & 0.45 &  $90_{-1.1}^{+5}$ & -- & 5338710083450233216 & 11.09 & 0.42 &  $188_{-5}^{+7}$  & Star 4$^{\mathrm{*}}$\\
5338702421228391936 & 11.30 & 0.47 &  $30_{-2.9}^{+0.1}$ & --& 5338661636226329216 & 11.18 & 0.46 &  $220_{-3}^{+4}$& Star 5\\ 
5338648854402511232 & 11.31 & 0.45 &  $55_{-0.8}^{+1.4}$ & --&  5338660807243993216 & 11.14 & 0.41 &  $150_{-6}^{+6}$ & Star 6\\
5338709808571831808 & 11.32 & 0.46 &  $65_{-4}^{+0.5}$ & -- & 5338635758994803456 & 11.64 & 0.56 &  $135_{-6}^{+1.6}$ & Star 7\\
5338678949185892864 & 11.32 & 0.46 &  $35_{-4}^{+0.2}$ & --& 5338652045510742016 & 11.48 & 0.54 &  $165_{-4}^{+2.7}$ & Star 8\\
5338650602401738112 & 11.33 & 0.63 &  $10_{-0.1}^{+1.2}$ & -- &  5340158586938781440 & 11.17 & 0.43 &  $145_{-1.4}^{+5}$ & Star 9\\
5340147557462156544 & 11.33 & 0.48 &  $40_{-2.4}^{+0.0}$ & -- & 5338645654599400704 & 11.05 & 0.40 &  $115_{-1.2}^{+4}$ & Star 10\\
5338720116492694272 & 11.40 & 0.56 &  $25_{-2.6}^{+0.1}$ & -- & 5340161954193782784 & 11.06 & 0.41 &  $60_{-1.7}^{+0.6}$ & Star 11\\
5340149309808437888 & 11.43 & 0.53 &  $50_{-1.1}^{+0.3}$ & --\\
\hline
\end{tabular}
\begin{flushleft}
\scriptsize
\tablenotetext{*}{$\mathrm{H\alpha}$ emitters that have spectra observed at multiple epochs.}

\tablecomments{The errors of $v \sin i$ correspond to $1\sigma$ uncertainties. (1) \textit{Gaia} ID in EDR3; (2) \textit{Gaia} $G$ magnitude; (3) Color index in \textit{Gaia} bands; (4) Projected rotational velocities; (5) Star labels; Stars with $\mathrm{H\alpha}$ emissions are labeled with 'Star 1' to 'Star 11', while other stars have no labels.}

\end{flushleft}

\end{table*}

\begin{figure*}[ht!]
\centering
\begin{tabular}{ccc}
    \includegraphics[width=0.31\textwidth]{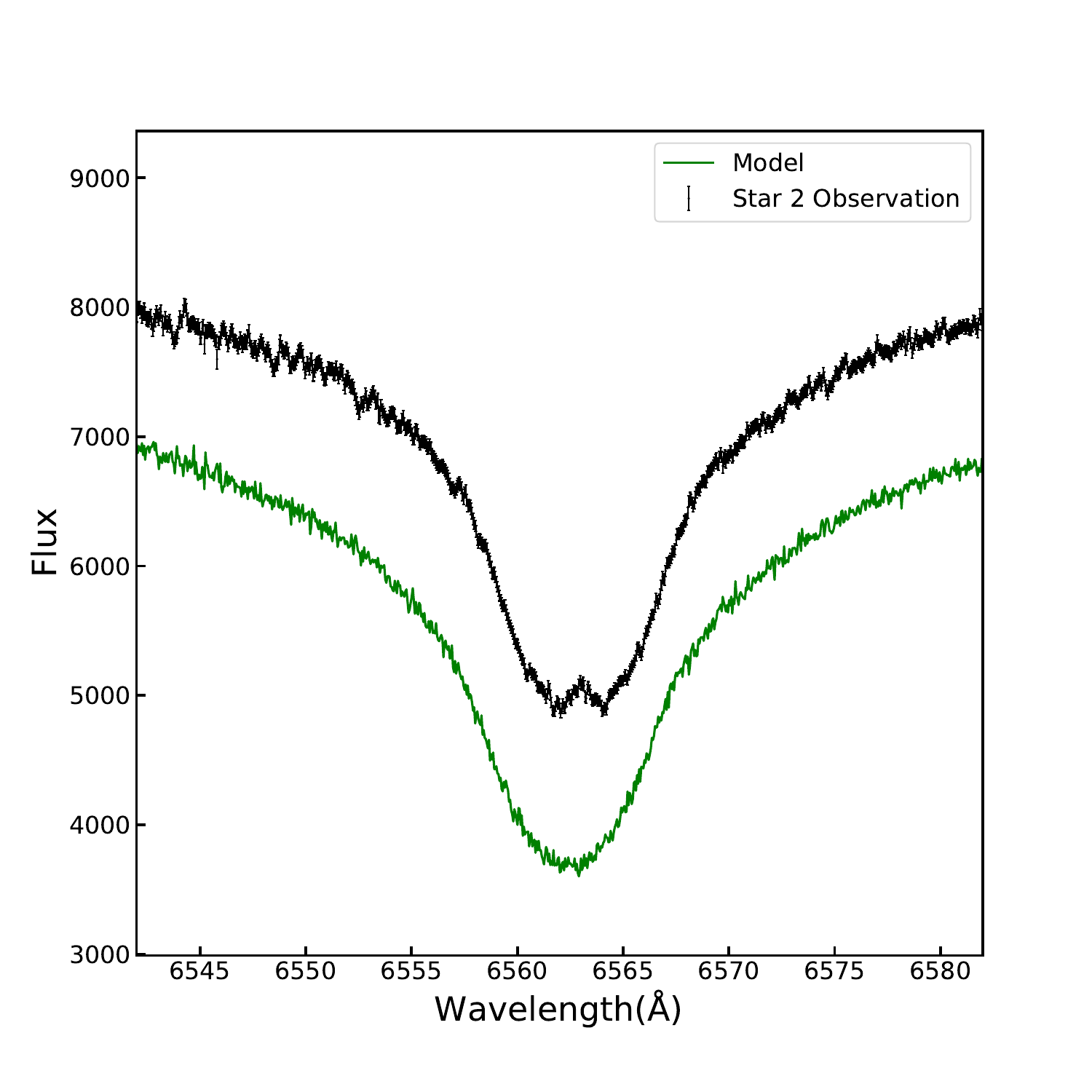}  & 
    \includegraphics[width=0.31\textwidth]{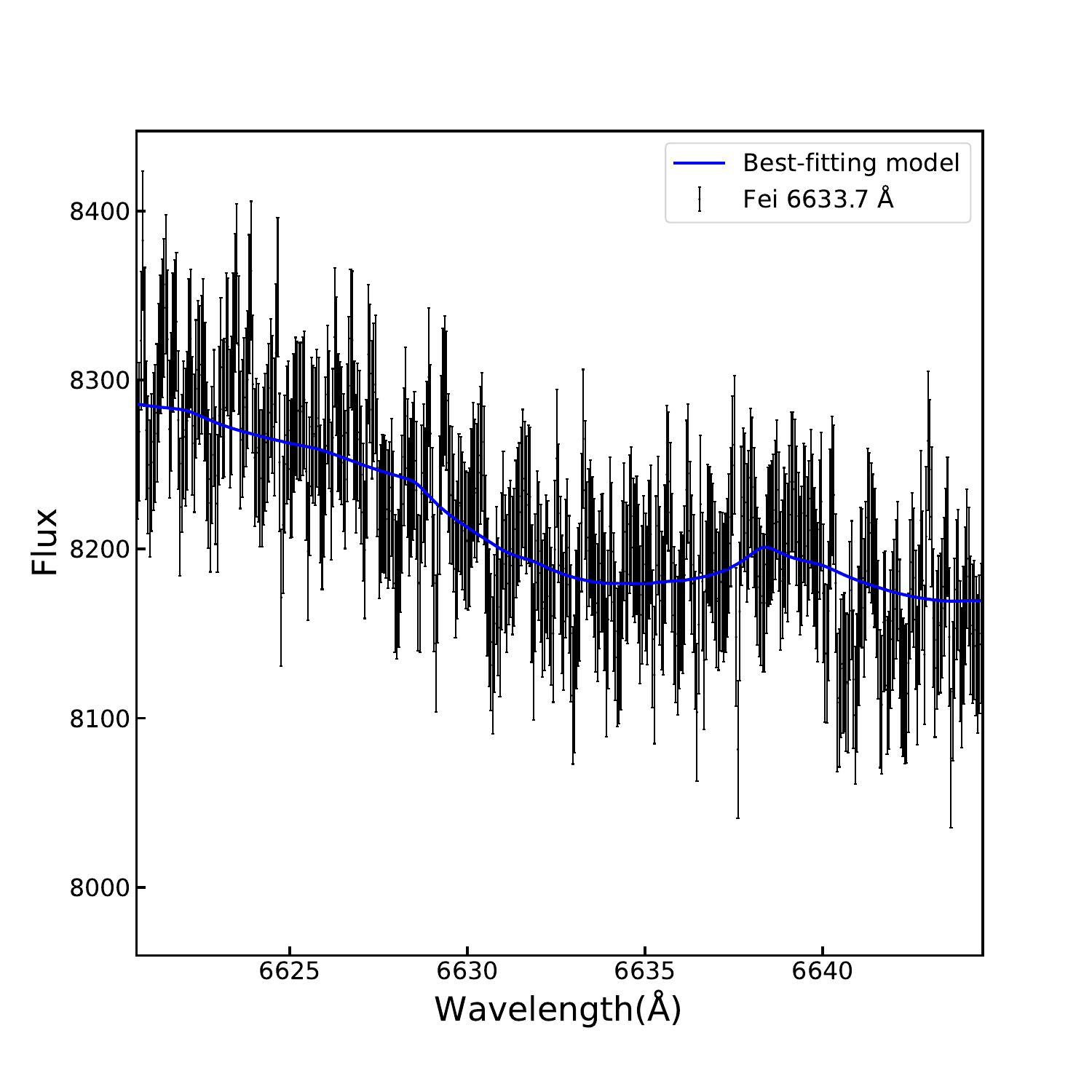}  & 
    \includegraphics[width=0.31\textwidth]{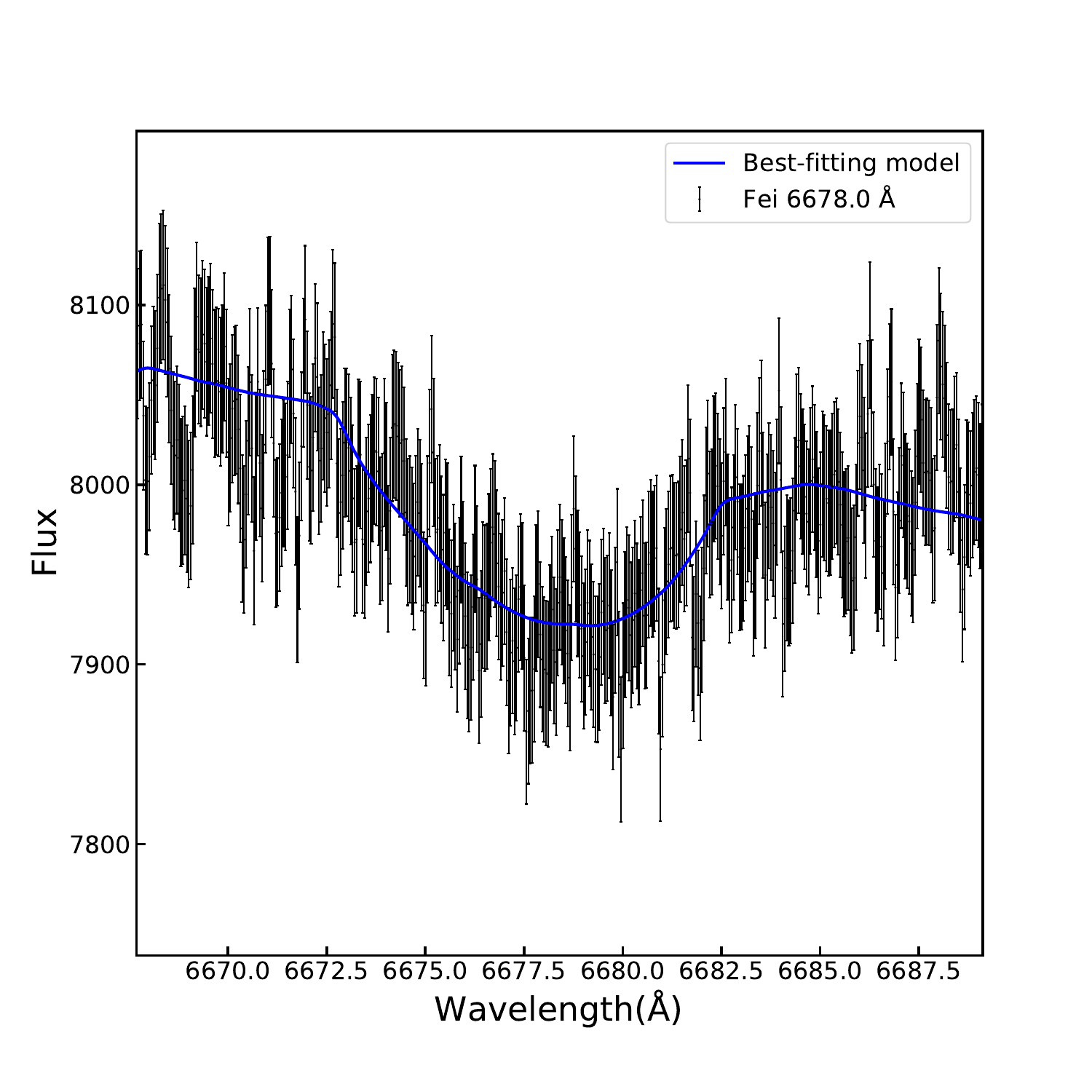}\\
    \includegraphics[width=0.31\textwidth]{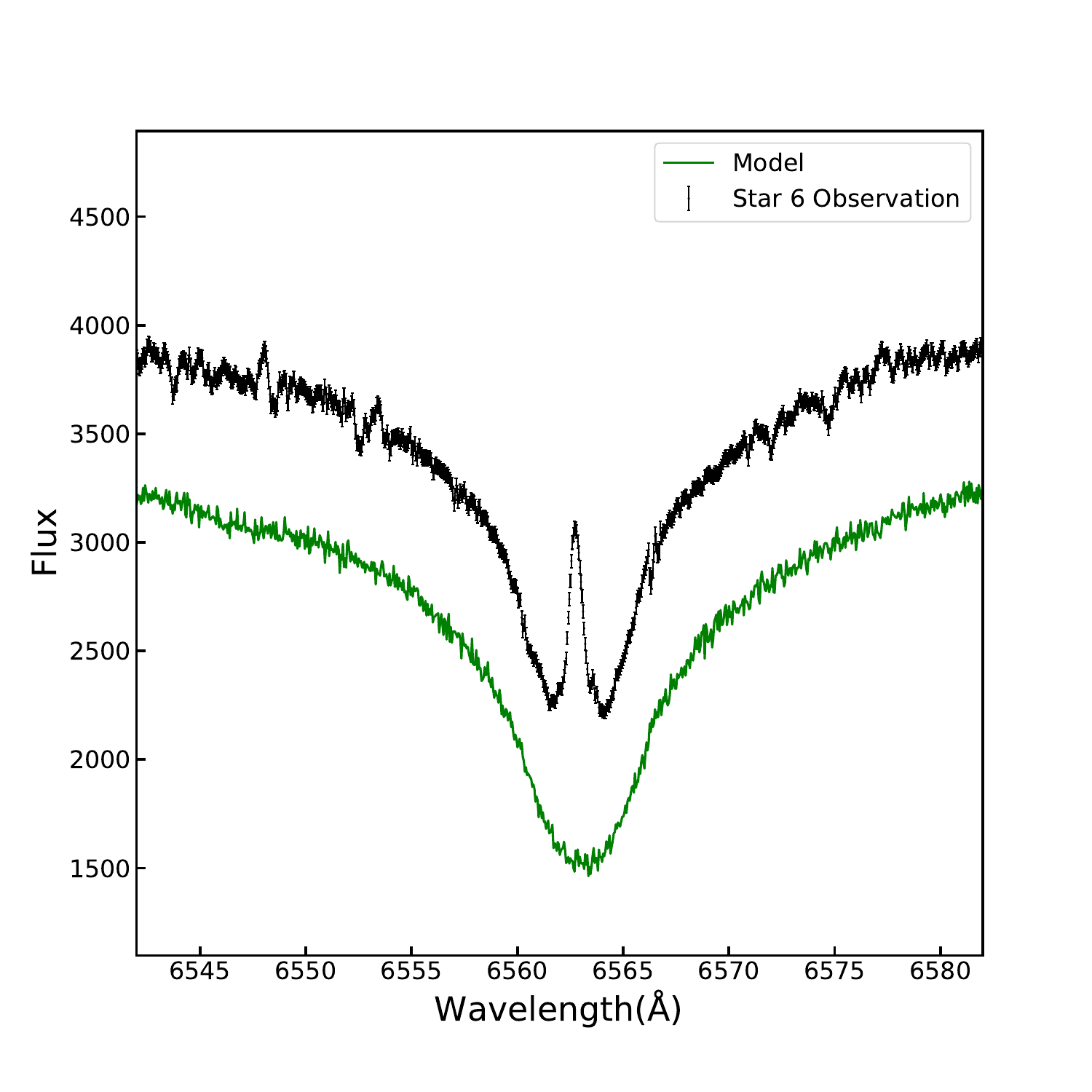}  & 
    \includegraphics[width=0.31\textwidth]{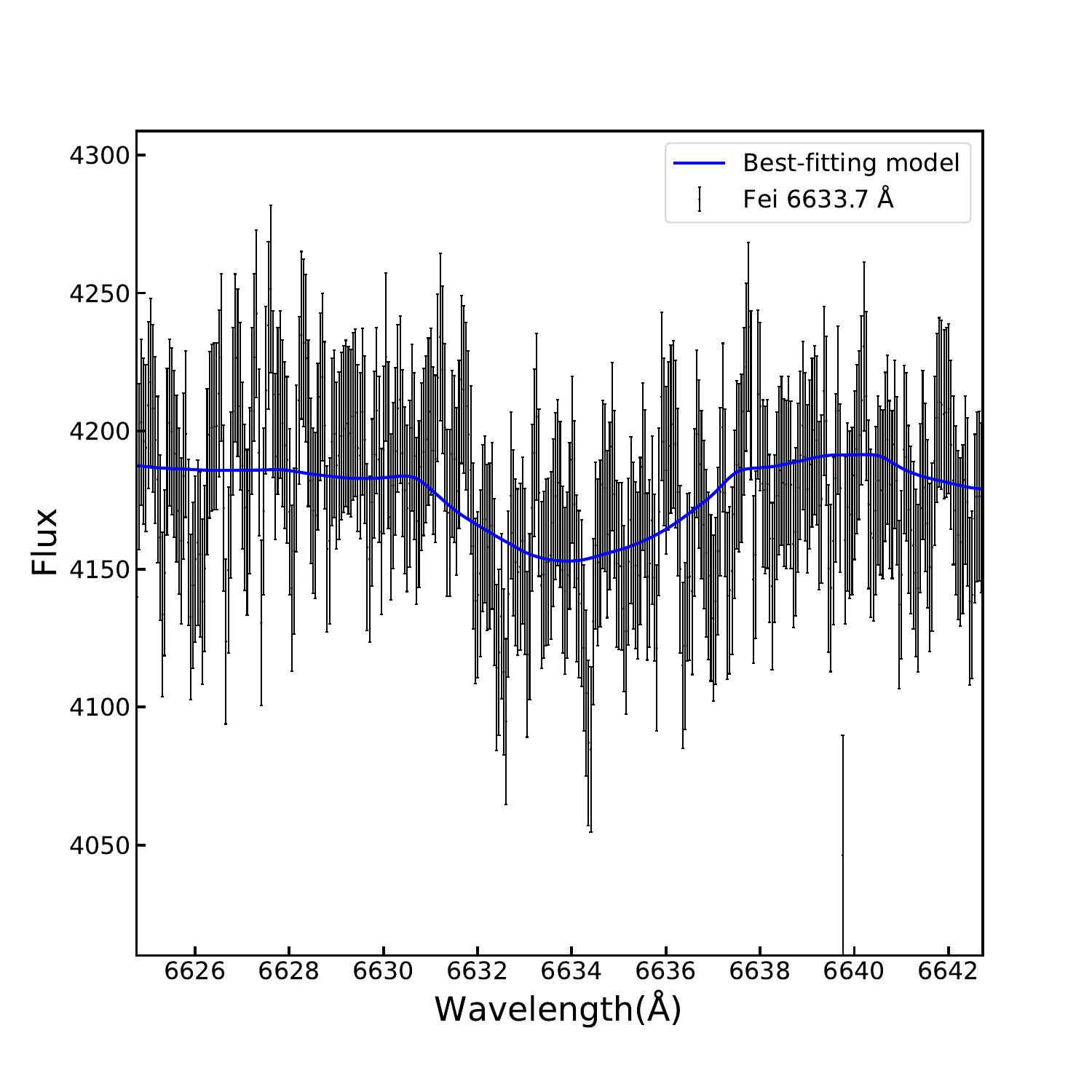 }  & 
    \includegraphics[width=0.31\textwidth]{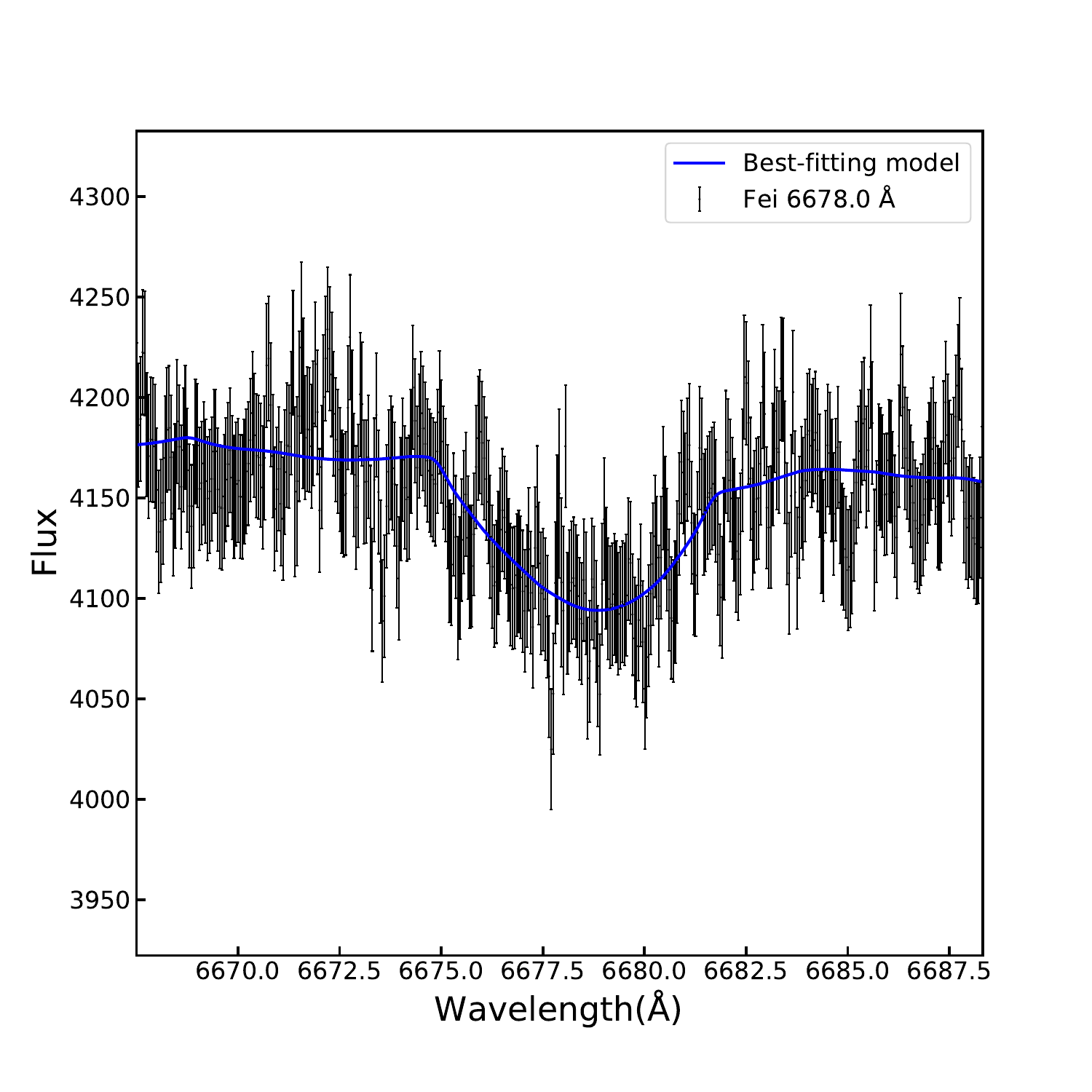}\\
\end{tabular}

	\caption{The spectra of $H_{\alpha}$, Fe I 6633.7{\AA} and Fe I 6678.0{\AA} lines of Star 2 (upper panels) and Star 6 (lower panels) in Table \ref{tab:3532 results}. The middle and right panels show the observed Fe I 6633.7{\AA} and Fe I 6678.0{\AA} profiles (black dots) with their best-fitting models (blue lines). The left panels show their observed spectra in the $H_{\alpha}$ lines (black dots), where the green lines are the model spectra generated using the best-fitting parameters for the Fe I 6633.7{\AA} and Fe I 6678.0{\AA} absorption profiles. Gaussian-distributed errors of flux corresponding to those of the observations are added in the model spectra of $H_{\alpha}$ lines. To separate the model and observed spectra in the left panels, the model spectra are shifted along the ordinate. In all the panels, the error bars show the measurement uncertainty of the flux of the observations.}
 \label{fig:3532 star 2 and 6}
\end{figure*}

\begin{figure*}[ht!]
\centering
\begin{tabular}{ccc}

    \includegraphics[width=0.3\textwidth]{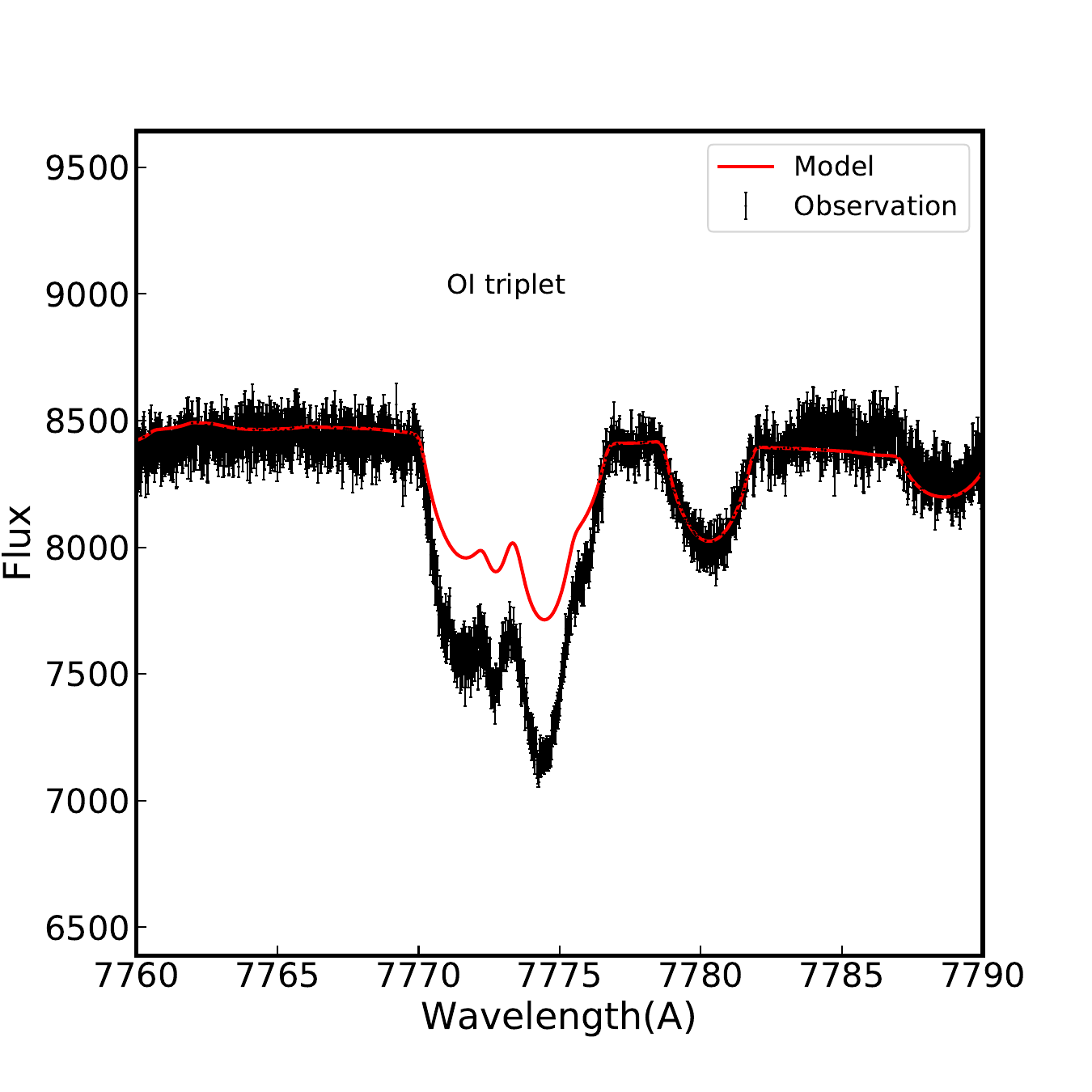}  & \includegraphics[width=0.3\textwidth]{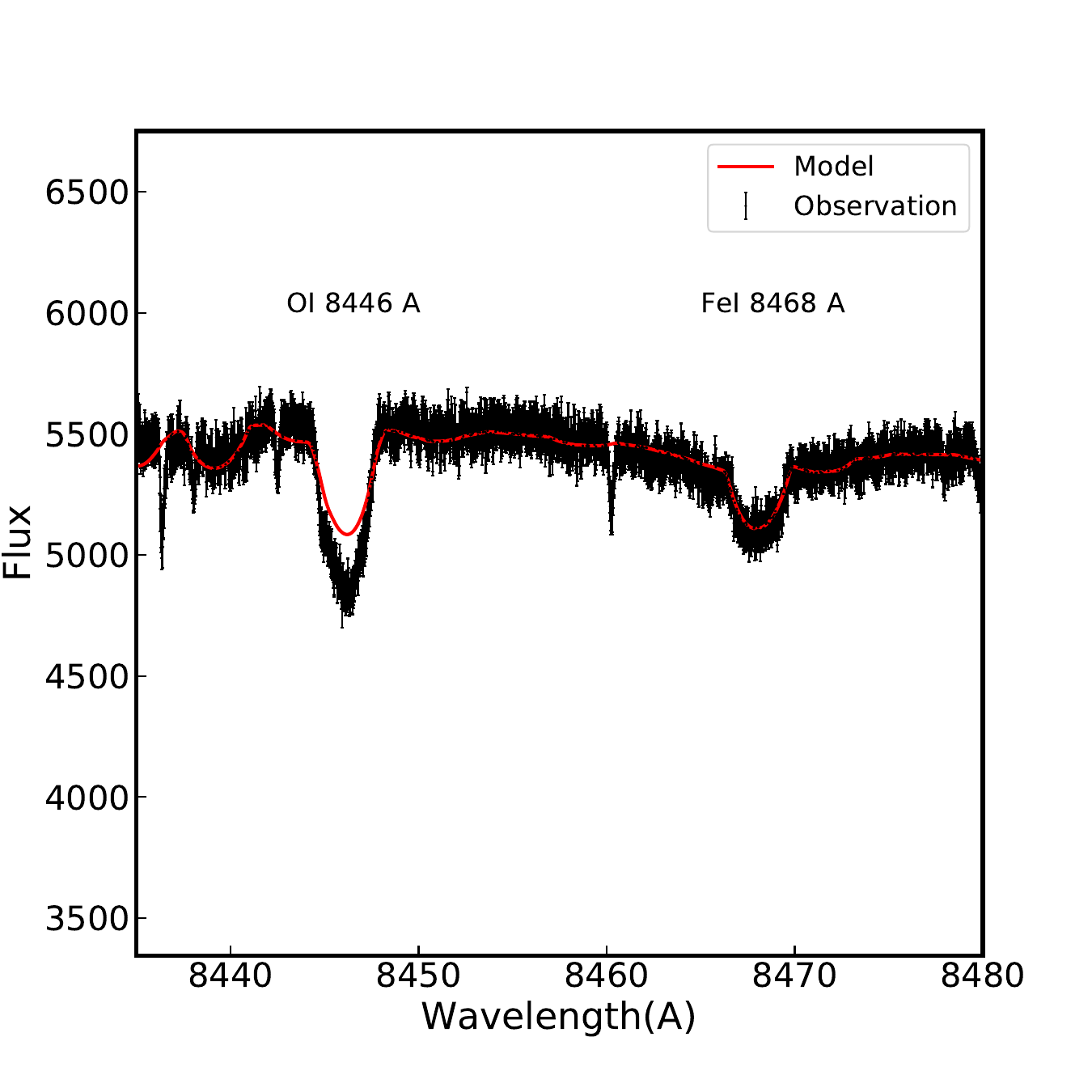} & \includegraphics[width=0.3\textwidth]{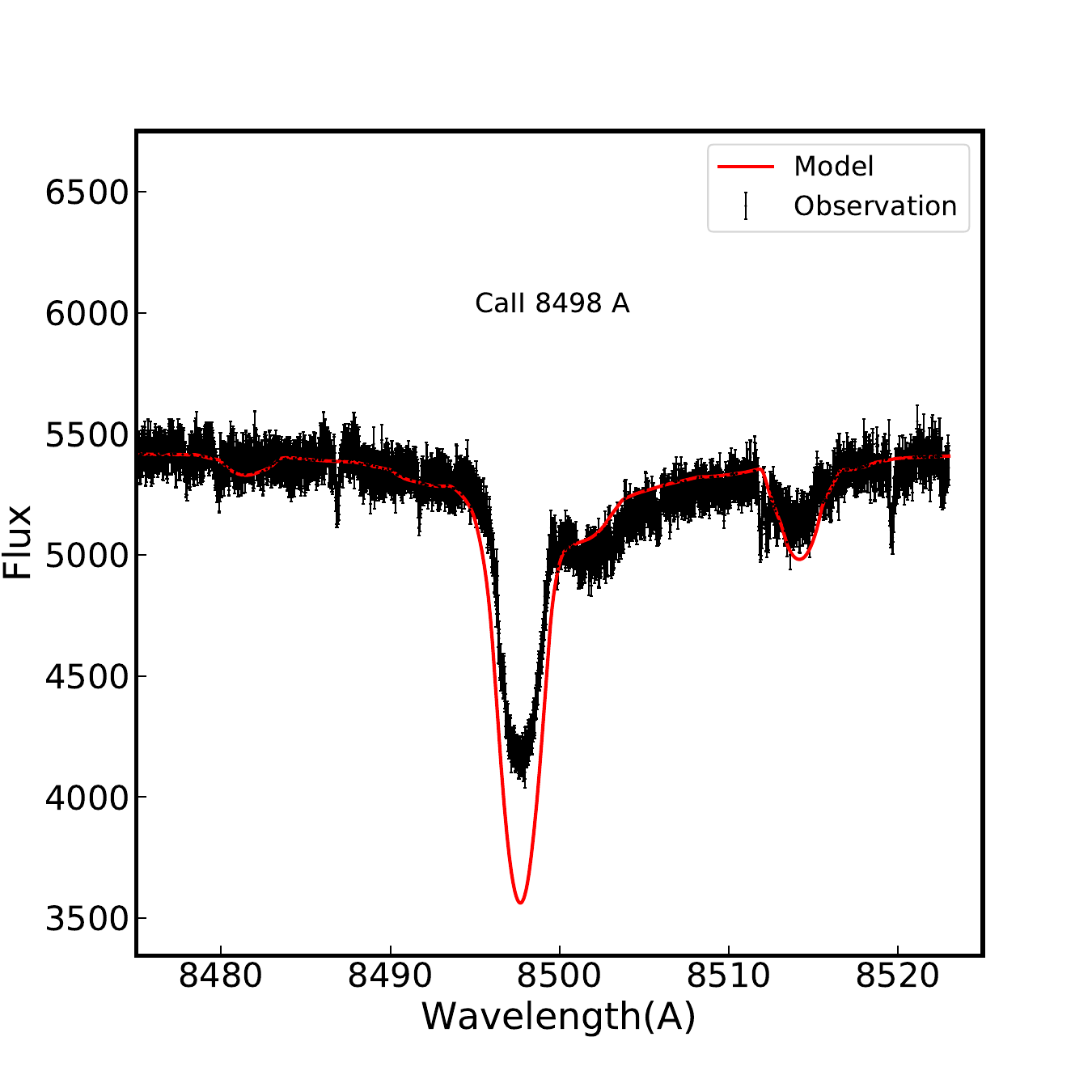}  \\
\end{tabular}
	\caption{Observed spectrum (black dots) of Star 11 in the O {\sc I} \unit[7774]{\AA} triplet (left), O {\sc I} \unit[8446]{\AA} line (middle) and Ca {\sc II} \unit[8498]{\AA} (right) lines, along with the model (red lines) generated using the best-fitting parameters for the Fe {\sc I} \unit[8498]{\AA} line shown in the middle panel. The error bars show the observation uncertainty of the flux.}\label{fig:Star 11}
\end{figure*}

The $\mathrm{H\alpha}$ emission features indicate that these stars may have gaseous material that is optically thin around them that emit $\mathrm{H\alpha}$ light due to the heating of the central stars. As shell star have additional absorption than normal stars in the O {\sc I} \unit[7774]{{\AA}} triplet (\unit[7772]{{\AA}}, \unit[7774]{\AA} and \unit[7775]{\AA}), O {\sc I} \unit[8446]{\AA} line, and Ca {\sc II} triplet (\unit[8498]{\AA}, \unit[8542]{\AA} and \unit[8662]{\AA}) due to the absorption of discs \citep{1986PASP...98..867S}, we explored whether there are shell stars in the $\mathrm{H\alpha}$ emitters through these near-infrared absorption features. In the ESO archive, only Star 11 has been spectroscopically observed in the corresponding infrared wavelengths. This spectrum was observed using the UVE spectrograph equipped on the VLT ( Program ID: 074.C-0399; PI: Jeffries), which has a wavelength range of 665--1043\,nm, $R=34540$ and SNR of $\sim 73$. Like the measurement of the $v\sin{i}$ in Section \ref{sec:Data R}, we fitted the observed absorption profile of Fe {\sc I} \unit[8498]{\AA} of this spectrum to get the best-fitting spectrum model. Figure \ref{fig:Star 11} shows the observed O {\sc I} \unit[7774]{\AA} triplet, O {\sc I} \unit[8446]{\AA} and Ca {\sc II} \unit[8498]{\AA} lines of Star 11, along with the best-fitting model for the Fe {\sc I} \unit[8498]{\AA}. The spectrum of Star 11 exhibits evident additional absorption in the O {\sc I} \unit[7774]{\AA} triplet and O {\sc I} \unit[8446]{\AA} line compared with the synthetic spectrum, indicating the appearance of a disc around it. However, this star has weaker absorption profile in the Ca {\sc II} \unit[8498]{\AA} compared with that of the model. One possible reason might be that Star 11 is a chemically peculiar star who has an underabundance of light elements such as Ca and O \citep{2022AJ....164..255C}. Chemically peculiar stars are characterised with slow rotation rates \citep{1974ARA&A..12..257P}, which the $v\sin{i}$ ($\unit[60]{km\,s^{-1}}$) of Star 11 is consistent with. If this is on the right track, the photospheric lines of Star 11 may show both weaker O {\sc I} \unit[7774]{\AA} triplet and O {\sc I} \unit[8446] absorption profiles than the model when it also has an underabundance of O. It indicates that the decreasing of the flux due to absorption of the disc might be larger than that shown in Figure \ref{fig:Star 11}. Since to test whether this star is chemically peculiar is beyond the topic of this paper, we leave it in the future work. We noted that the measured RV ($\unit[-11]{km\,s^{-1}}$) of this spectrum has large difference with that ($\unit[5]{km\,s^{-1}}$) of the spectrum measured in Section \ref{sec:SF} for the same star. This implies that Star 11 might be in a close binary. However, we did not find evident split patterns and additional absorption in other lines of both spectrum with wavelengths of 644--$\unit[682]{nm}$ and 665--1043 nm. Therefore, the contribution of flux from the companion should be small if it is a close binary system. 

To check whether these $\mathrm{H\alpha}$ emitters are fast-rotating stars, we plot the cumulative number distribution of the $v\sin{i}$ of all the spectroscopic sample stars and that of the stars with $\mathrm{H\alpha}$ emissions in the right panel of Figure \ref{fig:spectrum samples and dust_vsini_Dist}. Since the stars around the kink of the eMSs (corresponding to $\unit[11.4]{mag}<G<$\unit[11.6]{mag}), which corresponds to the mass where the magnetic braking on stellar rotation start to be effective \citep{1967ApJ...150..551K, 2018MNRAS.477.2640M}, are generally slow rotators, we also compare the $v\sin{i}$ distributions of stars with and without $\mathrm{H\alpha}$ emissions above the kink (within the grey dashed parallelogram in the left panel). Figure \ref{fig:spectrum samples and dust_vsini_Dist} shows that most (10/11) $\mathrm{H\alpha}$ emitters are fast rotators among the spectroscopic samples, with larger $v\sin{i}$ than the median $v\sin{i}$ value of the whole spectroscopic samples. For the stars above the eMS kink, 9 out of 11 $\mathrm{H\alpha}$ emitters have larger $v\sin{i}$ than the median $v\sin{i}$ value of the population, and the $v\sin{i}$ of $\sim 80\%$ of stars without $\mathrm{H\alpha}$ emissions.

\section{Discussion\label{sec:discussion}}

\subsection{$\mathrm{H\alpha}$ emissions from the sky background}
One fact that may influence the result of our work is the flux from the sky background. We note that the sky background spectra of the observation show weak $\mathrm{H\alpha}$ emissions based on the files provided by ESO archive. Although the flux from the sky background have been substracted from the final public spectra, the shapes of the $\mathrm{H\alpha}$ lines in the public spectra of single targets may be influenced, as the sky background spectra used in the substraction are the mean of those obtained for a field using multiple fibres, not designed for the neighboring of single targets \citep{2022A&A...666A.120G}. However, we argue that the light contamination from the sky background could not dominate the formation of the $\mathrm{H\alpha}$ emissions detected in this paper, as the appearance of the $\mathrm{H\alpha}$ emissions should not be strongly correlated with fast rotation (see Figure \ref{fig:spectrum samples and dust_vsini_Dist}) if they are mainly caused by background light contamination. The $\mathrm{H\alpha}$ emission features of most $\mathrm{H\alpha}$ emitters should be intrinsic.

\subsection{Implication for dust extinction scenario}
The spectrum fitting results in Section \ref{sec:R} indicate that $\sim 30\%$ A- and early F-type stars in NGC 3532 may have decretion discs which are formed due to fast rotation. As the radiation ionization of A and F stars for discs is thought to be weak \citep{2013A&ARv..21...69R}, the fraction of the A- and F-type stars who have expelled discs may be higher than $\sim 30\%$. This indicates that star clusters can harbor a significant population of fast-rotating A-type and F-type stars with excretion discs, thus, provide observational evidence to support the dust self-extinction scenario for the UV-dim stars that appear in intermediate-age clusters. NGC 1783 is a typical intermediate-age cluster where a large population of UV-dim stars was detected \citep{2023A&A...672A.161M}. The distribution of the UV-dim star was well reproduced by dusty star models \citep{2023MNRAS.521.4462D}. The masses of of the eMSTO stars of NGC 1783 is $\sim \unit[1.5]{M_\odot}$ \citep{2023MNRAS.521.4462D}, with which the masses of the $\mathrm{H\alpha}$ emitters (1.4--$\unit[1.6]{M_\odot}$) are well consistent. The $\mathrm{H\alpha}$ emitters found in NGC 3532 might be the counterparts of the UV-dim stars detected in the intermediate-age MC cluster NGC 1783, if the discs can survive more than $\unit[1]{Gyr}$. 

\subsection{Other mechanisms of excretion discs formation}
In the 39 spectroscopic sample stars, some fast rotating stars do not show $\mathrm{H\alpha}$ emissions. The Star 11 have $\mathrm{H\alpha}$ emission and the absorption profile of shell stars, meanwhile, a small $v\sin{i}$ value. As the discs of shell stars are along the line of sight, their spin axis should tend to be vertical to the line of sight, assuming that the spin axis align with the normal of the discs. Therefore, the inclination angle of the rotation axis of Star 11 should be large, implying its small $v\sin{i}$ should be caused by intrinsic slow rotation. In the young LMC cluster NGC 1850, the $v\sin{i}$ of the Be stars did not extend to higher values than normal MSTO stars, and the Be stars did not rotate at rates close to the break-up rotation velocities, implying that fast rotation might not be the exclusive mechanism to form the excretion discs \citep{2023MNRAS.518.1505K}. The shell and slowly rotating features of Star 11 also imply that other mechanism may play a role in the formation of excretion discs. 

The formation of excretion discs may be linked to stellar pulsation. With the advent of space-based photometry missions such as CoRoT \citep{CoRoT_2009A&A}, Kepler \citep{Kepler_2010Sci}, and TESS \citep{Ricker_2015}, nearly all Be stars have been found to exhibit multiperiodic light variations \citep{2013A&ARv..21...69R}, which are attributed to stellar pulsations caused by internal waves. The amplitudes of certain pulsation modes vary over time and correlate with outbursts \citep[e.g.,][]{Rivinius2003A&A, Huat2009, Goss2011MNRAS, Papics2017A&A}, supporting the hypothesis that stellar pulsations and the Be phenomenon are interconnected \citep[e.g.,][]{Baade1988}. However, whether outbursts trigger the pulsations or the pulsations expel material to form circumstellar disks remains under debate.

In particular, \cite{Huat2009} discovered that for HD\,49330, observed by CoRoT, the amplitudes of its gravity modes (restored by gravity) peaked during the outburst phase and remained weak during the quiescent phase. Conversely, the amplitudes of pressure modes (restored by pressure) showed the opposite trend. \cite{Neiner2020} successfully explained these anti-correlated amplitude variations, suggesting that the gravity modes were stochastically excited and transported angular momentum to the surface, thereby increasing the surface rotation rate and triggering the outburst. The outburst, in turn, broke the pressure mode cavity, leading to the suppression of pressure modes.

Following the simulation by \cite{Neiner2020}, we sought to identify such anti-correlated amplitude variations in NGC\,3532. Using TESS photometric data and following the steps similar to \cite{Ligang2024_NGC2516}, we conducted aperture photometry and Fourier analysis on early-type stars in NGC\,3532, which included two months of consecutive observations in 2019, one month in 2021, and another two months in 2023. While we observed some amplitude variations, we found no clear evidence of outbursts or anti-correlated amplitude variations. Data quality may be the main limiting factor, as TESS data are sporadically covered, and contamination from nearby stars’ light, as well as scattered light from Earth and the Moon, is unavoidable and significantly impacts amplitude measurements. We emphasize that continuous, high-precision photometric data (such as archived Kepler data or TESS data in the continuous viewing zone) are essential for revealing the pulsation-driven shell star phenomenon.

The high fraction of the A- and F-type $\mathrm{H\alpha}$ emitters in NGC 3532 also gives some puzzling on the fraction of A and F stars to have excretion discs, and the mechanism to drive $\mathrm{H\alpha}$ emissions in circumstellar discs. The radiation ionization for the gases in the discs of A and F stars are thought to be weak \citep{2013A&ARv..21...69R}, therefore, discs surrounding A and F type stars are more easier to be detected through the absorption features of discs in near-infrared spectra \citep[e.g.,][]{1986PASP...98..867S} when they are shell stars. Whether the high fraction of A and F stars with $\mathrm{H\alpha}$ emissions is common in young star clusters is not clear. Searching for such stars in more star clusters would be essential to test it. It would also help constrain the models of star-disc interactions of high-mass MS rotating stars, which is important for the planetary formation theory around early-type stars.

\subsection{Challenge for the dust extinction scenario}
The CMD positions of the UV-dim stars found in young MC star clusters may challenge the dust extinction scenario  \citep{2023MNRAS.524.6149M}. In these clusters, the stars with most evident UV-dim features were found to be located in the blue MSs, while blue MSs are thought to be populated by slowly rotating stars \citep{2023MNRAS.524.6149M}. Additionally, Be stars identified through their high flux in the \textit{HST} $F656N$ band \citep{2018MNRAS.477.2640M} were found to have different color distributions from those of UV-dim stars \citep[for details, please refer to this reference]{2023MNRAS.524.6149M} \footnote{It should be noted that A- and F-type $\mathrm{H_\alpha}$ emitters may not be able to be separated from other normal stars using narrow filters centered on the $\mathrm{H_\alpha}$ line like Be stars, since their $\mathrm{H_\alpha}$ emissions are generally weak. Therefore, they may not be classified into the population which have strong $\mathrm{H_\alpha}$ emissions like Be stars. They may have different color features from those of Be stars identified using narrow filters centered on the $\mathrm{H_\alpha}$ line.}. These results imply that the correlation between the UV-dim phenomenon and fast stellar rotation still needs to be explored. 

Future UV-band photometry for A- and F-type $\mathrm{H_\alpha}$ emitters, and spectroscopic measurements for UV-dim stars in young star clusters would be helpful. If most $\mathrm{H_\alpha}$ emitters are found not dim in the UV-bands, the UV-dim phenomenon may not be caused by circumstellar dust extinction. If the A- and F-type $\mathrm{H_\alpha}$ emitters are found indeed UV-dim, and the spectroscopic observations reveal that UV-dim stars in young star clusters are slow rotators with circumstellar gas, the stellar discs in young and intermediate-age clusters may be formed through different mechanisms. For young star clusters, recent intense mass-loss events and gas remnant after recent binary merging were suggested to account for the formation of UV-dim stars \citep{2023MNRAS.524.6149M}. For Star 11, the discrepancy between the wavelengths corresponding to the peak of the emission and the minimum of the $\mathrm{H_\alpha}$ absorption of the best-fitting model (see the bottom panels of Figure \ref{fig:3532 star 9}) possibly indicates a strong mass-loss event which takes a lot of angular momentum away from the star and makes the star rotate slowly. This star may be an example to form UV-dim stars through intense mass lose which results in slow rotation suggested by \cite{2023MNRAS.524.6149M}. We also note that UV-dim stars are rare for stars later than B type in those young MC clusters \citep{2023MNRAS.524.6149M}, while the intermediate-age cluster NGC 1783 were found to harbor a large population of A- and F-type UV-dim stars \citep{2023A&A...672A.161M}. It seems that the number of UV-dim stars with different spectral types may change as the clusters evolve, in particular for the stars later than B type. A- and F-type stars may lose masses more easily when they expand during MS time. In summary, the mechanisms to cause the UV-dim phenomenon in early-type stars may be complicated. The dust extinction model may be improved in the future.

\section{Conclusions\label{sec:con}}
In this paper, we report the detection of a population of A and F stars whose spectra show $\mathrm{H\alpha}$ emission features in the young Galactic star cluster NGC 3532, which account for $\sim 30\%$ of the spectroscopic samples explored in this paper. We found that most of these stars are fast rotating stars, implying that they have excretion discs that are formed due to fast rotation. Their appearance supports the scenario of dust extinction for eMSs, that a significant fraction of A and F stars in star clusters may rotate fast with expelled discs, which introduce extending of MSs and perform as UV-dim stars in photometric observations. They might be the counterparts of the UV-dim stars detected in the MC star clusters, in particular those in the intermediate-age cluster NGC 1783 \citep{2023A&A...672A.161M, 2023MNRAS.521.4462D}. Future UV photometric observations would help explore whether they are indeed dim in UV passbands, and shed more light on the star-disc models for early-type stars.

\begin{acknowledgements}
We acknowledge the anonymous referee and the editors of the American Astronomical Society Journals for their very useful comments and suggestions. This work was supported by the National Natural Science Foundation of China (NSFC) through grant 12233013. Gang Li has received funding from the KU\,Leuven Research Council (grant C16/18/005: PARADISE).

This work is Based on data obtained from the ESO Science Archive Facility with DOI: \url{https://doi.org/10.18727/archive/27} \citep{ESO_27}, \url{https://doi.org/10.18727/archive/25} \citep{ESO_25} and \url{https://doi.org/10.18727/archive/50} \citep{ESO_50}. This work has made use of data from the European Space Agency (ESA) mission {\it Gaia} (\url{https://www.cosmos.esa.int/gaia}), processed by the {\it Gaia} Data Processing and Analysis Consortium (DPAC, \url{https://www.cosmos.esa.int/web/gaia/dpac/consortium}). Funding for the DPAC has been provided by national institutions, in particular the institutions participating in the {\it Gaia} Multilateral Agreement. This research has used the POLLUX database (http://pollux.oreme.org), operated at LUPM (UniversitÃ© Montpellier--CNRS, France, with the support of the PNPS and INSU. 
\end{acknowledgements}

\software{PARSEC \citep[1.2S;][]{2017ApJ...835...77M}, Astropy
 \citep{2013A&A...558A..33A, 2018AJ....156..123A,2022ApJ...935..167A}, Matplotlib \citep{2007CSE.....9...90H}, SciPy \citep{2020SciPy-NMeth}, Pollux \citep{2010A&A...516A..13P}, SYNSPEC \citep{1992A&A...262..501H}, PyAstronomy \citep[][https://github.com/sczesla/PyAstronomy]{pya}, Astrolib PySynphot \citep{2013ascl.soft03023S}, TOPCAT \citep{2005ASPC..347...29T}}

\bibliographystyle{aasjournal}
\bibliography{paper.bib}

\begin{thebibliography}{}
\expandafter\ifx\csname natexlab\endcsname\relax\def\natexlab#1{#1}\fi
\providecommand{\url}[1]{\href{#1}{#1}}
\providecommand{\dodoi}[1]{doi:~\href{http://doi.org/#1}{\nolinkurl{#1}}}
\providecommand{\doeprint}[1]{\href{http://ascl.net/#1}{\nolinkurl{http://ascl.net/#1}}}
\providecommand{\doarXiv}[1]{\href{https://arxiv.org/abs/#1}{\nolinkurl{https://arxiv.org/abs/#1}}}

\bibitem[{{Astropy Collaboration} {et~al.}(2013){Astropy Collaboration}, {Robitaille}, {Tollerud}, {Greenfield}, {Droettboom}, {Bray}, {Aldcroft}, {Davis}, {Ginsburg}, {Price-Whelan}, {Kerzendorf}, {Conley}, {Crighton}, {Barbary}, {Muna}, {Ferguson}, {Grollier}, {Parikh}, {Nair}, {Unther}, {Deil}, {Woillez}, {Conseil}, {Kramer}, {Turner}, {Singer}, {Fox}, {Weaver}, {Zabalza}, {Edwards}, {Azalee Bostroem}, {Burke}, {Casey}, {Crawford}, {Dencheva}, {Ely}, {Jenness}, {Labrie}, {Lim}, {Pierfederici}, {Pontzen}, {Ptak}, {Refsdal}, {Servillat}, \& {Streicher}}]{2013A&A...558A..33A}
{Astropy Collaboration}, {Robitaille}, T.~P., {Tollerud}, E.~J., {et~al.} 2013, \aap, 558, A33, \dodoi{10.1051/0004-6361/201322068}

\bibitem[{{Astropy Collaboration} {et~al.}(2018){Astropy Collaboration}, {Price-Whelan}, {Sip{\H{o}}cz}, {G{\"u}nther}, {Lim}, {Crawford}, {Conseil}, {Shupe}, {Craig}, {Dencheva}, {Ginsburg}, {VanderPlas}, {Bradley}, {P{\'e}rez-Su{\'a}rez}, {de Val-Borro}, {Aldcroft}, {Cruz}, {Robitaille}, {Tollerud}, {Ardelean}, {Babej}, {Bach}, {Bachetti}, {Bakanov}, {Bamford}, {Barentsen}, {Barmby}, {Baumbach}, {Berry}, {Biscani}, {Boquien}, {Bostroem}, {Bouma}, {Brammer}, {Bray}, {Breytenbach}, {Buddelmeijer}, {Burke}, {Calderone}, {Cano Rodr{\'\i}guez}, {Cara}, {Cardoso}, {Cheedella}, {Copin}, {Corrales}, {Crichton}, {D'Avella}, {Deil}, {Depagne}, {Dietrich}, {Donath}, {Droettboom}, {Earl}, {Erben}, {Fabbro}, {Ferreira}, {Finethy}, {Fox}, {Garrison}, {Gibbons}, {Goldstein}, {Gommers}, {Greco}, {Greenfield}, {Groener}, {Grollier}, {Hagen}, {Hirst}, {Homeier}, {Horton}, {Hosseinzadeh}, {Hu}, {Hunkeler}, {Ivezi{\'c}}, {Jain}, {Jenness}, {Kanarek}, {Kendrew}, {Kern}, {Kerzendorf}, {Khvalko}, {King}, {Kirkby}, {Kulkarni},
  {Kumar}, {Lee}, {Lenz}, {Littlefair}, {Ma}, {Macleod}, {Mastropietro}, {McCully}, {Montagnac}, {Morris}, {Mueller}, {Mumford}, {Muna}, {Murphy}, {Nelson}, {Nguyen}, {Ninan}, {N{\"o}the}, {Ogaz}, {Oh}, {Parejko}, {Parley}, {Pascual}, {Patil}, {Patil}, {Plunkett}, {Prochaska}, {Rastogi}, {Reddy Janga}, {Sabater}, {Sakurikar}, {Seifert}, {Sherbert}, {Sherwood-Taylor}, {Shih}, {Sick}, {Silbiger}, {Singanamalla}, {Singer}, {Sladen}, {Sooley}, {Sornarajah}, {Streicher}, {Teuben}, {Thomas}, {Tremblay}, {Turner}, {Terr{\'o}n}, {van Kerkwijk}, {de la Vega}, {Watkins}, {Weaver}, {Whitmore}, {Woillez}, {Zabalza}, \& {Astropy Contributors}}]{2018AJ....156..123A}
{Astropy Collaboration}, {Price-Whelan}, A.~M., {Sip{\H{o}}cz}, B.~M., {et~al.} 2018, \aj, 156, 123, \dodoi{10.3847/1538-3881/aabc4f}

\bibitem[{{Astropy Collaboration} {et~al.}(2022){Astropy Collaboration}, {Price-Whelan}, {Lim}, {Earl}, {Starkman}, {Bradley}, {Shupe}, {Patil}, {Corrales}, {Brasseur}, {N{\"o}the}, {Donath}, {Tollerud}, {Morris}, {Ginsburg}, {Vaher}, {Weaver}, {Tocknell}, {Jamieson}, {van Kerkwijk}, {Robitaille}, {Merry}, {Bachetti}, {G{\"u}nther}, {Aldcroft}, {Alvarado-Montes}, {Archibald}, {B{\'o}di}, {Bapat}, {Barentsen}, {Baz{\'a}n}, {Biswas}, {Boquien}, {Burke}, {Cara}, {Cara}, {Conroy}, {Conseil}, {Craig}, {Cross}, {Cruz}, {D'Eugenio}, {Dencheva}, {Devillepoix}, {Dietrich}, {Eigenbrot}, {Erben}, {Ferreira}, {Foreman-Mackey}, {Fox}, {Freij}, {Garg}, {Geda}, {Glattly}, {Gondhalekar}, {Gordon}, {Grant}, {Greenfield}, {Groener}, {Guest}, {Gurovich}, {Handberg}, {Hart}, {Hatfield-Dodds}, {Homeier}, {Hosseinzadeh}, {Jenness}, {Jones}, {Joseph}, {Kalmbach}, {Karamehmetoglu}, {Ka{\l}uszy{\'n}ski}, {Kelley}, {Kern}, {Kerzendorf}, {Koch}, {Kulumani}, {Lee}, {Ly}, {Ma}, {MacBride}, {Maljaars}, {Muna}, {Murphy}, {Norman},
  {O'Steen}, {Oman}, {Pacifici}, {Pascual}, {Pascual-Granado}, {Patil}, {Perren}, {Pickering}, {Rastogi}, {Roulston}, {Ryan}, {Rykoff}, {Sabater}, {Sakurikar}, {Salgado}, {Sanghi}, {Saunders}, {Savchenko}, {Schwardt}, {Seifert-Eckert}, {Shih}, {Jain}, {Shukla}, {Sick}, {Simpson}, {Singanamalla}, {Singer}, {Singhal}, {Sinha}, {Sip{\H{o}}cz}, {Spitler}, {Stansby}, {Streicher}, {{\v{S}}umak}, {Swinbank}, {Taranu}, {Tewary}, {Tremblay}, {de Val-Borro}, {Van Kooten}, {Vasovi{\'c}}, {Verma}, {de Miranda Cardoso}, {Williams}, {Wilson}, {Winkel}, {Wood-Vasey}, {Xue}, {Yoachim}, {Zhang}, {Zonca}, \& {Astropy Project Contributors}}]{2022ApJ...935..167A}
{Astropy Collaboration}, {Price-Whelan}, A.~M., {Lim}, P.~L., {et~al.} 2022, \apj, 935, 167, \dodoi{10.3847/1538-4357/ac7c74}

\bibitem[{{Auvergne} {et~al.}(2009){Auvergne}, {Bodin}, {Boisnard}, {Buey}, {Chaintreuil}, {Epstein}, {Jouret}, {Lam-Trong}, {Levacher}, {Magnan}, {Perez}, {Plasson}, {Plesseria}, {Peter}, {Steller}, {Tiph{\`e}ne}, {Baglin}, {Agogu{\'e}}, {Appourchaux}, {Barbet}, {Beaufort}, {Bellenger}, {Berlin}, {Bernardi}, {Blouin}, {Boumier}, {Bonneau}, {Briet}, {Butler}, {Cautain}, {Chiavassa}, {Costes}, {Cuvilho}, {Cunha-Parro}, {de Oliveira Fialho}, {Decaudin}, {Defise}, {Djalal}, {Docclo}, {Drummond}, {Dupuis}, {Exil}, {Faur{\'e}}, {Gaboriaud}, {Gamet}, {Gavalda}, {Grolleau}, {Gueguen}, {Guivarc'h}, {Guterman}, {Hasiba}, {Huntzinger}, {Hustaix}, {Imbert}, {Jeanville}, {Johlander}, {Jorda}, {Journoud}, {Karioty}, {Kerjean}, {Lafond}, {Lapeyrere}, {Landiech}, {Larqu{\'e}}, {Laudet}, {Le Merrer}, {Leporati}, {Leruyet}, {Levieuge}, {Llebaria}, {Martin}, {Mazy}, {Mesnager}, {Michel}, {Moalic}, {Monjoin}, {Naudet}, {Neukirchner}, {Nguyen-Kim}, {Ollivier}, {Orcesi}, {Ottacher}, {Oulali}, {Parisot}, {Perruchot}, {Piacentino},
  {Pinheiro da Silva}, {Platzer}, {Pontet}, {Pradines}, {Quentin}, {Rohbeck}, {Rolland}, {Rollenhagen}, {Romagnan}, {Russ}, {Samadi}, {Schmidt}, {Schwartz}, {Sebbag}, {Smit}, {Sunter}, {Tello}, {Toulouse}, {Ulmer}, {Vandermarcq}, {Vergnault}, {Wallner}, {Waultier}, \& {Zanatta}}]{CoRoT_2009A&A}
{Auvergne}, M., {Bodin}, P., {Boisnard}, L., {et~al.} 2009, \aap, 506, 411, \dodoi{10.1051/0004-6361/200810860}

\bibitem[{{Avni}(1976)}]{1976ApJ...210..642A}
{Avni}, Y. 1976, \apj, 210, 642, \dodoi{10.1086/154870}

\bibitem[{{Baade}(1988)}]{Baade1988}
{Baade}, D. 1988, in IAU Symposium, Vol. 132, The Impact of Very High S/N Spectroscopy on Stellar Physics, ed. G.~{Cayrel de Strobel} \& M.~{Spite}, 217

\bibitem[{{Bastian} \& {de Mink}(2009)}]{2009MNRAS.398L..11B}
{Bastian}, N., \& {de Mink}, S.~E. 2009, \mnras, 398, L11, \dodoi{10.1111/j.1745-3933.2009.00696.x}

\bibitem[{{Bastian} {et~al.}(2020){Bastian}, {Kamann}, {Amard}, {Charbonnel}, {Haemmerl{\'e}}, \& {Matt}}]{2020MNRAS.495.1978B}
{Bastian}, N., {Kamann}, S., {Amard}, L., {et~al.} 2020, \mnras, 495, 1978, \dodoi{10.1093/mnras/staa1332}

\bibitem[{{Bastian} {et~al.}(2017){Bastian}, {Cabrera-Ziri}, {Niederhofer}, {de Mink}, {Georgy}, {Baade}, {Correnti}, {Usher}, \& {Romaniello}}]{2017MNRAS.465.4795B}
{Bastian}, N., {Cabrera-Ziri}, I., {Niederhofer}, F., {et~al.} 2017, \mnras, 465, 4795, \dodoi{10.1093/mnras/stw3042}

\bibitem[{{Borucki} {et~al.}(2010){Borucki}, {Koch}, {Basri}, {Batalha}, {Brown}, {Caldwell}, {Caldwell}, {Christensen-Dalsgaard}, {Cochran}, {DeVore}, {Dunham}, {Dupree}, {Gautier}, {Geary}, {Gilliland}, {Gould}, {Howell}, {Jenkins}, {Kondo}, {Latham}, {Marcy}, {Meibom}, {Kjeldsen}, {Lissauer}, {Monet}, {Morrison}, {Sasselov}, {Tarter}, {Boss}, {Brownlee}, {Owen}, {Buzasi}, {Charbonneau}, {Doyle}, {Fortney}, {Ford}, {Holman}, {Seager}, {Steffen}, {Welsh}, {Rowe}, {Anderson}, {Buchhave}, {Ciardi}, {Walkowicz}, {Sherry}, {Horch}, {Isaacson}, {Everett}, {Fischer}, {Torres}, {Johnson}, {Endl}, {MacQueen}, {Bryson}, {Dotson}, {Haas}, {Kolodziejczak}, {Van Cleve}, {Chandrasekaran}, {Twicken}, {Quintana}, {Clarke}, {Allen}, {Li}, {Wu}, {Tenenbaum}, {Verner}, {Bruhweiler}, {Barnes}, \& {Prsa}}]{Kepler_2010Sci}
{Borucki}, W.~J., {Koch}, D., {Basri}, G., {et~al.} 2010, Science, 327, 977, \dodoi{10.1126/science.1185402}

\bibitem[{{Bu} {et~al.}(2024{\natexlab{a}}){Bu}, {He}, {Fang}, \& {Li}}]{2024arXiv241200520B}
{Bu}, Y., {He}, C., {Fang}, M., \& {Li}, C. 2024{\natexlab{a}}, arXiv e-prints, arXiv:2412.00520, \dodoi{10.48550/arXiv.2412.00520}

\bibitem[{{Bu} {et~al.}(2024{\natexlab{b}}){Bu}, {He}, {Wang}, {Lin}, \& {Li}}]{2024ApJ...968...22B}
{Bu}, Y., {He}, C., {Wang}, L., {Lin}, J., \& {Li}, C. 2024{\natexlab{b}}, \apj, 968, 22, \dodoi{10.3847/1538-4357/ad3e6e}

\bibitem[{{Cardelli} {et~al.}(1989){Cardelli}, {Clayton}, \& {Mathis}}]{1989ApJ...345..245C}
{Cardelli}, J.~A., {Clayton}, G.~C., \& {Mathis}, J.~S. 1989, \apj, 345, 245, \dodoi{10.1086/167900}

\bibitem[{{Casamiquela} {et~al.}(2022){Casamiquela}, {Gebran}, {Ag{\"u}eros}, {Bouy}, \& {Soubiran}}]{2022AJ....164..255C}
{Casamiquela}, L., {Gebran}, M., {Ag{\"u}eros}, M.~A., {Bouy}, H., \& {Soubiran}, C. 2022, \aj, 164, 255, \dodoi{10.3847/1538-3881/ac9c56}

\bibitem[{{Cordoni} {et~al.}(2018){Cordoni}, {Milone}, {Marino}, {Di Criscienzo}, {D'Antona}, {Dotter}, {Lagioia}, \& {Tailo}}]{2018ApJ...869..139C}
{Cordoni}, G., {Milone}, A.~P., {Marino}, A.~F., {et~al.} 2018, \apj, 869, 139, \dodoi{10.3847/1538-4357/aaedc1}

\bibitem[{{Correnti} {et~al.}(2017){Correnti}, {Goudfrooij}, {Bellini}, {Kalirai}, \& {Puzia}}]{2017MNRAS.467.3628C}
{Correnti}, M., {Goudfrooij}, P., {Bellini}, A., {Kalirai}, J.~S., \& {Puzia}, T.~H. 2017, \mnras, 467, 3628, \dodoi{10.1093/mnras/stx010}

\bibitem[{{Czesla} {et~al.}(2019){Czesla}, {Schr{\"o}ter}, {Schneider}, {Huber}, {Pfeifer}, {Andreasen}, \& {Zechmeister}}]{pya}
{Czesla}, S., {Schr{\"o}ter}, S., {Schneider}, C.~P., {et~al.} 2019, {PyA: Python astronomy-related packages}.
\newblock \doeprint{1906.010}

\bibitem[{{D'Antona} {et~al.}(2015){D'Antona}, {Di Criscienzo}, {Decressin}, {Milone}, {Vesperini}, \& {Ventura}}]{2015MNRAS.453.2637D}
{D'Antona}, F., {Di Criscienzo}, M., {Decressin}, T., {et~al.} 2015, \mnras, 453, 2637, \dodoi{10.1093/mnras/stv1794}

\bibitem[{{D'Antona} {et~al.}(2017){D'Antona}, {Milone}, {Tailo}, {Ventura}, {Vesperini}, \& {di Criscienzo}}]{2017NatAs...1E.186D}
{D'Antona}, F., {Milone}, A.~P., {Tailo}, M., {et~al.} 2017, Nature Astronomy, 1, 0186, \dodoi{10.1038/s41550-017-0186}

\bibitem[{{D'Antona} {et~al.}(2023){D'Antona}, {Dell'Agli}, {Tailo}, {Milone}, {Ventura}, {Vesperini}, {Cordoni}, {Dotter}, \& {Marino}}]{2023MNRAS.521.4462D}
{D'Antona}, F., {Dell'Agli}, F., {Tailo}, M., {et~al.} 2023, \mnras, 521, 4462, \dodoi{10.1093/mnras/stad851}

\bibitem[{{de Grijs} {et~al.}(2013){de Grijs}, {Li}, {Zheng}, {Deng}, {Hu}, {Kouwenhoven}, \& {Wicker}}]{2013ApJ...765....4D}
{de Grijs}, R., {Li}, C., {Zheng}, Y., {et~al.} 2013, \apj, 765, 4, \dodoi{10.1088/0004-637X/765/1/4}

\bibitem[{{Dufton} {et~al.}(2013){Dufton}, {Langer}, {Dunstall}, {Evans}, {Brott}, {de Mink}, {Howarth}, {Kennedy}, {McEvoy}, {Potter}, {Ram{\'\i}rez-Agudelo}, {Sana}, {Sim{\'o}n-D{\'\i}az}, {Taylor}, \& {Vink}}]{2013A&A...550A.109D}
{Dufton}, P.~L., {Langer}, N., {Dunstall}, P.~R., {et~al.} 2013, \aap, 550, A109, \dodoi{10.1051/0004-6361/201220273}

\bibitem[{{Espinosa Lara} \& {Rieutord}(2011)}]{2011A&A...533A..43E}
{Espinosa Lara}, F., \& {Rieutord}, M. 2011, \aap, 533, A43, \dodoi{10.1051/0004-6361/201117252}

\bibitem[{{Fuller}(2021)}]{Fuller2021_inverse_tides}
{Fuller}, J. 2021, \mnras, 501, 483, \dodoi{10.1093/mnras/staa3636}

\bibitem[{{Gaia Collaboration} {et~al.}(2016){Gaia Collaboration}, {Prusti}, {de Bruijne}, {Brown}, {Vallenari}, {Babusiaux}, {Bailer-Jones}, {Bastian}, {Biermann}, {Evans}, {Eyer}, {Jansen}, {Jordi}, {Klioner}, {Lammers}, {Lindegren}, {Luri}, {Mignard}, {Milligan}, {Panem}, {Poinsignon}, {Pourbaix}, {Randich}, {Sarri}, {Sartoretti}, {Siddiqui}, {Soubiran}, {Valette}, {van Leeuwen}, {Walton}, {Aerts}, {Arenou}, {Cropper}, {Drimmel}, {H{\o}g}, {Katz}, {Lattanzi}, {O'Mullane}, {Grebel}, {Holland}, {Huc}, {Passot}, {Bramante}, {Cacciari}, {Casta{\~n}eda}, {Chaoul}, {Cheek}, {De Angeli}, {Fabricius}, {Guerra}, {Hern{\'a}ndez}, {Jean-Antoine-Piccolo}, {Masana}, {Messineo}, {Mowlavi}, {Nienartowicz}, {Ord{\'o}{\~n}ez-Blanco}, {Panuzzo}, {Portell}, {Richards}, {Riello}, {Seabroke}, {Tanga}, {Th{\'e}venin}, {Torra}, {Els}, {Gracia-Abril}, {Comoretto}, {Garcia-Reinaldos}, {Lock}, {Mercier}, {Altmann}, {Andrae}, {Astraatmadja}, {Bellas-Velidis}, {Benson}, {Berthier}, {Blomme}, {Busso}, {Carry}, {Cellino}, {Clementini},
  {Cowell}, {Creevey}, {Cuypers}, {Davidson}, {De Ridder}, {de Torres}, {Delchambre}, {Dell'Oro}, {Ducourant}, {Fr{\'e}mat}, {Garc{\'\i}a-Torres}, {Gosset}, {Halbwachs}, {Hambly}, {Harrison}, {Hauser}, {Hestroffer}, {Hodgkin}, {Huckle}, {Hutton}, {Jasniewicz}, {Jordan}, {Kontizas}, {Korn}, {Lanzafame}, {Manteiga}, {Moitinho}, {Muinonen}, {Osinde}, {Pancino}, {Pauwels}, {Petit}, {Recio-Blanco}, {Robin}, {Sarro}, {Siopis}, {Smith}, {Smith}, {Sozzetti}, {Thuillot}, {van Reeven}, {Viala}, {Abbas}, {Abreu Aramburu}, {Accart}, {Aguado}, {Allan}, {Allasia}, {Altavilla}, {{\'A}lvarez}, {Alves}, {Anderson}, {Andrei}, {Anglada Varela}, {Antiche}, {Antoja}, {Ant{\'o}n}, {Arcay}, {Atzei}, {Ayache}, {Bach}, {Baker}, {Balaguer-N{\'u}{\~n}ez}, {Barache}, {Barata}, {Barbier}, {Barblan}, {Baroni}, {Barrado y Navascu{\'e}s}, {Barros}, {Barstow}, {Becciani}, {Bellazzini}, {Bellei}, {Bello Garc{\'\i}a}, {Belokurov}, {Bendjoya}, {Berihuete}, {Bianchi}, {Bienaym{\'e}}, {Billebaud}, {Blagorodnova}, {Blanco-Cuaresma}, {Boch},
  {Bombrun}, {Borrachero}, {Bouquillon}, {Bourda}, {Bouy}, {Bragaglia}, {Breddels}, {Brouillet}, {Br{\"u}semeister}, {Bucciarelli}, {Budnik}, {Burgess}, {Burgon}, {Burlacu}, {Busonero}, {Buzzi}, {Caffau}, {Cambras}, {Campbell}, {Cancelliere}, {Cantat-Gaudin}, {Carlucci}, {Carrasco}, {Castellani}, {Charlot}, {Charnas}, {Charvet}, {Chassat}, {Chiavassa}, {Clotet}, {Cocozza}, {Collins}, {Collins}, {Costigan}, {Crifo}, {Cross}, {Crosta}, {Crowley}, {Dafonte}, {Damerdji}, {Dapergolas}, {David}, {David}, {De Cat}, {de Felice}, {de Laverny}, {De Luise}, {De March}, {de Martino}, {de Souza}, {Debosscher}, {del Pozo}, {Delbo}, {Delgado}, {Delgado}, {di Marco}, {Di Matteo}, {Diakite}, {Distefano}, {Dolding}, {Dos Anjos}, {Drazinos}, {Dur{\'a}n}, {Dzigan}, {Ecale}, {Edvardsson}, {Enke}, {Erdmann}, {Escolar}, {Espina}, {Evans}, {Eynard Bontemps}, {Fabre}, {Fabrizio}, {Faigler}, {Falc{\~a}o}, {Farr{\`a}s Casas}, {Faye}, {Federici}, {Fedorets}, {Fern{\'a}ndez-Hern{\'a}ndez}, {Fernique}, {Fienga}, {Figueras}, {Filippi},
  {Findeisen}, {Fonti}, {Fouesneau}, {Fraile}, {Fraser}, {Fuchs}, {Furnell}, {Gai}, {Galleti}, {Galluccio}, {Garabato}, {Garc{\'\i}a-Sedano}, {Gar{\'e}}, {Garofalo}, {Garralda}, {Gavras}, {Gerssen}, {Geyer}, {Gilmore}, {Girona}, {Giuffrida}, {Gomes}, {Gonz{\'a}lez-Marcos}, {Gonz{\'a}lez-N{\'u}{\~n}ez}, {Gonz{\'a}lez-Vidal}, {Granvik}, {Guerrier}, {Guillout}, {Guiraud}, {G{\'u}rpide}, {Guti{\'e}rrez-S{\'a}nchez}, {Guy}, {Haigron}, {Hatzidimitriou}, {Haywood}, {Heiter}, {Helmi}, {Hobbs}, {Hofmann}, {Holl}, {Holland}, {Hunt}, {Hypki}, {Icardi}, {Irwin}, {Jevardat de Fombelle}, {Jofr{\'e}}, {Jonker}, {Jorissen}, {Julbe}, {Karampelas}, {Kochoska}, {Kohley}, {Kolenberg}, {Kontizas}, {Koposov}, {Kordopatis}, {Koubsky}, {Kowalczyk}, {Krone-Martins}, {Kudryashova}, {Kull}, {Bachchan}, {Lacoste-Seris}, {Lanza}, {Lavigne}, {Le Poncin-Lafitte}, {Lebreton}, {Lebzelter}, {Leccia}, {Leclerc}, {Lecoeur-Taibi}, {Lemaitre}, {Lenhardt}, {Leroux}, {Liao}, {Licata}, {Lindstr{\o}m}, {Lister}, {Livanou}, {Lobel}, {L{\"o}ffler},
  {L{\'o}pez}, {Lopez-Lozano}, {Lorenz}, {Loureiro}, {MacDonald}, {Magalh{\~a}es Fernandes}, {Managau}, {Mann}, {Mantelet}, {Marchal}, {Marchant}, {Marconi}, {Marie}, {Marinoni}, {Marrese}, {Marschalk{\'o}}, {Marshall}, {Mart{\'\i}n-Fleitas}, {Martino}, {Mary}, {Matijevi{\v{c}}}, {Mazeh}, {McMillan}, {Messina}, {Mestre}, {Michalik}, {Millar}, {Miranda}, {Molina}, {Molinaro}, {Molinaro}, {Moln{\'a}r}, {Moniez}, {Montegriffo}, {Monteiro}, {Mor}, {Mora}, {Morbidelli}, {Morel}, {Morgenthaler}, {Morley}, {Morris}, {Mulone}, {Muraveva}, {Musella}, {Narbonne}, {Nelemans}, {Nicastro}, {Noval}, {Ord{\'e}novic}, {Ordieres-Mer{\'e}}, {Osborne}, {Pagani}, {Pagano}, {Pailler}, {Palacin}, {Palaversa}, {Parsons}, {Paulsen}, {Pecoraro}, {Pedrosa}, {Pentik{\"a}inen}, {Pereira}, {Pichon}, {Piersimoni}, {Pineau}, {Plachy}, {Plum}, {Poujoulet}, {Pr{\v{s}}a}, {Pulone}, {Ragaini}, {Rago}, {Rambaux}, {Ramos-Lerate}, {Ranalli}, {Rauw}, {Read}, {Regibo}, {Renk}, {Reyl{\'e}}, {Ribeiro}, {Rimoldini}, {Ripepi}, {Riva}, {Rixon},
  {Roelens}, {Romero-G{\'o}mez}, {Rowell}, {Royer}, {Rudolph}, {Ruiz-Dern}, {Sadowski}, {Sagrist{\`a} Sell{\'e}s}, {Sahlmann}, {Salgado}, {Salguero}, {Sarasso}, {Savietto}, {Schnorhk}, {Schultheis}, {Sciacca}, {Segol}, {Segovia}, {Segransan}, {Serpell}, {Shih}, {Smareglia}, {Smart}, {Smith}, {Solano}, {Solitro}, {Sordo}, {Soria Nieto}, {Souchay}, {Spagna}, {Spoto}, {Stampa}, {Steele}, {Steidelm{\"u}ller}, {Stephenson}, {Stoev}, {Suess}, {S{\"u}veges}, {Surdej}, {Szabados}, {Szegedi-Elek}, {Tapiador}, {Taris}, {Tauran}, {Taylor}, {Teixeira}, {Terrett}, {Tingley}, {Trager}, {Turon}, {Ulla}, {Utrilla}, {Valentini}, {van Elteren}, {Van Hemelryck}, {van Leeuwen}, {Varadi}, {Vecchiato}, {Veljanoski}, {Via}, {Vicente}, {Vogt}, {Voss}, {Votruba}, {Voutsinas}, {Walmsley}, {Weiler}, {Weingrill}, {Werner}, {Wevers}, {Whitehead}, {Wyrzykowski}, {Yoldas}, {{\v{Z}}erjal}, {Zucker}, {Zurbach}, {Zwitter}, {Alecu}, {Allen}, {Allende Prieto}, {Amorim}, {Anglada-Escud{\'e}}, {Arsenijevic}, {Azaz}, {Balm}, {Beck}, {Bernstein},
  {Bigot}, {Bijaoui}, {Blasco}, {Bonfigli}, {Bono}, {Boudreault}, {Bressan}, {Brown}, {Brunet}, {Bunclark}, {Buonanno}, {Butkevich}, {Carret}, {Carrion}, {Chemin}, {Ch{\'e}reau}, {Corcione}, {Darmigny}, {de Boer}, {de Teodoro}, {de Zeeuw}, {Delle Luche}, {Domingues}, {Dubath}, {Fodor}, {Fr{\'e}zouls}, {Fries}, {Fustes}, {Fyfe}, {Gallardo}, {Gallegos}, {Gardiol}, {Gebran}, {Gomboc}, {G{\'o}mez}, {Grux}, {Gueguen}, {Heyrovsky}, {Hoar}, {Iannicola}, {Isasi Parache}, {Janotto}, {Joliet}, {Jonckheere}, {Keil}, {Kim}, {Klagyivik}, {Klar}, {Knude}, {Kochukhov}, {Kolka}, {Kos}, {Kutka}, {Lainey}, {LeBouquin}, {Liu}, {Loreggia}, {Makarov}, {Marseille}, {Martayan}, {Martinez-Rubi}, {Massart}, {Meynadier}, {Mignot}, {Munari}, {Nguyen}, {Nordlander}, {Ocvirk}, {O'Flaherty}, {Olias Sanz}, {Ortiz}, {Osorio}, {Oszkiewicz}, {Ouzounis}, {Palmer}, {Park}, {Pasquato}, {Peltzer}, {Peralta}, {P{\'e}turaud}, {Pieniluoma}, {Pigozzi}, {Poels}, {Prat}, {Prod'homme}, {Raison}, {Rebordao}, {Risquez}, {Rocca-Volmerange}, {Rosen},
  {Ruiz-Fuertes}, {Russo}, {Sembay}, {Serraller Vizcaino}, {Short}, {Siebert}, {Silva}, {Sinachopoulos}, {Slezak}, {Soffel}, {Sosnowska}, {Strai{\v{z}}ys}, {ter Linden}, {Terrell}, {Theil}, {Tiede}, {Troisi}, {Tsalmantza}, {Tur}, {Vaccari}, {Vachier}, {Valles}, {Van Hamme}, {Veltz}, {Virtanen}, {Wallut}, {Wichmann}, {Wilkinson}, {Ziaeepour}, \& {Zschocke}}]{2016A&A...595A...1G}
{Gaia Collaboration}, {Prusti}, T., {de Bruijne}, J.~H.~J., {et~al.} 2016, \aap, 595, A1, \dodoi{10.1051/0004-6361/201629272}

\bibitem[{{Gaia Collaboration} {et~al.}(2018){Gaia Collaboration}, {Babusiaux}, {van Leeuwen}, {Barstow}, {Jordi}, {Vallenari}, {Bossini}, {Bressan}, {Cantat-Gaudin}, {van Leeuwen}, {Brown}, {Prusti}, {de Bruijne}, {Bailer-Jones}, {Biermann}, {Evans}, {Eyer}, {Jansen}, {Klioner}, {Lammers}, {Lindegren}, {Luri}, {Mignard}, {Panem}, {Pourbaix}, {Randich}, {Sartoretti}, {Siddiqui}, {Soubiran}, {Walton}, {Arenou}, {Bastian}, {Cropper}, {Drimmel}, {Katz}, {Lattanzi}, {Bakker}, {Cacciari}, {Casta{\~n}eda}, {Chaoul}, {Cheek}, {De Angeli}, {Fabricius}, {Guerra}, {Holl}, {Masana}, {Messineo}, {Mowlavi}, {Nienartowicz}, {Panuzzo}, {Portell}, {Riello}, {Seabroke}, {Tanga}, {Th{\'e}venin}, {Gracia-Abril}, {Comoretto}, {Garcia-Reinaldos}, {Teyssier}, {Altmann}, {Andrae}, {Audard}, {Bellas-Velidis}, {Benson}, {Berthier}, {Blomme}, {Burgess}, {Busso}, {Carry}, {Cellino}, {Clementini}, {Clotet}, {Creevey}, {Davidson}, {De Ridder}, {Delchambre}, {Dell'Oro}, {Ducourant}, {Fern{\'a}ndez-Hern{\'a}ndez}, {Fouesneau},
  {Fr{\'e}mat}, {Galluccio}, {Garc{\'\i}a-Torres}, {Gonz{\'a}lez-N{\'u}{\~n}ez}, {Gonz{\'a}lez-Vidal}, {Gosset}, {Guy}, {Halbwachs}, {Hambly}, {Harrison}, {Hern{\'a}ndez}, {Hestroffer}, {Hodgkin}, {Hutton}, {Jasniewicz}, {Jean-Antoine-Piccolo}, {Jordan}, {Korn}, {Krone-Martins}, {Lanzafame}, {Lebzelter}, {L{\"o}ffler}, {Manteiga}, {Marrese}, {Mart{\'\i}n-Fleitas}, {Moitinho}, {Mora}, {Muinonen}, {Osinde}, {Pancino}, {Pauwels}, {Petit}, {Recio-Blanco}, {Richards}, {Rimoldini}, {Robin}, {Sarro}, {Siopis}, {Smith}, {Sozzetti}, {S{\"u}veges}, {Torra}, {van Reeven}, {Abbas}, {Abreu Aramburu}, {Accart}, {Aerts}, {Altavilla}, {{\'A}lvarez}, {Alvarez}, {Alves}, {Anderson}, {Andrei}, {Anglada Varela}, {Antiche}, {Antoja}, {Arcay}, {Astraatmadja}, {Bach}, {Baker}, {Balaguer-N{\'u}{\~n}ez}, {Balm}, {Barache}, {Barata}, {Barbato}, {Barblan}, {Barklem}, {Barrado}, {Barros}, {Bartholom{\'e} Mu{\~n}oz}, {Bassilana}, {Becciani}, {Bellazzini}, {Berihuete}, {Bertone}, {Bianchi}, {Bienaym{\'e}}, {Blanco-Cuaresma}, {Boch},
  {Boeche}, {Bombrun}, {Borrachero}, {Bouquillon}, {Bourda}, {Bragaglia}, {Bramante}, {Breddels}, {Brouillet}, {Br{\"u}semeister}, {Brugaletta}, {Bucciarelli}, {Burlacu}, {Busonero}, {Butkevich}, {Buzzi}, {Caffau}, {Cancelliere}, {Cannizzaro}, {Carballo}, {Carlucci}, {Carrasco}, {Casamiquela}, {Castellani}, {Castro-Ginard}, {Charlot}, {Chemin}, {Chiavassa}, {Cocozza}, {Costigan}, {Cowell}, {Crifo}, {Crosta}, {Crowley}, {Cuypers}, {Dafonte}, {Damerdji}, {Dapergolas}, {David}, {David}, {de Laverny}, {De Luise}, {De March}, {de Martino}, {de Souza}, {de Torres}, {Debosscher}, {del Pozo}, {Delbo}, {Delgado}, {Delgado}, {Diakite}, {Diener}, {Distefano}, {Dolding}, {Drazinos}, {Dur{\'a}n}, {Edvardsson}, {Enke}, {Eriksson}, {Esquej}, {Eynard Bontemps}, {Fabre}, {Fabrizio}, {Faigler}, {Falc{\~a}o}, {Farr{\`a}s Casas}, {Federici}, {Fedorets}, {Fernique}, {Figueras}, {Filippi}, {Findeisen}, {Fonti}, {Fraile}, {Fraser}, {Fr{\'e}zouls}, {Gai}, {Galleti}, {Garabato}, {Garc{\'\i}a-Sedano}, {Garofalo}, {Garralda}, {Gavel},
  {Gavras}, {Gerssen}, {Geyer}, {Giacobbe}, {Gilmore}, {Girona}, {Giuffrida}, {Glass}, {Gomes}, {Granvik}, {Gueguen}, {Guerrier}, {Guiraud}, {Guti{\'e}}, {Haigron}, {Hatzidimitriou}, {Hauser}, {Haywood}, {Heiter}, {Helmi}, {Heu}, {Hilger}, {Hobbs}, {Hofmann}, {Holland}, {Huckle}, {Hypki}, {Icardi}, {Jan{\ss}en}, {Jevardat de Fombelle}, {Jonker}, {Juh{\'a}sz}, {Julbe}, {Karampelas}, {Kewley}, {Klar}, {Kochoska}, {Kohley}, {Kolenberg}, {Kontizas}, {Kontizas}, {Koposov}, {Kordopatis}, {Kostrzewa-Rutkowska}, {Koubsky}, {Lambert}, {Lanza}, {Lasne}, {Lavigne}, {Le Fustec}, {Le Poncin-Lafitte}, {Lebreton}, {Leccia}, {Leclerc}, {Lecoeur-Taibi}, {Lenhardt}, {Leroux}, {Liao}, {Licata}, {Lindstr{\o}m}, {Lister}, {Livanou}, {Lobel}, {L{\'o}pez}, {Managau}, {Mann}, {Mantelet}, {Marchal}, {Marchant}, {Marconi}, {Marinoni}, {Marschalk{\'o}}, {Marshall}, {Martino}, {Marton}, {Mary}, {Massari}, {Matijevi{\v{c}}}, {Mazeh}, {McMillan}, {Messina}, {Michalik}, {Millar}, {Molina}, {Molinaro}, {Moln{\'a}r}, {Montegriffo}, {Mor},
  {Morbidelli}, {Morel}, {Morris}, {Mulone}, {Muraveva}, {Musella}, {Nelemans}, {Nicastro}, {Noval}, {O'Mullane}, {Ord{\'e}novic}, {Ord{\'o}{\~n}ez-Blanco}, {Osborne}, {Pagani}, {Pagano}, {Pailler}, {Palacin}, {Palaversa}, {Panahi}, {Pawlak}, {Piersimoni}, {Pineau}, {Plachy}, {Plum}, {Poggio}, {Poujoulet}, {Pr{\v{s}}a}, {Pulone}, {Racero}, {Ragaini}, {Rambaux}, {Ramos-Lerate}, {Regibo}, {Reyl{\'e}}, {Riclet}, {Ripepi}, {Riva}, {Rivard}, {Rixon}, {Roegiers}, {Roelens}, {Romero-G{\'o}mez}, {Rowell}, {Royer}, {Ruiz-Dern}, {Sadowski}, {Sagrist{\`a} Sell{\'e}s}, {Sahlmann}, {Salgado}, {Salguero}, {Sanna}, {Santana-Ros}, {Sarasso}, {Savietto}, {Schultheis}, {Sciacca}, {Segol}, {Segovia}, {S{\'e}gransan}, {Shih}, {Siltala}, {Silva}, {Smart}, {Smith}, {Solano}, {Solitro}, {Sordo}, {Soria Nieto}, {Souchay}, {Spagna}, {Spoto}, {Stampa}, {Steele}, {Steidelm{\"u}ller}, {Stephenson}, {Stoev}, {Suess}, {Surdej}, {Szabados}, {Szegedi-Elek}, {Tapiador}, {Taris}, {Tauran}, {Taylor}, {Teixeira}, {Terrett}, {Teyssandier},
  {Thuillot}, {Titarenko}, {Torra Clotet}, {Turon}, {Ulla}, {Utrilla}, {Uzzi}, {Vaillant}, {Valentini}, {Valette}, {van Elteren}, {Van Hemelryck}, {Vaschetto}, {Vecchiato}, {Veljanoski}, {Viala}, {Vicente}, {Vogt}, {von Essen}, {Voss}, {Votruba}, {Voutsinas}, {Walmsley}, {Weiler}, {Wertz}, {Wevers}, {Wyrzykowski}, {Yoldas}, {{\v{Z}}erjal}, {Ziaeepour}, {Zorec}, {Zschocke}, {Zucker}, {Zurbach}, \& {Zwitter}}]{2018A&A...616A..10G}
{Gaia Collaboration}, {Babusiaux}, C., {van Leeuwen}, F., {et~al.} 2018, \aap, 616, A10, \dodoi{10.1051/0004-6361/201832843}

\bibitem[{{Gaia Collaboration} {et~al.}(2021){Gaia Collaboration}, {Brown}, {Vallenari}, {Prusti}, {de Bruijne}, {Babusiaux}, {Biermann}, {Creevey}, {Evans}, {Eyer}, {Hutton}, {Jansen}, {Jordi}, {Klioner}, {Lammers}, {Lindegren}, {Luri}, {Mignard}, {Panem}, {Pourbaix}, {Randich}, {Sartoretti}, {Soubiran}, {Walton}, {Arenou}, {Bailer-Jones}, {Bastian}, {Cropper}, {Drimmel}, {Katz}, {Lattanzi}, {van Leeuwen}, {Bakker}, {Cacciari}, {Casta{\~n}eda}, {De Angeli}, {Ducourant}, {Fabricius}, {Fouesneau}, {Fr{\'e}mat}, {Guerra}, {Guerrier}, {Guiraud}, {Jean-Antoine Piccolo}, {Masana}, {Messineo}, {Mowlavi}, {Nicolas}, {Nienartowicz}, {Pailler}, {Panuzzo}, {Riclet}, {Roux}, {Seabroke}, {Sordo}, {Tanga}, {Th{\'e}venin}, {Gracia-Abril}, {Portell}, {Teyssier}, {Altmann}, {Andrae}, {Bellas-Velidis}, {Benson}, {Berthier}, {Blomme}, {Brugaletta}, {Burgess}, {Busso}, {Carry}, {Cellino}, {Cheek}, {Clementini}, {Damerdji}, {Davidson}, {Delchambre}, {Dell'Oro}, {Fern{\'a}ndez-Hern{\'a}ndez}, {Galluccio}, {Garc{\'\i}a-Lario},
  {Garcia-Reinaldos}, {Gonz{\'a}lez-N{\'u}{\~n}ez}, {Gosset}, {Haigron}, {Halbwachs}, {Hambly}, {Harrison}, {Hatzidimitriou}, {Heiter}, {Hern{\'a}ndez}, {Hestroffer}, {Hodgkin}, {Holl}, {Jan{\ss}en}, {Jevardat de Fombelle}, {Jordan}, {Krone-Martins}, {Lanzafame}, {L{\"o}ffler}, {Lorca}, {Manteiga}, {Marchal}, {Marrese}, {Moitinho}, {Mora}, {Muinonen}, {Osborne}, {Pancino}, {Pauwels}, {Petit}, {Recio-Blanco}, {Richards}, {Riello}, {Rimoldini}, {Robin}, {Roegiers}, {Rybizki}, {Sarro}, {Siopis}, {Smith}, {Sozzetti}, {Ulla}, {Utrilla}, {van Leeuwen}, {van Reeven}, {Abbas}, {Abreu Aramburu}, {Accart}, {Aerts}, {Aguado}, {Ajaj}, {Altavilla}, {{\'A}lvarez}, {{\'A}lvarez Cid-Fuentes}, {Alves}, {Anderson}, {Anglada Varela}, {Antoja}, {Audard}, {Baines}, {Baker}, {Balaguer-N{\'u}{\~n}ez}, {Balbinot}, {Balog}, {Barache}, {Barbato}, {Barros}, {Barstow}, {Bartolom{\'e}}, {Bassilana}, {Bauchet}, {Baudesson-Stella}, {Becciani}, {Bellazzini}, {Bernet}, {Bertone}, {Bianchi}, {Blanco-Cuaresma}, {Boch}, {Bombrun}, {Bossini},
  {Bouquillon}, {Bragaglia}, {Bramante}, {Breedt}, {Bressan}, {Brouillet}, {Bucciarelli}, {Burlacu}, {Busonero}, {Butkevich}, {Buzzi}, {Caffau}, {Cancelliere}, {C{\'a}novas}, {Cantat-Gaudin}, {Carballo}, {Carlucci}, {Carnerero}, {Carrasco}, {Casamiquela}, {Castellani}, {Castro-Ginard}, {Castro Sampol}, {Chaoul}, {Charlot}, {Chemin}, {Chiavassa}, {Cioni}, {Comoretto}, {Cooper}, {Cornez}, {Cowell}, {Crifo}, {Crosta}, {Crowley}, {Dafonte}, {Dapergolas}, {David}, {David}, {de Laverny}, {De Luise}, {De March}, {De Ridder}, {de Souza}, {de Teodoro}, {de Torres}, {del Peloso}, {del Pozo}, {Delbo}, {Delgado}, {Delgado}, {Delisle}, {Di Matteo}, {Diakite}, {Diener}, {Distefano}, {Dolding}, {Eappachen}, {Edvardsson}, {Enke}, {Esquej}, {Fabre}, {Fabrizio}, {Faigler}, {Fedorets}, {Fernique}, {Fienga}, {Figueras}, {Fouron}, {Fragkoudi}, {Fraile}, {Franke}, {Gai}, {Garabato}, {Garcia-Gutierrez}, {Garc{\'\i}a-Torres}, {Garofalo}, {Gavras}, {Gerlach}, {Geyer}, {Giacobbe}, {Gilmore}, {Girona}, {Giuffrida}, {Gomel}, {Gomez},
  {Gonzalez-Santamaria}, {Gonz{\'a}lez-Vidal}, {Granvik}, {Guti{\'e}rrez-S{\'a}nchez}, {Guy}, {Hauser}, {Haywood}, {Helmi}, {Hidalgo}, {Hilger}, {H{\l}adczuk}, {Hobbs}, {Holland}, {Huckle}, {Jasniewicz}, {Jonker}, {Juaristi Campillo}, {Julbe}, {Karbevska}, {Kervella}, {Khanna}, {Kochoska}, {Kontizas}, {Kordopatis}, {Korn}, {Kostrzewa-Rutkowska}, {Kruszy{\'n}ska}, {Lambert}, {Lanza}, {Lasne}, {Le Campion}, {Le Fustec}, {Lebreton}, {Lebzelter}, {Leccia}, {Leclerc}, {Lecoeur-Taibi}, {Liao}, {Licata}, {Lindstr{\o}m}, {Lister}, {Livanou}, {Lobel}, {Madrero Pardo}, {Managau}, {Mann}, {Marchant}, {Marconi}, {Marcos Santos}, {Marinoni}, {Marocco}, {Marshall}, {Martin Polo}, {Mart{\'\i}n-Fleitas}, {Masip}, {Massari}, {Mastrobuono-Battisti}, {Mazeh}, {McMillan}, {Messina}, {Michalik}, {Millar}, {Mints}, {Molina}, {Molinaro}, {Moln{\'a}r}, {Montegriffo}, {Mor}, {Morbidelli}, {Morel}, {Morris}, {Mulone}, {Munoz}, {Muraveva}, {Murphy}, {Musella}, {Noval}, {Ord{\'e}novic}, {Orr{\`u}}, {Osinde}, {Pagani}, {Pagano},
  {Palaversa}, {Palicio}, {Panahi}, {Pawlak}, {Pe{\~n}alosa Esteller}, {Penttil{\"a}}, {Piersimoni}, {Pineau}, {Plachy}, {Plum}, {Poggio}, {Poretti}, {Poujoulet}, {Pr{\v{s}}a}, {Pulone}, {Racero}, {Ragaini}, {Rainer}, {Raiteri}, {Rambaux}, {Ramos}, {Ramos-Lerate}, {Re Fiorentin}, {Regibo}, {Reyl{\'e}}, {Ripepi}, {Riva}, {Rixon}, {Robichon}, {Robin}, {Roelens}, {Rohrbasser}, {Romero-G{\'o}mez}, {Rowell}, {Royer}, {Rybicki}, {Sadowski}, {Sagrist{\`a} Sell{\'e}s}, {Sahlmann}, {Salgado}, {Salguero}, {Samaras}, {Sanchez Gimenez}, {Sanna}, {Santove{\~n}a}, {Sarasso}, {Schultheis}, {Sciacca}, {Segol}, {Segovia}, {S{\'e}gransan}, {Semeux}, {Shahaf}, {Siddiqui}, {Siebert}, {Siltala}, {Slezak}, {Smart}, {Solano}, {Solitro}, {Souami}, {Souchay}, {Spagna}, {Spoto}, {Steele}, {Steidelm{\"u}ller}, {Stephenson}, {S{\"u}veges}, {Szabados}, {Szegedi-Elek}, {Taris}, {Tauran}, {Taylor}, {Teixeira}, {Thuillot}, {Tonello}, {Torra}, {Torra}, {Turon}, {Unger}, {Vaillant}, {van Dillen}, {Vanel}, {Vecchiato}, {Viala}, {Vicente},
  {Voutsinas}, {Weiler}, {Wevers}, {Wyrzykowski}, {Yoldas}, {Yvard}, {Zhao}, {Zorec}, {Zucker}, {Zurbach}, \& {Zwitter}}]{2021A&A...649A...1G}
{Gaia Collaboration}, {Brown}, A.~G.~A., {Vallenari}, A., {et~al.} 2021, \aap, 649, A1, \dodoi{10.1051/0004-6361/202039657}

\bibitem[{{Gilmore} \& {Randich}(2015)}]{ESO_25}
{Gilmore}, G., \& {Randich}, S. 2015, {Gaia-ESO spectroscopic survey},  European Southern Observatory, \dodoi{10.18727/archive/25}

\bibitem[{{Gilmore} {et~al.}(2012){Gilmore}, {Randich}, {Asplund}, {Binney}, {Bonifacio}, {Drew}, {Feltzing}, {Ferguson}, {Jeffries}, {Micela}, {Negueruela}, {Prusti}, {Rix}, {Vallenari}, {Alfaro}, {Allende-Prieto}, {Babusiaux}, {Bensby}, {Blomme}, {Bragaglia}, {Flaccomio}, {Fran{\c{c}}ois}, {Irwin}, {Koposov}, {Korn}, {Lanzafame}, {Pancino}, {Paunzen}, {Recio-Blanco}, {Sacco}, {Smiljanic}, {Van Eck}, {Walton}, {Aden}, {Aerts}, {Affer}, {Alcala}, {Altavilla}, {Alves}, {Antoja}, {Arenou}, {Argiroffi}, {Asensio Ramos}, {Bailer-Jones}, {Balaguer-Nunez}, {Bayo}, {Barbuy}, {Barisevicius}, {Barrado y Navascues}, {Battistini}, {Bellas Velidis}, {Bellazzini}, {Belokurov}, {Bergemann}, {Bertelli}, {Biazzo}, {Bienayme}, {Bland-Hawthorn}, {Boeche}, {Bonito}, {Boudreault}, {Bouvier}, {Brandao}, {Brown}, {de Bruijne}, {Burleigh}, {Caballero}, {Caffau}, {Calura}, {Capuzzo-Dolcetta}, {Caramazza}, {Carraro}, {Casagrande}, {Casewell}, {Chapman}, {Chiappini}, {Chorniy}, {Christlieb}, {Cignoni}, {Cocozza}, {Colless}, {Collet},
  {Collins}, {Correnti}, {Covino}, {Crnojevic}, {Cropper}, {Cunha}, {Damiani}, {David}, {Delgado}, {Duffau}, {Edvardsson}, {Eldridge}, {Enke}, {Eriksson}, {Evans}, {Eyer}, {Famaey}, {Fellhauer}, {Ferreras}, {Figueras}, {Fiorentino}, {Flynn}, {Folha}, {Franciosini}, {Frasca}, {Freeman}, {Fremat}, {Friel}, {Gaensicke}, {Gameiro}, {Garzon}, {Geier}, {Geisler}, {Gerhard}, {Gibson}, {Gomboc}, {Gomez}, {Gonzalez-Fernandez}, {Gonzalez Hernandez}, {Gosset}, {Grebel}, {Greimel}, {Groenewegen}, {Grundahl}, {Guarcello}, {Gustafsson}, {Hadrava}, {Hatzidimitriou}, {Hambly}, {Hammersley}, {Hansen}, {Haywood}, {Heber}, {Heiter}, {Held}, {Helmi}, {Hensler}, {Herrero}, {Hill}, {Hodgkin}, {Huelamo}, {Huxor}, {Ibata}, {Jackson}, {de Jong}, {Jonker}, {Jordan}, {Jordi}, {Jorissen}, {Katz}, {Kawata}, {Keller}, {Kharchenko}, {Klement}, {Klutsch}, {Knude}, {Koch}, {Kochukhov}, {Kontizas}, {Koubsky}, {Lallement}, {de Laverny}, {van Leeuwen}, {Lemasle}, {Lewis}, {Lind}, {Lindstrom}, {Lobel}, {Lopez Santiago}, {Lucas}, {Ludwig},
  {Lueftinger}, {Magrini}, {Maiz Apellaniz}, {Maldonado}, {Marconi}, {Marino}, {Martayan}, {Martinez-Valpuesta}, {Matijevic}, {McMahon}, {Messina}, {Meyer}, {Miglio}, {Mikolaitis}, {Minchev}, {Minniti}, {Moitinho}, {Momany}, {Monaco}, {Montalto}, {Monteiro}, {Monier}, {Montes}, {Mora}, {Moraux}, {Morel}, {Mowlavi}, {Mucciarelli}, {Munari}, {Napiwotzki}, {Nardetto}, {Naylor}, {Naze}, {Nelemans}, {Okamoto}, {Ortolani}, {Pace}, {Palla}, {Palous}, {Parker}, {Penarrubia}, {Pillitteri}, {Piotto}, {Posbic}, {Prisinzano}, {Puzeras}, {Quirrenbach}, {Ragaini}, {Read}, {Read}, {Reyle}, {De Ridder}, {Robichon}, {Robin}, {Roeser}, {Romano}, {Royer}, {Ruchti}, {Ruzicka}, {Ryan}, {Ryde}, {Santos}, {Sanz Forcada}, {Sarro Baro}, {Sbordone}, {Schilbach}, {Schmeja}, {Schnurr}, {Schoenrich}, {Scholz}, {Seabroke}, {Sharma}, {De Silva}, {Smith}, {Solano}, {Sordo}, {Soubiran}, {Sousa}, {Spagna}, {Steffen}, {Steinmetz}, {Stelzer}, {Stempels}, {Tabernero}, {Tautvaisiene}, {Thevenin}, {Torra}, {Tosi}, {Tolstoy}, {Turon}, {Walker},
  {Wambsganss}, {Worley}, {Venn}, {Vink}, {Wyse}, {Zaggia}, {Zeilinger}, {Zoccali}, {Zorec}, {Zucker}, {Zwitter}, \& {Gaia-ESO Survey Team}}]{2012Msngr.147...25G}
{Gilmore}, G., {Randich}, S., {Asplund}, M., {et~al.} 2012, The Messenger, 147, 25

\bibitem[{{Gilmore} {et~al.}(2022){Gilmore}, {Randich}, {Worley}, {Hourihane}, {Gonneau}, {Sacco}, {Lewis}, {Magrini}, {Fran{\c{c}}ois}, {Jeffries}, {Koposov}, {Bragaglia}, {Alfaro}, {Allende Prieto}, {Blomme}, {Korn}, {Lanzafame}, {Pancino}, {Recio-Blanco}, {Smiljanic}, {Van Eck}, {Zwitter}, {Bensby}, {Flaccomio}, {Irwin}, {Franciosini}, {Morbidelli}, {Damiani}, {Bonito}, {Friel}, {Vink}, {Prisinzano}, {Abbas}, {Hatzidimitriou}, {Held}, {Jordi}, {Paunzen}, {Spagna}, {Jackson}, {Ma{\'\i}z Apell{\'a}niz}, {Asplund}, {Bonifacio}, {Feltzing}, {Binney}, {Drew}, {Ferguson}, {Micela}, {Negueruela}, {Prusti}, {Rix}, {Vallenari}, {Bergemann}, {Casey}, {de Laverny}, {Frasca}, {Hill}, {Lind}, {Sbordone}, {Sousa}, {Adibekyan}, {Caffau}, {Daflon}, {Feuillet}, {Gebran}, {Gonzalez Hernandez}, {Guiglion}, {Herrero}, {Lobel}, {Merle}, {Mikolaitis}, {Montes}, {Morel}, {Ruchti}, {Soubiran}, {Tabernero}, {Tautvai{\v{s}}ien{\.{e}}}, {Traven}, {Valentini}, {Van der Swaelmen}, {Villanova}, {Viscasillas V{\'a}zquez}, {Bayo},
  {Biazzo}, {Carraro}, {Edvardsson}, {Heiter}, {Jofr{\'e}}, {Marconi}, {Martayan}, {Masseron}, {Monaco}, {Walton}, {Zaggia}, {Aguirre B{\o}rsen-Koch}, {Alves}, {Balaguer-Nunez}, {Barklem}, {Barrado}, {Bellazzini}, {Berlanas}, {Binks}, {Bressan}, {Capuzzo-Dolcetta}, {Casagrande}, {Casamiquela}, {Collins}, {D'Orazi}, {Dantas}, {Debattista}, {Delgado-Mena}, {Di Marcantonio}, {Drazdauskas}, {Evans}, {Famaey}, {Franchini}, {Fr{\'e}mat}, {Fu}, {Geisler}, {Gerhard}, {Gonz{\'a}lez Solares}, {Grebel}, {Guti{\'e}rrez Albarr{\'a}n}, {Jim{\'e}nez-Esteban}, {J{\"o}nsson}, {Khachaturyants}, {Kordopatis}, {Kos}, {Lagarde}, {Ludwig}, {Mahy}, {Mapelli}, {Marfil}, {Martell}, {Messina}, {Miglio}, {Minchev}, {Moitinho}, {Montalban}, {Monteiro}, {Morossi}, {Mowlavi}, {Mucciarelli}, {Murphy}, {Nardetto}, {Ortolani}, {Paletou}, {Palou{\v{s}}}, {Pickering}, {Quirrenbach}, {Re Fiorentin}, {Read}, {Romano}, {Ryde}, {Sanna}, {Santos}, {Seabroke}, {Spina}, {Steinmetz}, {Stonkut{\'e}}, {Sutorius}, {Th{\'e}venin}, {Tosi}, {Tsantaki},
  {Wright}, {Wyse}, {Zoccali}, {Zorec}, \& {Zucker}}]{2022A&A...666A.120G}
{Gilmore}, G., {Randich}, S., {Worley}, C.~C., {et~al.} 2022, \aap, 666, A120, \dodoi{10.1051/0004-6361/202243134}

\bibitem[{{Goss} {et~al.}(2011){Goss}, {Karoff}, {Chaplin}, {Elsworth}, \& {Stevens}}]{Goss2011MNRAS}
{Goss}, K.~J.~F., {Karoff}, C., {Chaplin}, W.~J., {Elsworth}, Y., \& {Stevens}, I.~R. 2011, \mnras, 411, 162, \dodoi{10.1111/j.1365-2966.2010.17665.x}

\bibitem[{{Goudfrooij} {et~al.}(2011){Goudfrooij}, {Puzia}, {Chandar}, \& {Kozhurina-Platais}}]{2011ApJ...737....4G}
{Goudfrooij}, P., {Puzia}, T.~H., {Chandar}, R., \& {Kozhurina-Platais}, V. 2011, \apj, 737, 4, \dodoi{10.1088/0004-637X/737/1/4}

\bibitem[{{Goudfrooij} {et~al.}(2009){Goudfrooij}, {Puzia}, {Kozhurina-Platais}, \& {Chandar}}]{2009AJ....137.4988G}
{Goudfrooij}, P., {Puzia}, T.~H., {Kozhurina-Platais}, V., \& {Chandar}, R. 2009, \aj, 137, 4988, \dodoi{10.1088/0004-6256/137/6/4988}

\bibitem[{{He} {et~al.}(2022){He}, {Sun}, {Li}, {Li}, {Shao}, {Zhong}, {Chen}, {Grijs}, {Tang}, {Qin}, \& {Randriamanakoto}}]{2022ApJ...938...42H}
{He}, C., {Sun}, W., {Li}, C., {et~al.} 2022, \apj, 938, 42, \dodoi{10.3847/1538-4357/ac8b08}

\bibitem[{{He} {et~al.}(2023){He}, {Li}, {Sun}, {de Grijs}, {Li}, {Zhong}, {Qin}, {Chen}, {Wang}, {Tang}, {Shao}, \& {Xu}}]{2023MNRAS.525.5880H}
{He}, C., {Li}, C., {Sun}, W., {et~al.} 2023, \mnras, 525, 5880, \dodoi{10.1093/mnras/stad2674}

\bibitem[{{Heger} \& {Langer}(2000)}]{Heger2000ApJ.II}
{Heger}, A., \& {Langer}, N. 2000, \apj, 544, 1016, \dodoi{10.1086/317239}

\bibitem[{{Heger} {et~al.}(2005){Heger}, {Woosley}, \& {Spruit}}]{Heger2005ApJ}
{Heger}, A., {Woosley}, S.~E., \& {Spruit}, H.~C. 2005, \apj, 626, 350, \dodoi{10.1086/429868}

\bibitem[{{Huat} {et~al.}(2009){Huat}, {Hubert}, {Baudin}, {Floquet}, {Neiner}, {Fr{\'e}mat}, {Guti{\'e}rrez-Soto}, {Andrade}, {de Batz}, {Diago}, {Emilio}, {Espinosa Lara}, {Fabregat}, {Janot-Pacheco}, {Leroy}, {Martayan}, {Semaan}, {Suso}, {Auvergne}, {Catala}, {Michel}, \& {Samadi}}]{Huat2009}
{Huat}, A.~L., {Hubert}, A.~M., {Baudin}, F., {et~al.} 2009, \aap, 506, 95, \dodoi{10.1051/0004-6361/200911928}

\bibitem[{{Hubeny} \& {Lanz}(1992)}]{1992A&A...262..501H}
{Hubeny}, I., \& {Lanz}, T. 1992, \aap, 262, 501

\bibitem[{{Hunter}(2007)}]{2007CSE.....9...90H}
{Hunter}, J.~D. 2007, Computing in Science and Engineering, 9, 90, \dodoi{10.1109/MCSE.2007.55}

\bibitem[{{Johnston} {et~al.}(2019){Johnston}, {Aerts}, {Pedersen}, \& {Bastian}}]{2019A&A...632A..74J}
{Johnston}, C., {Aerts}, C., {Pedersen}, M.~G., \& {Bastian}, N. 2019, \aap, 632, A74, \dodoi{10.1051/0004-6361/201936549}

\bibitem[{{Kamann} {et~al.}(2021){Kamann}, {Bastian}, {Usher}, {Cabrera-Ziri}, \& {Saracino}}]{2021MNRAS.508.2302K}
{Kamann}, S., {Bastian}, N., {Usher}, C., {Cabrera-Ziri}, I., \& {Saracino}, S. 2021, \mnras, 508, 2302, \dodoi{10.1093/mnras/stab2643}

\bibitem[{{Kamann} {et~al.}(2020){Kamann}, {Bastian}, {Gossage}, {Baade}, {Cabrera-Ziri}, {Da Costa}, {de Mink}, {Georgy}, {Giesers}, {G{\"o}ttgens}, {Hilker}, {Husser}, {Lardo}, {Larsen}, {Mackey}, {Martocchia}, {Mucciarelli}, {Platais}, {Roth}, {Salaris}, {Usher}, \& {Yong}}]{2020MNRAS.492.2177K}
{Kamann}, S., {Bastian}, N., {Gossage}, S., {et~al.} 2020, \mnras, 492, 2177, \dodoi{10.1093/mnras/stz3583}

\bibitem[{{Kamann} {et~al.}(2023){Kamann}, {Saracino}, {Bastian}, {Gossage}, {Usher}, {Baade}, {Cabrera-Ziri}, {de Mink}, {Ekstrom}, {Georgy}, {Hilker}, {Larsen}, {Mackey}, {Niederhofer}, {Platais}, \& {Yong}}]{2023MNRAS.518.1505K}
{Kamann}, S., {Saracino}, S., {Bastian}, N., {et~al.} 2023, \mnras, 518, 1505, \dodoi{10.1093/mnras/stac3170}

\bibitem[{{Kraft}(1967)}]{1967ApJ...150..551K}
{Kraft}, R.~P. 1967, \apj, 150, 551, \dodoi{10.1086/149359}

\bibitem[{{Kurucz}(2005)}]{2005MSAIS...8...14K}
{Kurucz}, R.~L. 2005, Memorie della Societa Astronomica Italiana Supplementi, 8, 14

\bibitem[{{Li} {et~al.}(2014){Li}, {de Grijs}, \& {Deng}}]{2014Natur.516..367L}
{Li}, C., {de Grijs}, R., \& {Deng}, L. 2014, \nat, 516, 367, \dodoi{10.1038/nature13969}

\bibitem[{{Li} {et~al.}(2017){Li}, {de Grijs}, {Deng}, \& {Milone}}]{2017ApJ...844..119L}
{Li}, C., {de Grijs}, R., {Deng}, L., \& {Milone}, A.~P. 2017, \apj, 844, 119, \dodoi{10.3847/1538-4357/aa7b36}

\bibitem[{Li {et~al.}(2024)Li, Milone, Sun, \& {de Grijs}}]{LI2024}
Li, C., Milone, A.~P., Sun, W., \& {de Grijs}, R. 2024, Fundamental Research, \dodoi{https://doi.org/10.1016/j.fmre.2023.12.007}

\bibitem[{{Li} {et~al.}(2020{\natexlab{a}}){Li}, {Guo}, {Fuller}, {Bedding}, {Murphy}, {Colman}, \& {Hey}}]{Ligang_2020_35binaries}
{Li}, G., {Guo}, Z., {Fuller}, J., {et~al.} 2020{\natexlab{a}}, \mnras, 497, 4363, \dodoi{10.1093/mnras/staa2266}

\bibitem[{{Li} {et~al.}(2020{\natexlab{b}}){Li}, {Van Reeth}, {Bedding}, {Murphy}, {Antoci}, {Ouazzani}, \& {Barbara}}]{Ligang2020_611}
{Li}, G., {Van Reeth}, T., {Bedding}, T.~R., {et~al.} 2020{\natexlab{b}}, \mnras, 491, 3586, \dodoi{10.1093/mnras/stz2906}

\bibitem[{{Li} {et~al.}(2024){Li}, {Aerts}, {Bedding}, {Fritzewski}, {Murphy}, {Van Reeth}, {Montet}, {Jian}, {Mombarg}, {Gossage}, \& {Sreenivas}}]{Ligang2024_NGC2516}
{Li}, G., {Aerts}, C., {Bedding}, T.~R., {et~al.} 2024, \aap, 686, A142, \dodoi{10.1051/0004-6361/202348901}

\bibitem[{{Mackey} {et~al.}(2008){Mackey}, {Broby Nielsen}, {Ferguson}, \& {Richardson}}]{2008ApJ...681L..17M}
{Mackey}, A.~D., {Broby Nielsen}, P., {Ferguson}, A.~M.~N., \& {Richardson}, J.~C. 2008, \apjl, 681, L17, \dodoi{10.1086/590343}

\bibitem[{{Maeder} \& {Meynet}(2000)}]{2000ARA&A..38..143M}
{Maeder}, A., \& {Meynet}, G. 2000, \araa, 38, 143, \dodoi{10.1146/annurev.astro.38.1.143}

\bibitem[{{Marigo} {et~al.}(2017){Marigo}, {Girardi}, {Bressan}, {Rosenfield}, {Aringer}, {Chen}, {Dussin}, {Nanni}, {Pastorelli}, {Rodrigues}, {Trabucchi}, {Bladh}, {Dalcanton}, {Groenewegen}, {Montalb{\'a}n}, \& {Wood}}]{2017ApJ...835...77M}
{Marigo}, P., {Girardi}, L., {Bressan}, A., {et~al.} 2017, \apj, 835, 77, \dodoi{10.3847/1538-4357/835/1/77}

\bibitem[{{Marino} {et~al.}(2018{\natexlab{a}}){Marino}, {Milone}, {Casagrande}, {Przybilla}, {Balaguer-N{\'u}{\~n}ez}, {Di Criscienzo}, {Serenelli}, \& {Vilardell}}]{2018ApJ...863L..33M}
{Marino}, A.~F., {Milone}, A.~P., {Casagrande}, L., {et~al.} 2018{\natexlab{a}}, \apjl, 863, L33, \dodoi{10.3847/2041-8213/aad868}

\bibitem[{{Marino} {et~al.}(2018{\natexlab{b}}){Marino}, {Przybilla}, {Milone}, {Da Costa}, {D'Antona}, {Dotter}, \& {Dupree}}]{2018AJ....156..116M}
{Marino}, A.~F., {Przybilla}, N., {Milone}, A.~P., {et~al.} 2018{\natexlab{b}}, \aj, 156, 116, \dodoi{10.3847/1538-3881/aad3cd}

\bibitem[{{Milone} {et~al.}(2009){Milone}, {Bedin}, {Piotto}, \& {Anderson}}]{2009A&A...497..755M}
{Milone}, A.~P., {Bedin}, L.~R., {Piotto}, G., \& {Anderson}, J. 2009, \aap, 497, 755, \dodoi{10.1051/0004-6361/200810870}

\bibitem[{{Milone} {et~al.}(2016){Milone}, {Marino}, {D'Antona}, {Bedin}, {Da Costa}, {Jerjen}, \& {Mackey}}]{2016MNRAS.458.4368M}
{Milone}, A.~P., {Marino}, A.~F., {D'Antona}, F., {et~al.} 2016, \mnras, 458, 4368, \dodoi{10.1093/mnras/stw608}

\bibitem[{{Milone} {et~al.}(2015){Milone}, {Bedin}, {Piotto}, {Marino}, {Cassisi}, {Bellini}, {Jerjen}, {Pietrinferni}, {Aparicio}, \& {Rich}}]{2015MNRAS.450.3750M}
{Milone}, A.~P., {Bedin}, L.~R., {Piotto}, G., {et~al.} 2015, \mnras, 450, 3750, \dodoi{10.1093/mnras/stv829}

\bibitem[{{Milone} {et~al.}(2018){Milone}, {Marino}, {Di Criscienzo}, {D'Antona}, {Bedin}, {Da Costa}, {Piotto}, {Tailo}, {Dotter}, {Angeloni}, {Anderson}, {Jerjen}, {Li}, {Dupree}, {Granata}, {Lagioia}, {Mackey}, {Nardiello}, \& {Vesperini}}]{2018MNRAS.477.2640M}
{Milone}, A.~P., {Marino}, A.~F., {Di Criscienzo}, M., {et~al.} 2018, \mnras, 477, 2640, \dodoi{10.1093/mnras/sty661}

\bibitem[{{Milone} {et~al.}(2023{\natexlab{a}}){Milone}, {Cordoni}, {Marino}, {D'Antona}, {Bellini}, {Di Criscienzo}, {Dondoglio}, {Lagioia}, {Langer}, {Legnardi}, {Libralato}, {Baumgardt}, {Bettinelli}, {Cavecchi}, {de Grijs}, {Deng}, {Hastings}, {Li}, {Mohandasan}, {Renzini}, {Vesperini}, {Wang}, {Ziliotto}, {Carlos}, {Costa}, {Dell'Agli}, {Di Stefano}, {Jang}, {Martorano}, {Simioni}, {Tailo}, \& {Ventura}}]{2023A&A...672A.161M}
{Milone}, A.~P., {Cordoni}, G., {Marino}, A.~F., {et~al.} 2023{\natexlab{a}}, \aap, 672, A161, \dodoi{10.1051/0004-6361/202244798}

\bibitem[{{Milone} {et~al.}(2023{\natexlab{b}}){Milone}, {Cordoni}, {Marino}, {Muratore}, {D'Antona}, {Di Criscienzo}, {Dondoglio}, {Lagioia}, {Legnardi}, {Mohandasan}, {Ziliotto}, {Dell'Agli}, {Tailo}, \& {Ventura}}]{2023MNRAS.524.6149M}
---. 2023{\natexlab{b}}, \mnras, 524, 6149, \dodoi{10.1093/mnras/stad2242}

\bibitem[{{Mombarg} {et~al.}(2024){Mombarg}, {Aerts}, \& {Molenberghs}}]{Mombarg2024}
{Mombarg}, J.~S.~G., {Aerts}, C., \& {Molenberghs}, G. 2024, \aap, 685, A21, \dodoi{10.1051/0004-6361/202449213}

\bibitem[{{Mombarg} {et~al.}(2022){Mombarg}, {Dotter}, {Rieutord}, {Michielsen}, {Van Reeth}, \& {Aerts}}]{Mombarg2022ApJ}
{Mombarg}, J. S.~G., {Dotter}, A., {Rieutord}, M., {et~al.} 2022, \apj, 925, 154, \dodoi{10.3847/1538-4357/ac3dfb}

\bibitem[{{Muratore} {et~al.}(2024){Muratore}, {Milone}, {D'Antona}, {Nastasio}, {Cordoni}, {Legnardi}, {He}, {Ziliotto}, {Dondoglio}, {Bernizzoni}, {Tailo}, {Bortolan}, {Dell'Agli}, {Deng}, {Lagioia}, {Li}, {Marino}, \& {Ventura}}]{2024arXiv241102508M}
{Muratore}, F., {Milone}, A.~P., {D'Antona}, F., {et~al.} 2024, arXiv e-prints, arXiv:2411.02508, \dodoi{10.48550/arXiv.2411.02508}

\bibitem[{{Neiner} {et~al.}(2020){Neiner}, {Lee}, {Mathis}, {Saio}, {Lovekin}, \& {Augustson}}]{Neiner2020}
{Neiner}, C., {Lee}, U., {Mathis}, S., {et~al.} 2020, \aap, 644, A9, \dodoi{10.1051/0004-6361/201935858}

\bibitem[{{Nguyen} {et~al.}(2022){Nguyen}, {Costa}, {Girardi}, {Volpato}, {Bressan}, {Chen}, {Marigo}, {Fu}, \& {Goudfrooij}}]{2022A&A...665A.126N}
{Nguyen}, C.~T., {Costa}, G., {Girardi}, L., {et~al.} 2022, \aap, 665, A126, \dodoi{10.1051/0004-6361/202244166}

\bibitem[{Observatory(2015)}]{ESO_27}
Observatory, E.~S. 2015, {GIRAFFE/MEDUSA reduced data obtained by standard ESO pipeline processing},  European Southern Observatory, \dodoi{10.18727/archive/27}

\bibitem[{Observatory(2020)}]{ESO_50}
---. 2020, {UVES reduced data obtained by standard ESO pipeline processing},  European Southern Observatory, \dodoi{10.18727/archive/50}

\bibitem[{{O'Donnell}(1994)}]{1994ApJ...422..158O}
{O'Donnell}, J.~E. 1994, \apj, 422, 158, \dodoi{10.1086/173713}

\bibitem[{{Palacios} {et~al.}(2010){Palacios}, {Gebran}, {Josselin}, {Martins}, {Plez}, {Belmas}, \& {L{\`e}bre}}]{2010A&A...516A..13P}
{Palacios}, A., {Gebran}, M., {Josselin}, E., {et~al.} 2010, \aap, 516, A13, \dodoi{10.1051/0004-6361/200913932}

\bibitem[{{Pang} {et~al.}(2022){Pang}, {Tang}, {Li}, {Yu}, {Wang}, {Li}, {Li}, {Wang}, {Wang}, {Zhang}, {Pasquato}, \& {Kouwenhoven}}]{2022ApJ...931..156P}
{Pang}, X., {Tang}, S.-Y., {Li}, Y., {et~al.} 2022, \apj, 931, 156, \dodoi{10.3847/1538-4357/ac674e}

\bibitem[{{P{\'a}pics} {et~al.}(2017){P{\'a}pics}, {Tkachenko}, {Van Reeth}, {Aerts}, {Moravveji}, {Van de Sande}, {De Smedt}, {Bloemen}, {Southworth}, {Debosscher}, {Niemczura}, \& {Gameiro}}]{Papics2017A&A}
{P{\'a}pics}, P.~I., {Tkachenko}, A., {Van Reeth}, T., {et~al.} 2017, \aap, 598, A74, \dodoi{10.1051/0004-6361/201629814}

\bibitem[{{Paxton} {et~al.}(2013){Paxton}, {Cantiello}, {Arras}, {Bildsten}, {Brown}, {Dotter}, {Mankovich}, {Montgomery}, {Stello}, {Timmes}, \& {Townsend}}]{2013ApJS..208....4P}
{Paxton}, B., {Cantiello}, M., {Arras}, P., {et~al.} 2013, \apjs, 208, 4, \dodoi{10.1088/0067-0049/208/1/4}

\bibitem[{{Paxton} {et~al.}(2019){Paxton}, {Smolec}, {Schwab}, {Gautschy}, {Bildsten}, {Cantiello}, {Dotter}, {Farmer}, {Goldberg}, {Jermyn}, {Kanbur}, {Marchant}, {Thoul}, {Townsend}, {Wolf}, {Zhang}, \& {Timmes}}]{2019ApJS..243...10P}
{Paxton}, B., {Smolec}, R., {Schwab}, J., {et~al.} 2019, \apjs, 243, 10, \dodoi{10.3847/1538-4365/ab2241}

\bibitem[{{Platais} {et~al.}(2012){Platais}, {Melo}, {Quinn}, {Clem}, {de Mink}, {Dotter}, {Kozhurina-Platais}, {Latham}, \& {Bellini}}]{2012ApJ...751L...8P}
{Platais}, I., {Melo}, C., {Quinn}, S.~N., {et~al.} 2012, \apjl, 751, L8, \dodoi{10.1088/2041-8205/751/1/L8}

\bibitem[{{Preston}(1974)}]{1974ARA&A..12..257P}
{Preston}, G.~W. 1974, \araa, 12, 257, \dodoi{10.1146/annurev.aa.12.090174.001353}

\bibitem[{{Qin} {et~al.}(2023){Qin}, {Zhong}, {Tang}, \& {Chen}}]{2023ApJS..265...12Q}
{Qin}, S., {Zhong}, J., {Tang}, T., \& {Chen}, L. 2023, \apjs, 265, 12, \dodoi{10.3847/1538-4365/acadd6}

\bibitem[{{Ricker} {et~al.}(2015){Ricker}, {Winn}, {Vanderspek}, {Latham}, {Bakos}, {Bean}, {Berta-Thompson}, {Brown}, {Buchhave}, {Butler}, {Butler}, {Chaplin}, {Charbonneau}, {Christensen-Dalsgaard}, {Clampin}, {Deming}, {Doty}, {De Lee}, {Dressing}, {Dunham}, {Endl}, {Fressin}, {Ge}, {Henning}, {Holman}, {Howard}, {Ida}, {Jenkins}, {Jernigan}, {Johnson}, {Kaltenegger}, {Kawai}, {Kjeldsen}, {Laughlin}, {Levine}, {Lin}, {Lissauer}, {MacQueen}, {Marcy}, {McCullough}, {Morton}, {Narita}, {Paegert}, {Palle}, {Pepe}, {Pepper}, {Quirrenbach}, {Rinehart}, {Sasselov}, {Sato}, {Seager}, {Sozzetti}, {Stassun}, {Sullivan}, {Szentgyorgyi}, {Torres}, {Udry}, \& {Villasenor}}]{Ricker_2015}
{Ricker}, G.~R., {Winn}, J.~N., {Vanderspek}, R., {et~al.} 2015, Journal of Astronomical Telescopes, Instruments, and Systems, 1, 014003, \dodoi{10.1117/1.JATIS.1.1.014003}

\bibitem[{{Rivinius} {et~al.}(2003){Rivinius}, {Baade}, \& {{\v{S}}tefl}}]{Rivinius2003A&A}
{Rivinius}, T., {Baade}, D., \& {{\v{S}}tefl}, S. 2003, \aap, 411, 229, \dodoi{10.1051/0004-6361:20031285}

\bibitem[{{Rivinius} {et~al.}(2013){Rivinius}, {Carciofi}, \& {Martayan}}]{2013A&ARv..21...69R}
{Rivinius}, T., {Carciofi}, A.~C., \& {Martayan}, C. 2013, \aapr, 21, 69, \dodoi{10.1007/s00159-013-0069-0}

\bibitem[{{Rivinius} {et~al.}(2006){Rivinius}, {{\v{S}}tefl}, \& {Baade}}]{2006A&A...459..137R}
{Rivinius}, T., {{\v{S}}tefl}, S., \& {Baade}, D. 2006, \aap, 459, 137, \dodoi{10.1051/0004-6361:20053008}

\bibitem[{{Salinas} {et~al.}(2016){Salinas}, {Pajkos}, {Strader}, {Vivas}, \& {Contreras Ramos}}]{2016ApJ...832L..14S}
{Salinas}, R., {Pajkos}, M.~A., {Strader}, J., {Vivas}, A.~K., \& {Contreras Ramos}, R. 2016, \apjl, 832, L14, \dodoi{10.3847/2041-8205/832/1/L14}

\bibitem[{{Slettebak}(1986)}]{1986PASP...98..867S}
{Slettebak}, A. 1986, \pasp, 98, 867, \dodoi{10.1086/131836}

\bibitem[{{STScI Development Team}(2013)}]{2013ascl.soft03023S}
{STScI Development Team}. 2013, {pysynphot: Synthetic photometry software package}, Astrophysics Source Code Library, record ascl:1303.023.
\newblock \doeprint{1303.023}

\bibitem[{{Sun} {et~al.}(2019{\natexlab{a}}){Sun}, {de Grijs}, {Deng}, \& {Albrow}}]{2019ApJ...876..113S}
{Sun}, W., {de Grijs}, R., {Deng}, L., \& {Albrow}, M.~D. 2019{\natexlab{a}}, \apj, 876, 113, \dodoi{10.3847/1538-4357/ab16e4}

\bibitem[{{Sun} {et~al.}(2021{\natexlab{a}}){Sun}, {de Grijs}, {Deng}, \& {Albrow}}]{2021MNRAS.502.4350S}
---. 2021{\natexlab{a}}, \mnras, 502, 4350, \dodoi{10.1093/mnras/stab347}

\bibitem[{{Sun} {et~al.}(2021{\natexlab{b}}){Sun}, {Duan}, {Deng}, \& {de Grijs}}]{2021ApJ...921..145S}
{Sun}, W., {Duan}, X.-W., {Deng}, L., \& {de Grijs}, R. 2021{\natexlab{b}}, \apj, 921, 145, \dodoi{10.3847/1538-4357/ac1ad0}

\bibitem[{{Sun} {et~al.}(2019{\natexlab{b}}){Sun}, {Li}, {Deng}, \& {de Grijs}}]{2019ApJ...883..182S}
{Sun}, W., {Li}, C., {Deng}, L., \& {de Grijs}, R. 2019{\natexlab{b}}, \apj, 883, 182, \dodoi{10.3847/1538-4357/ab3cd0}

\bibitem[{{Tarricq} {et~al.}(2021){Tarricq}, {Soubiran}, {Casamiquela}, {Cantat-Gaudin}, {Chemin}, {Anders}, {Antoja}, {Romero-G{\'o}mez}, {Figueras}, {Jordi}, {Bragaglia}, {Balaguer-N{\'u}{\~n}ez}, {Carrera}, {Castro-Ginard}, {Moitinho}, {Ramos}, \& {Bossini}}]{2021A&A...647A..19T}
{Tarricq}, Y., {Soubiran}, C., {Casamiquela}, L., {et~al.} 2021, \aap, 647, A19, \dodoi{10.1051/0004-6361/202039388}

\bibitem[{{Taylor}(2005)}]{2005ASPC..347...29T}
{Taylor}, M.~B. 2005, in Astronomical Society of the Pacific Conference Series, Vol. 347, Astronomical Data Analysis Software and Systems XIV, ed. P.~{Shopbell}, M.~{Britton}, \& R.~{Ebert}, 29

\bibitem[{Virtanen {et~al.}(2020)Virtanen, Gommers, Oliphant, Haberland, Reddy, Cournapeau, Burovski, Peterson, Weckesser, Bright, {van der Walt}, Brett, Wilson, Millman, Mayorov, Nelson, Jones, Kern, Larson, Carey, Polat, Feng, Moore, {VanderPlas}, Laxalde, Perktold, Cimrman, Henriksen, Quintero, Harris, Archibald, Ribeiro, Pedregosa, {van Mulbregt}, \& {SciPy 1.0 Contributors}}]{2020SciPy-NMeth}
Virtanen, P., Gommers, R., Oliphant, T.~E., {et~al.} 2020, Nature Methods, 17, 261, \dodoi{10.1038/s41592-019-0686-2}

\bibitem[{{von Zeipel}(1924)}]{1924MNRAS..84..665V}
{von Zeipel}, H. 1924, \mnras, 84, 665, \dodoi{10.1093/mnras/84.9.665}

\bibitem[{{Wall}(1996)}]{1996QJRAS..37..519W}
{Wall}, J.~V. 1996, Quarterly Journal of the Royal Astronomical Society, 37, 519

\bibitem[{{Wang} {et~al.}(2022){Wang}, {Langer}, {Schootemeijer}, {Milone}, {Hastings}, {Xu}, {Bodensteiner}, {Sana}, {Castro}, {Lennon}, {Marchant}, {Koter}, \& {Mink}}]{2022NatAs...6..480W}
{Wang}, C., {Langer}, N., {Schootemeijer}, A., {et~al.} 2022, Nature Astronomy, 6, 480, \dodoi{10.1038/s41550-021-01597-5}

\bibitem[{Wang {et~al.}(2023)Wang, Li, Wang, He, \& Wang}]{Wang_2023}
Wang, L., Li, C., Wang, L., He, C., \& Wang, C. 2023, The Astrophysical Journal, 949, 53, \dodoi{10.3847/1538-4357/accae0}

\bibitem[{{Yang} {et~al.}(2021){Yang}, {Li}, {de Grijs}, \& {Deng}}]{2021ApJ...912...27Y}
{Yang}, Y., {Li}, C., {de Grijs}, R., \& {Deng}, L. 2021, \apj, 912, 27, \dodoi{10.3847/1538-4357/abec4b}

\bibitem[{{Zorec} \& {Royer}(2012)}]{2012A&A...537A.120Z}
{Zorec}, J., \& {Royer}, F. 2012, \aap, 537, A120, \dodoi{10.1051/0004-6361/201117691}

\end{thebibliography}

\appendix
\begin{figure*}[ht!]
\centering
\begin{tabular}{ccc}
    \includegraphics[width=0.31\textwidth]{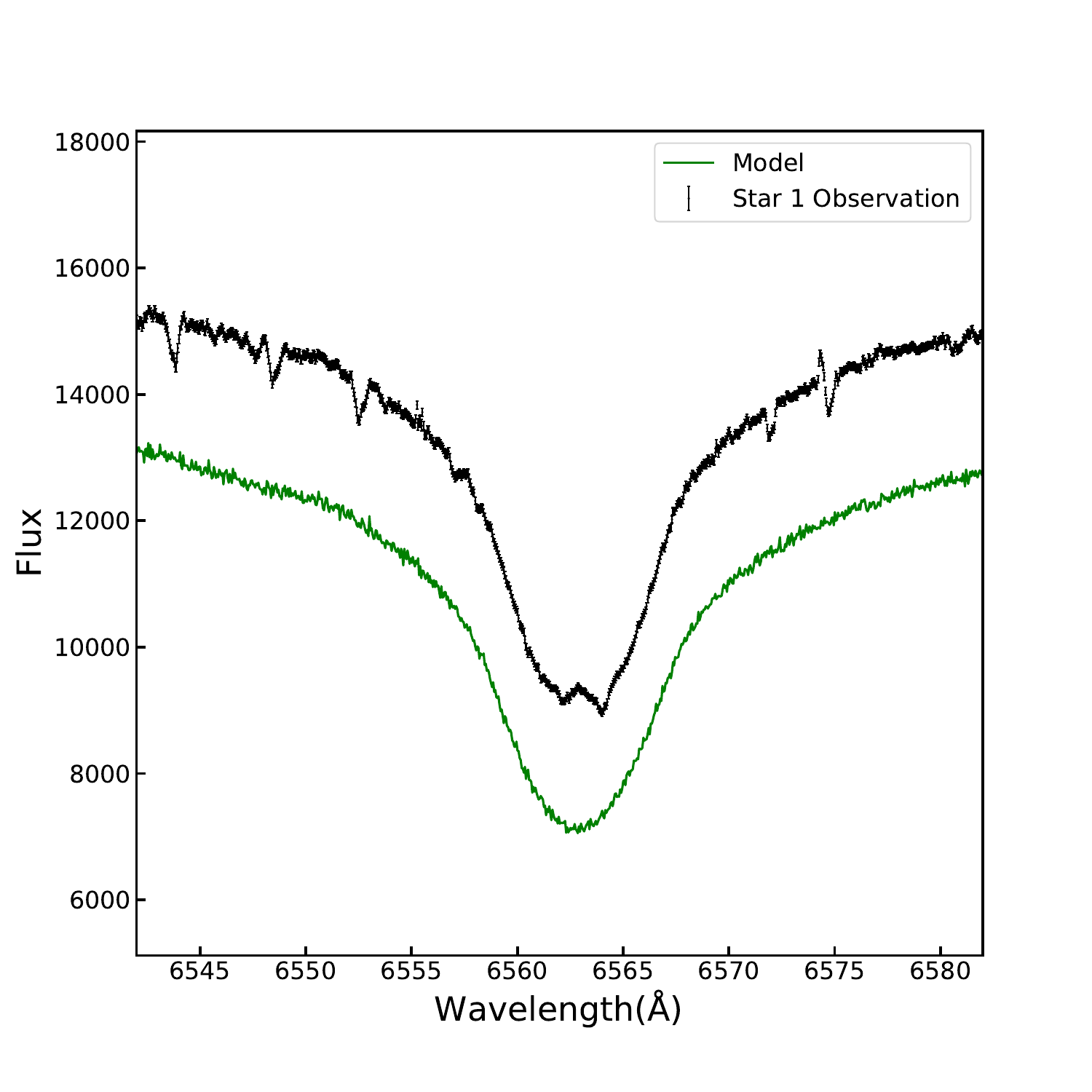}  & 
    \includegraphics[width=0.31\textwidth]{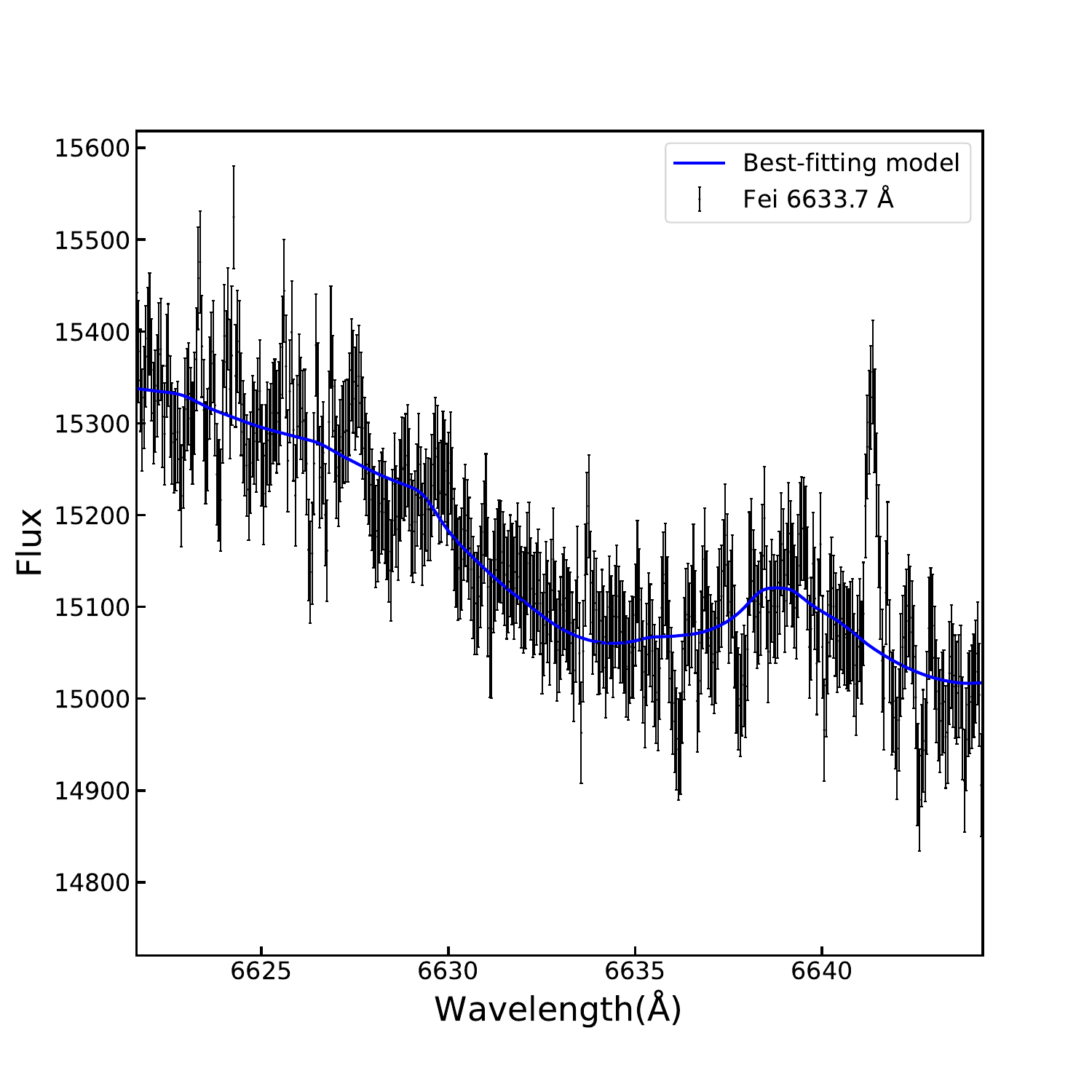}  & 
    \includegraphics[width=0.31\textwidth]{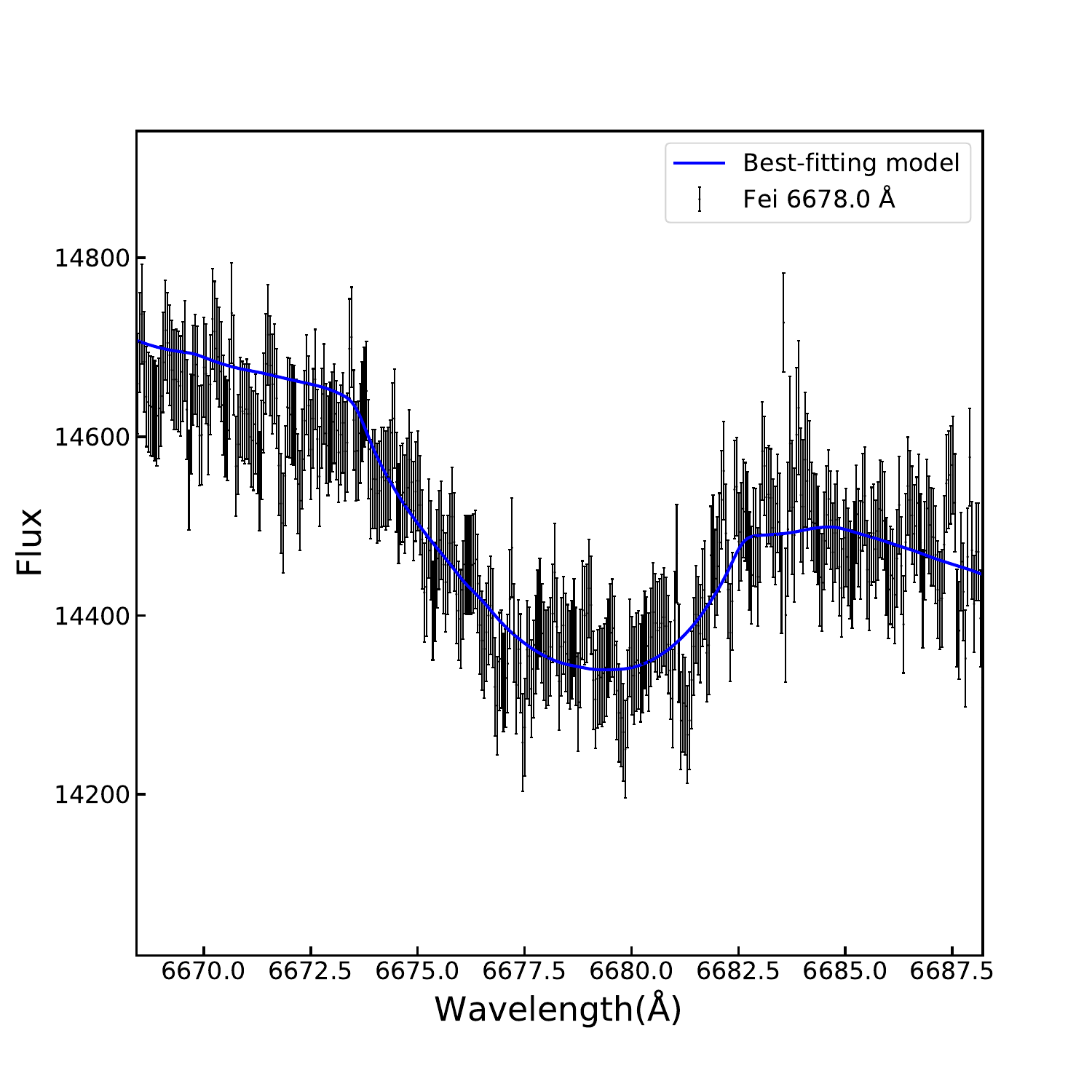}\\
    \includegraphics[width=0.31\textwidth]{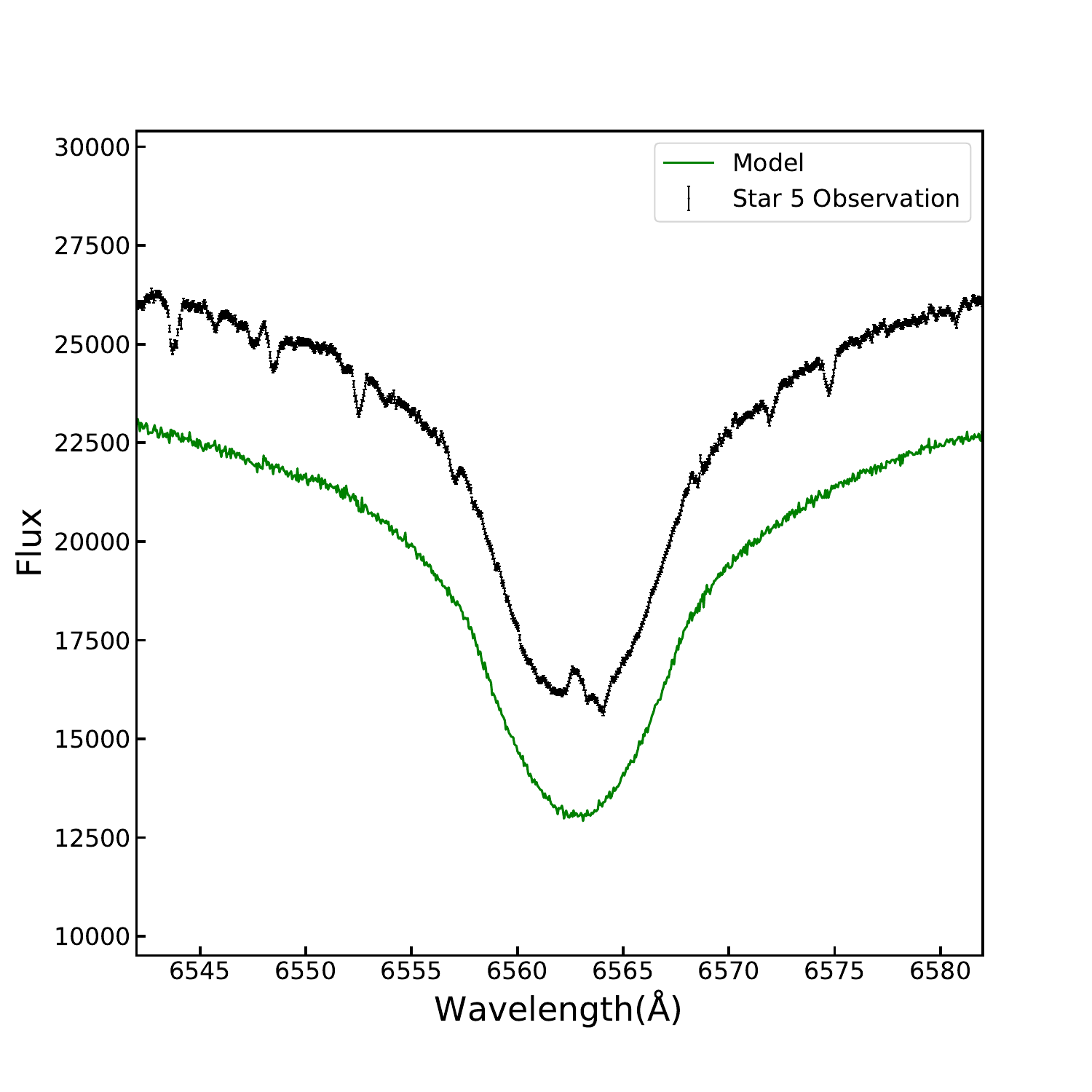}  & 
    \includegraphics[width=0.31\textwidth]{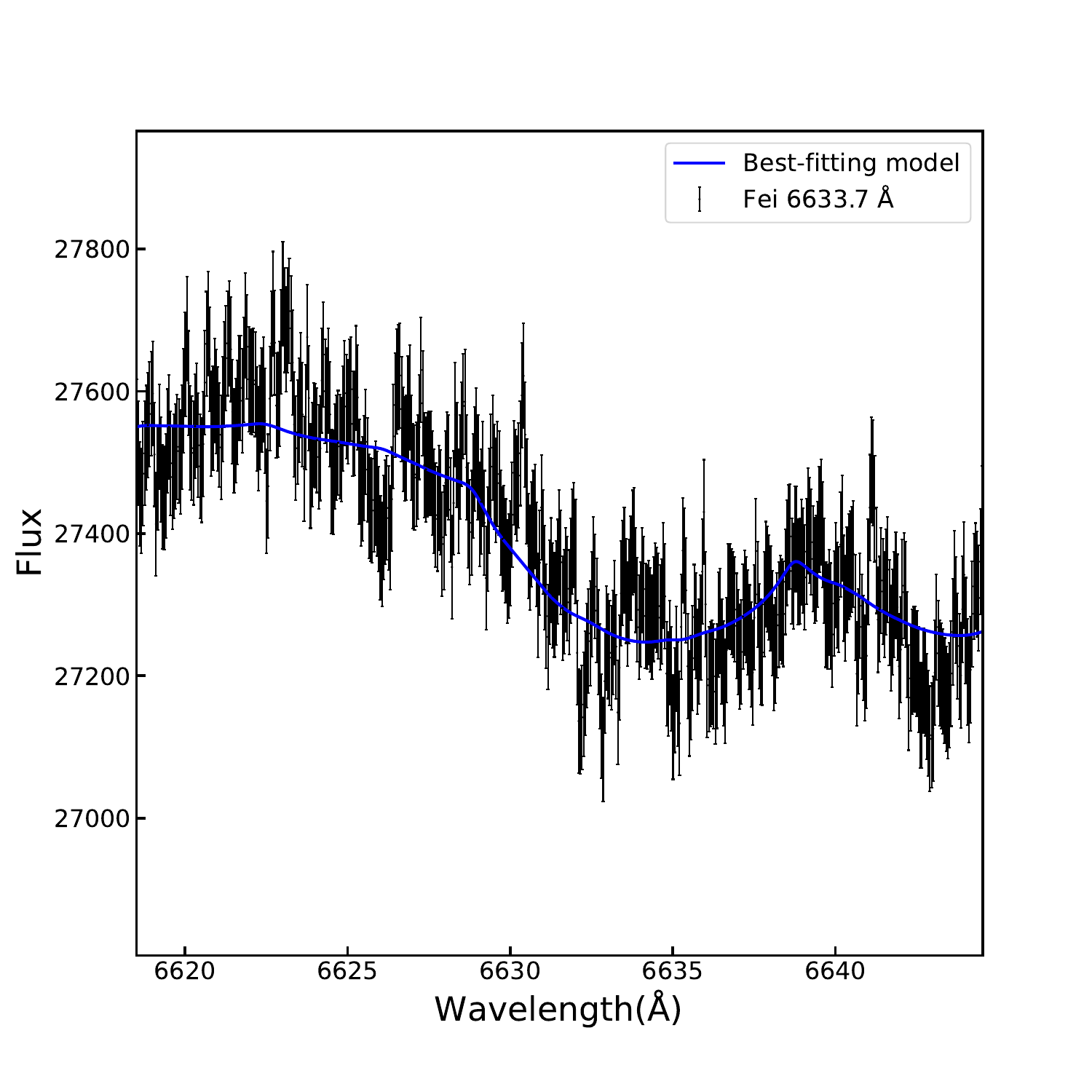 }  & 
    \includegraphics[width=0.31\textwidth]{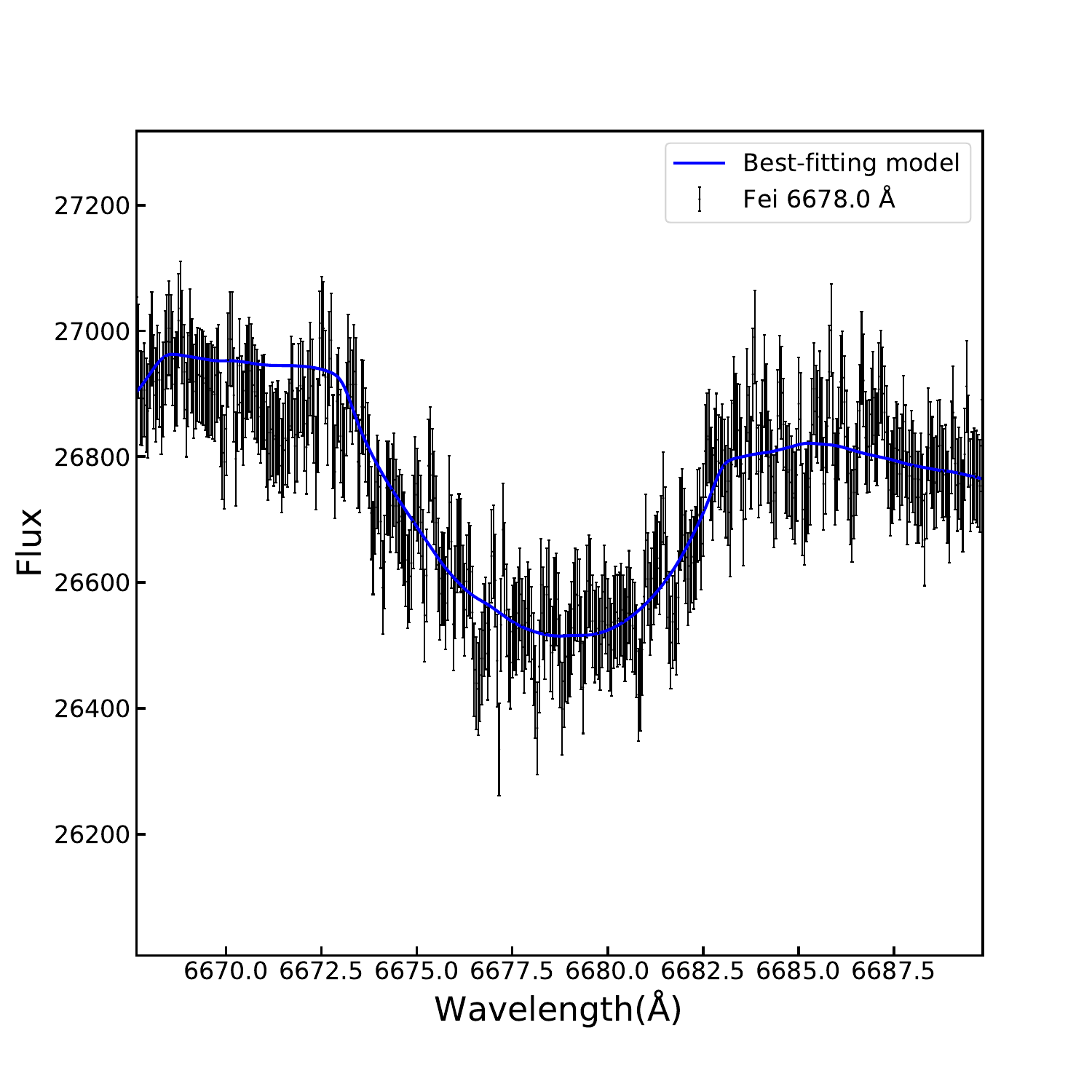}\\
     \includegraphics[width=0.31\textwidth]{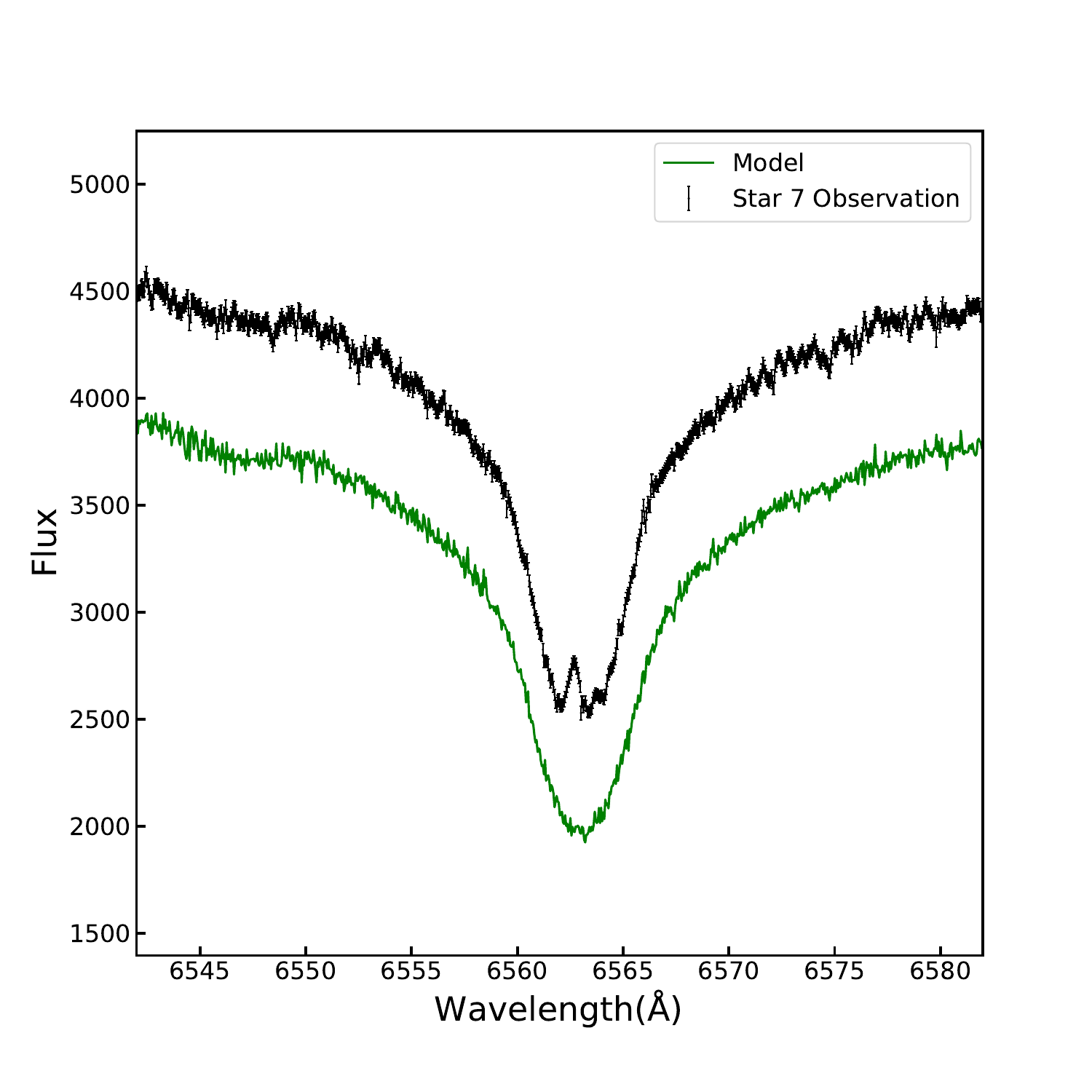}  & 
    \includegraphics[width=0.31\textwidth]{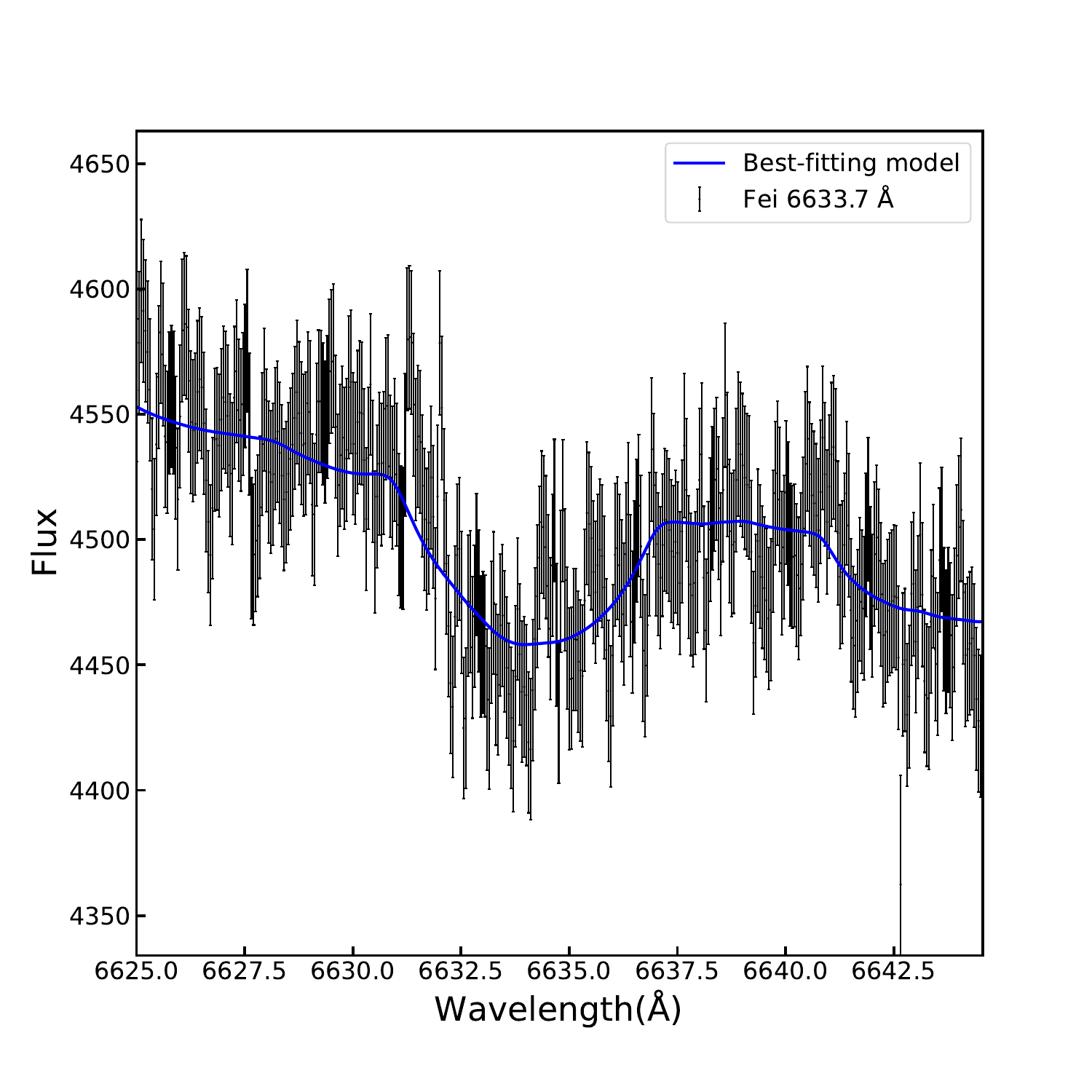 }  & 
    \includegraphics[width=0.31\textwidth]{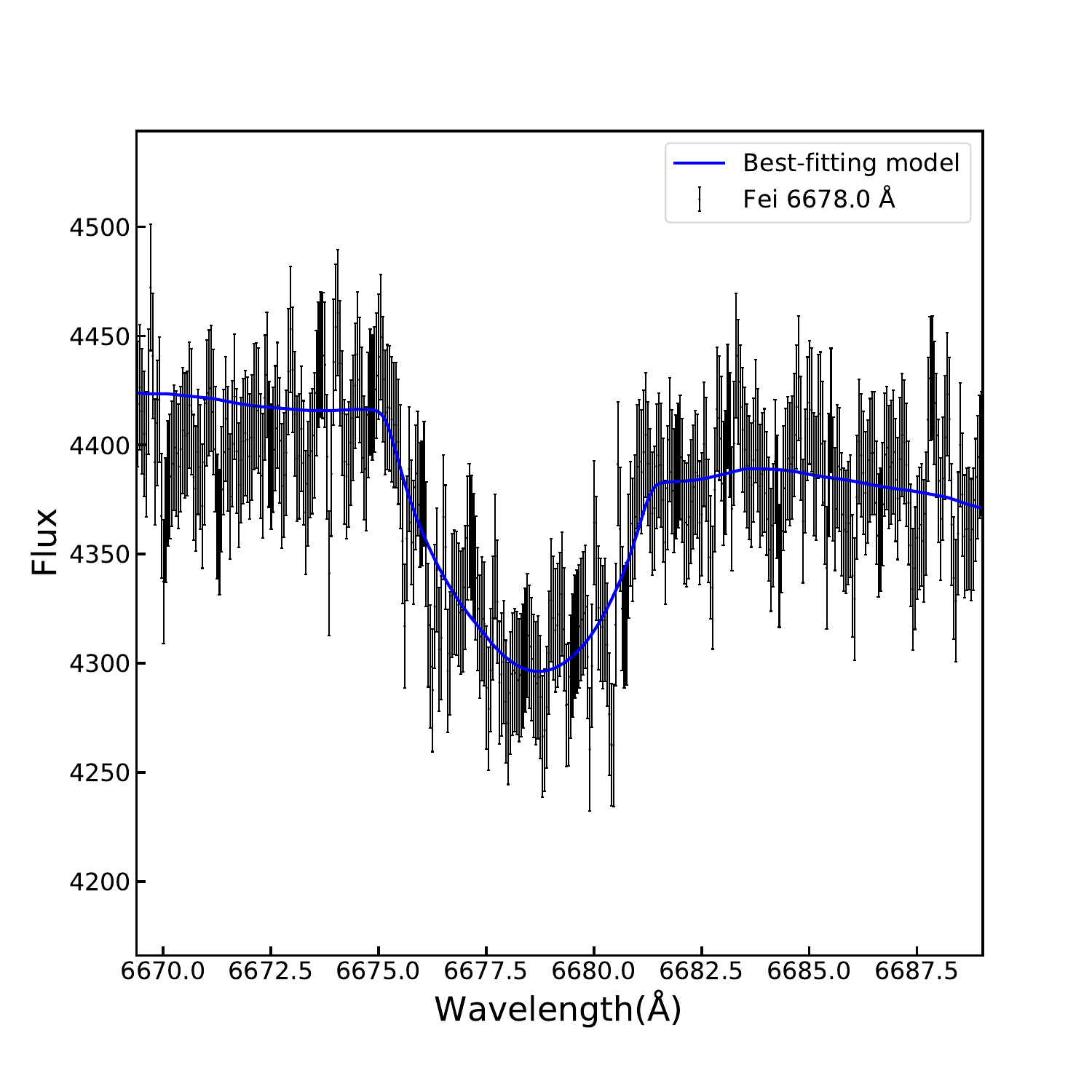}\\
     
\end{tabular}

	\caption{As Figure \ref{fig:3532 star 2 and 6}, from the top to the bottom, the panels of each row for Star 1, Star 5 and Star 7 listed in Table \ref{tab:3532 results}, respectively.}
 \label{fig:3532 star 1}
\end{figure*}

\newpage
\begin{figure*}
\centering
\begin{tabular}{ccc}
    \includegraphics[width=0.31\textwidth]{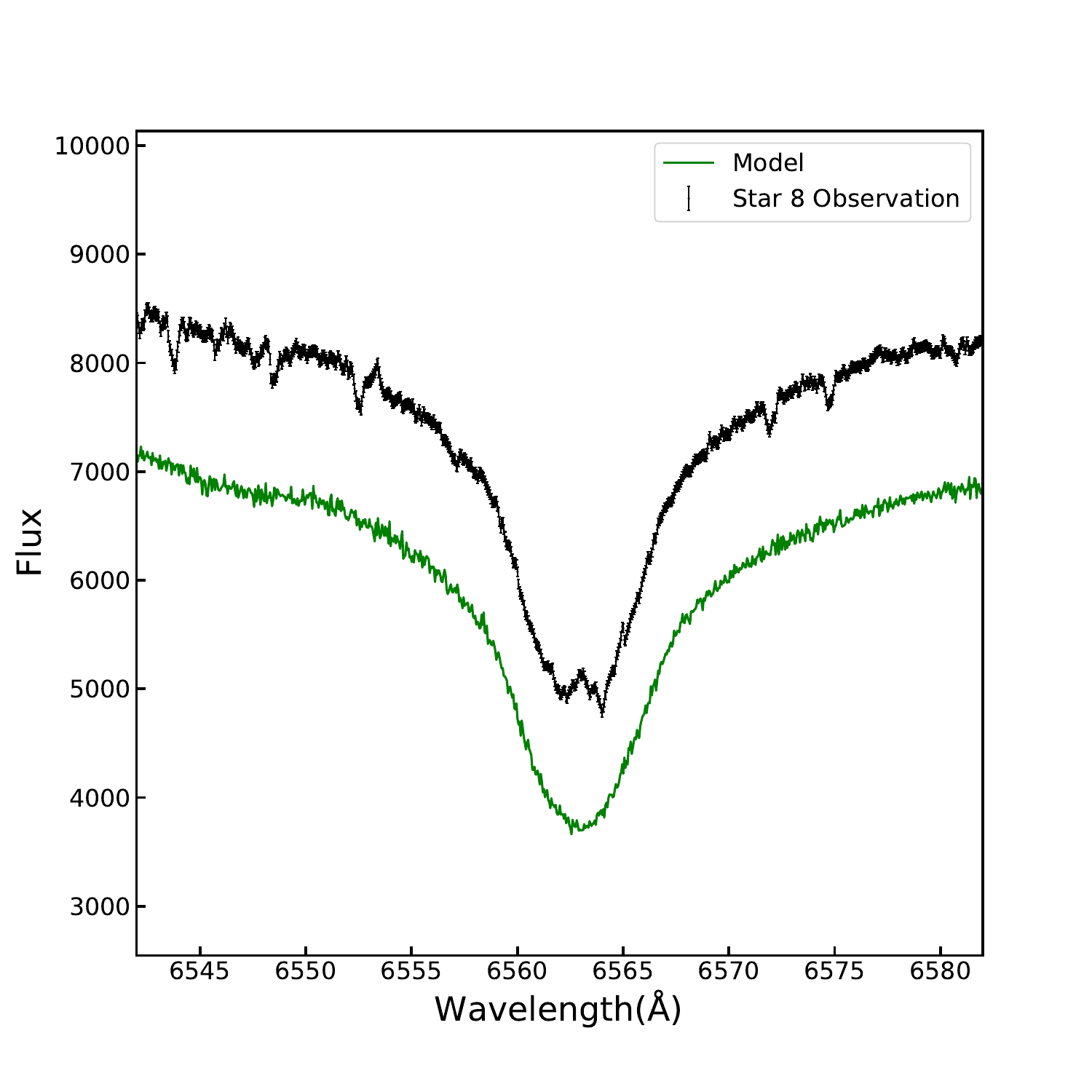}  & 
    \includegraphics[width=0.31\textwidth]{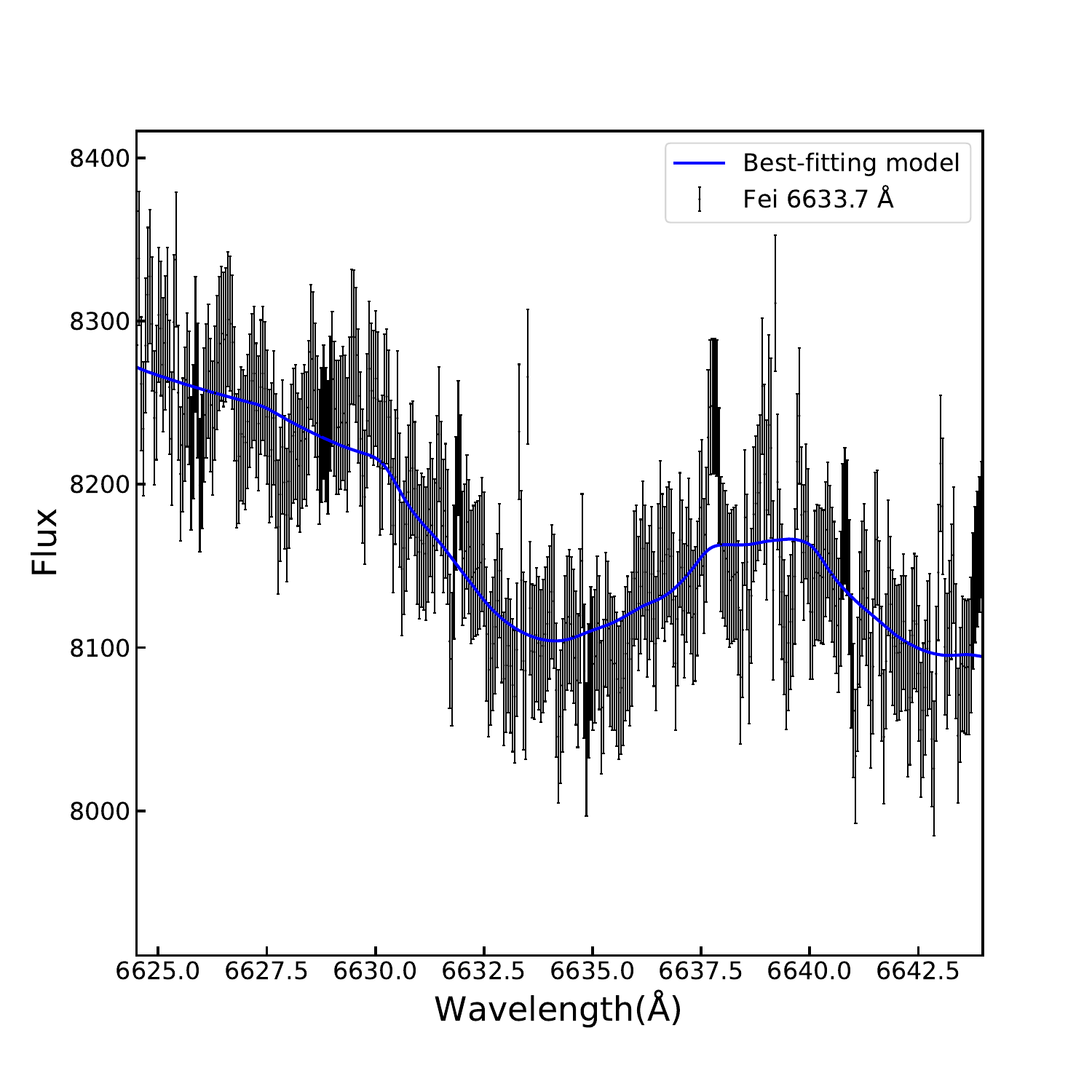}  & 
    \includegraphics[width=0.31\textwidth]{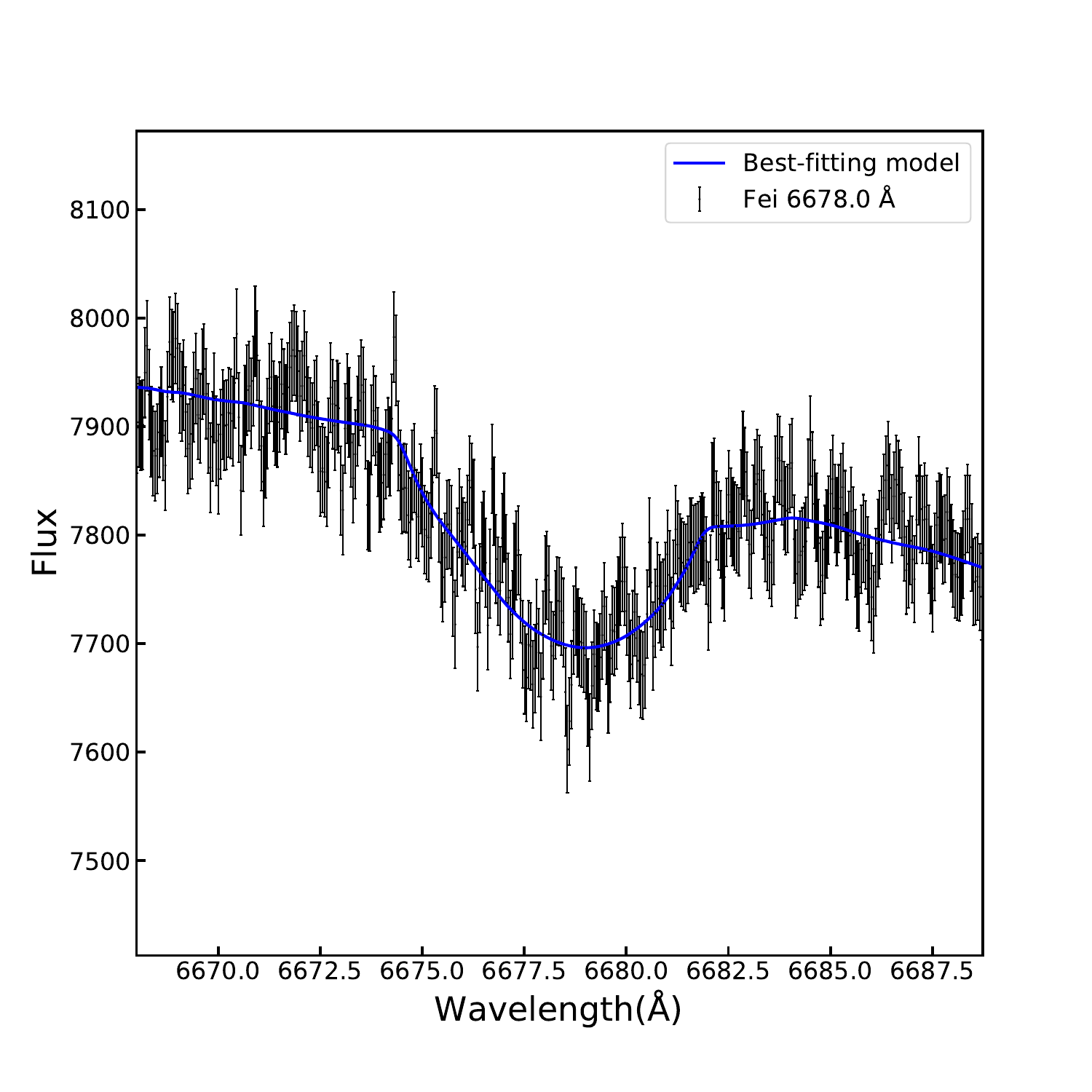}\\
    \includegraphics[width=0.31\textwidth]{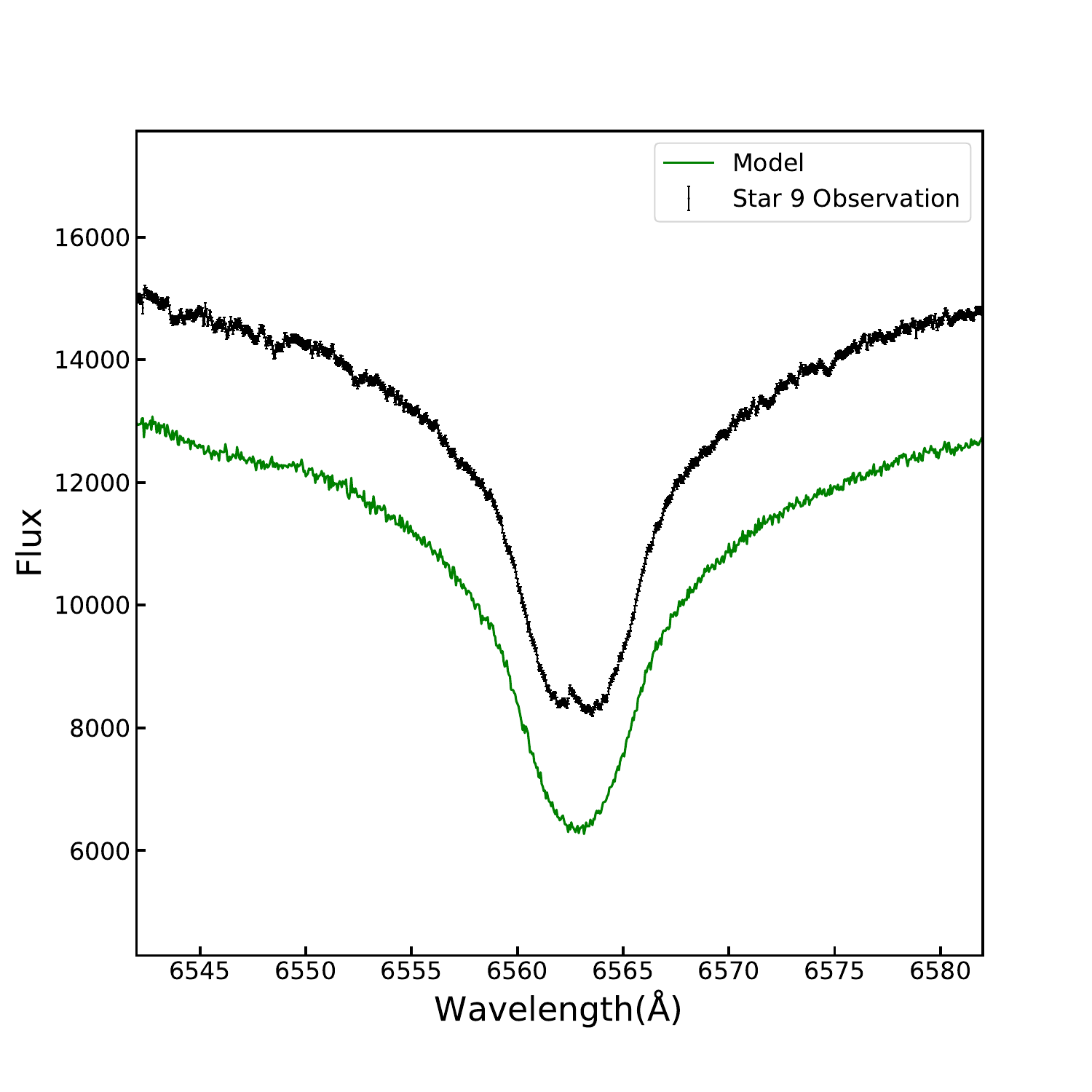}  & 
    \includegraphics[width=0.31\textwidth]{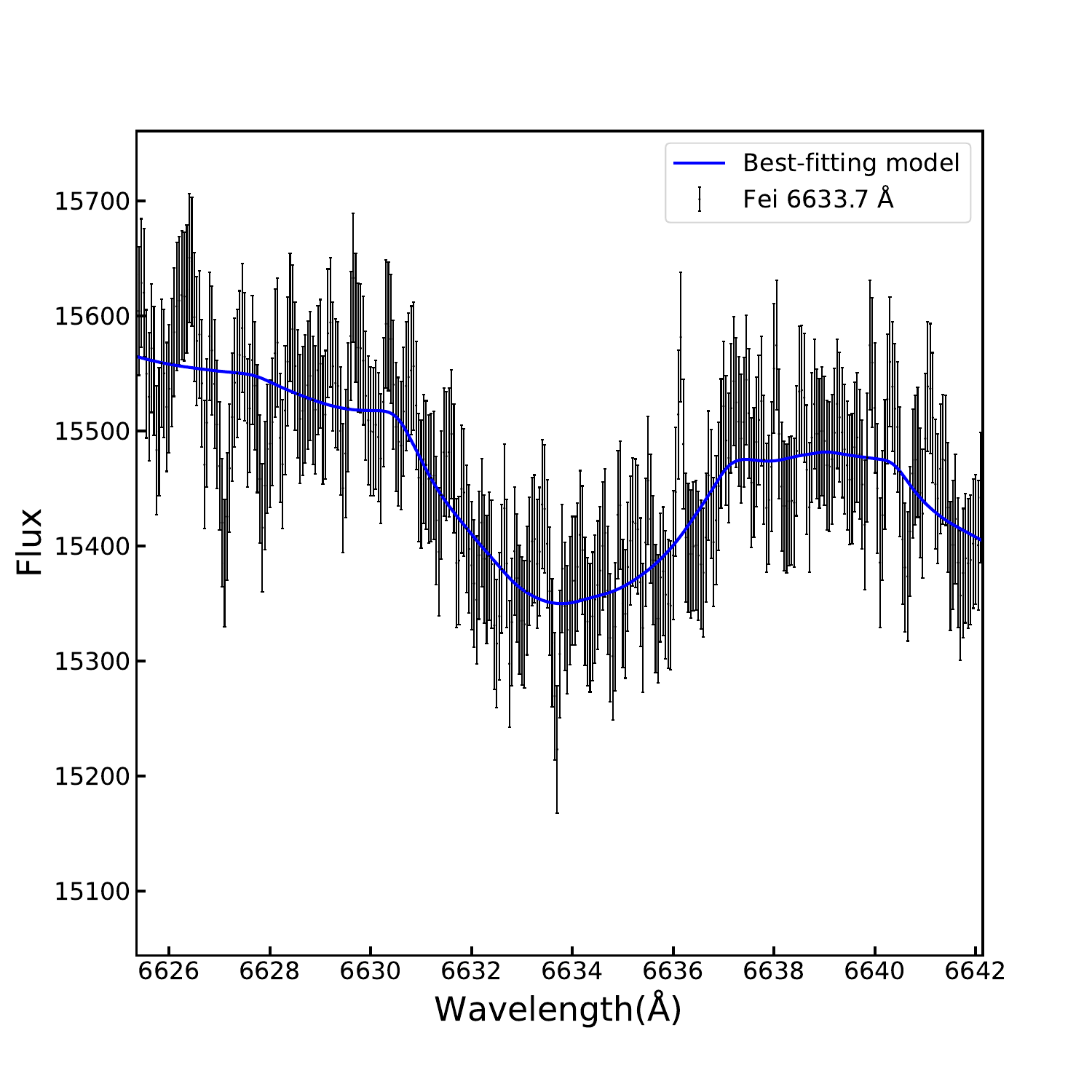 }  & 
    \includegraphics[width=0.31\textwidth]{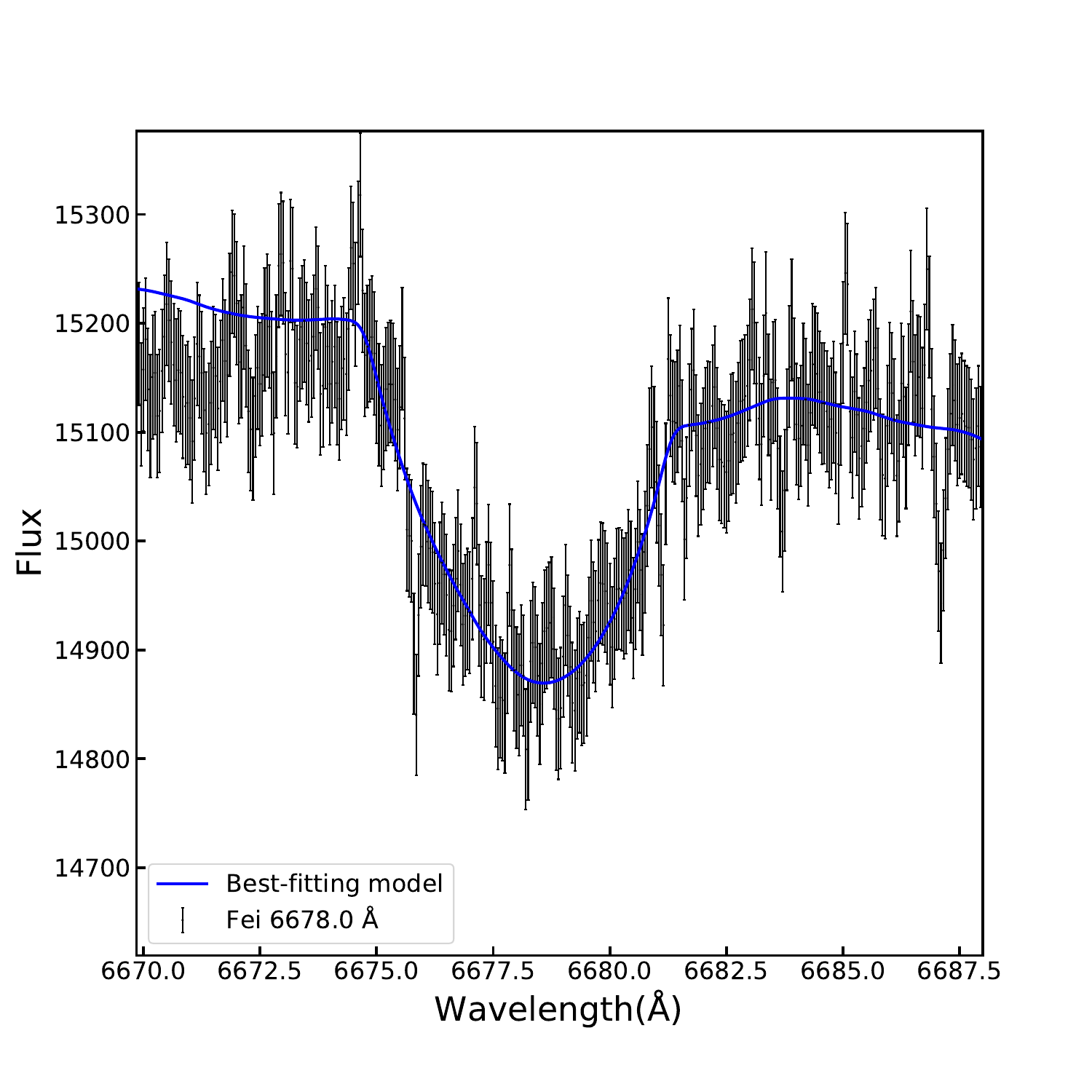}\\
     \includegraphics[width=0.31\textwidth]{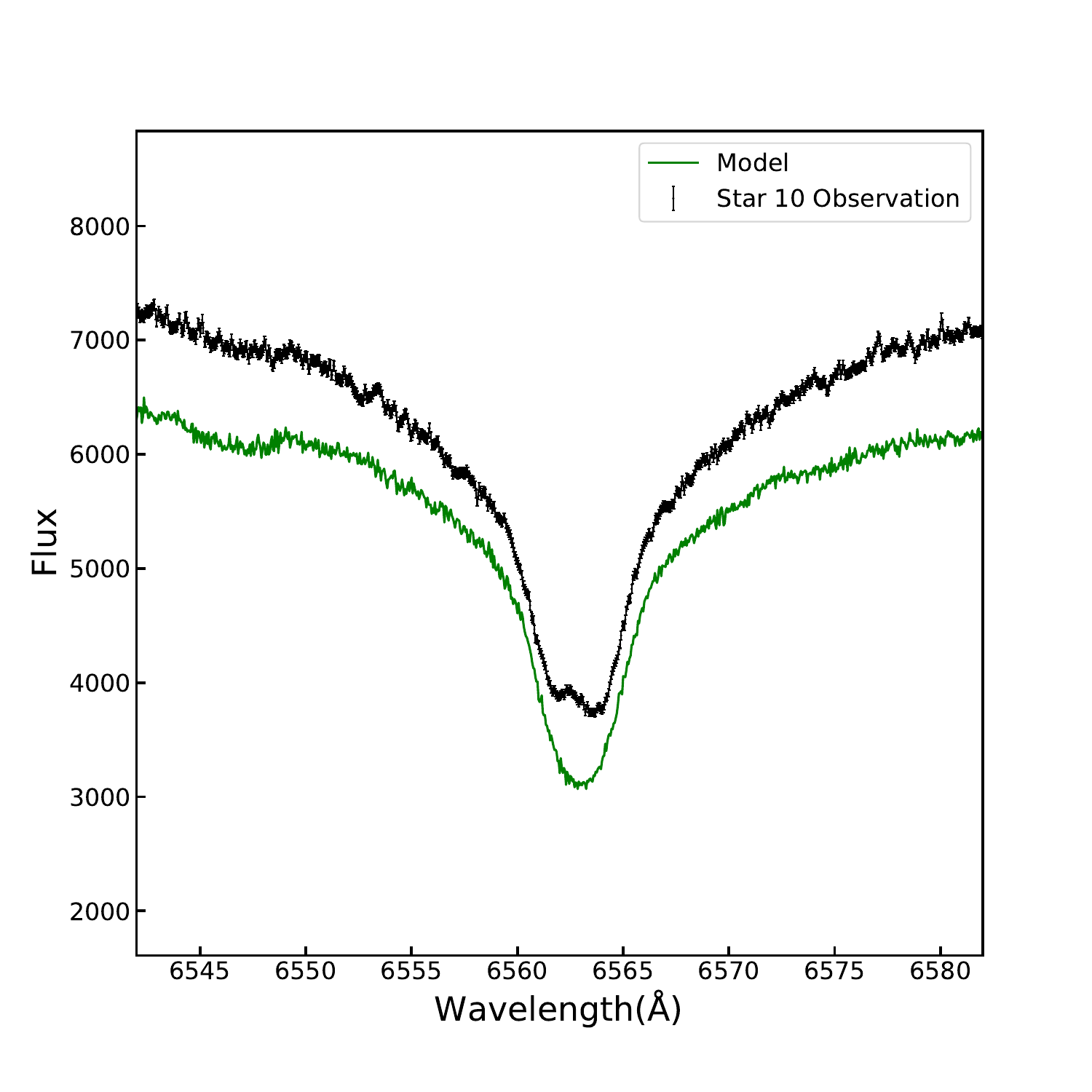}  & 
    \includegraphics[width=0.31\textwidth]{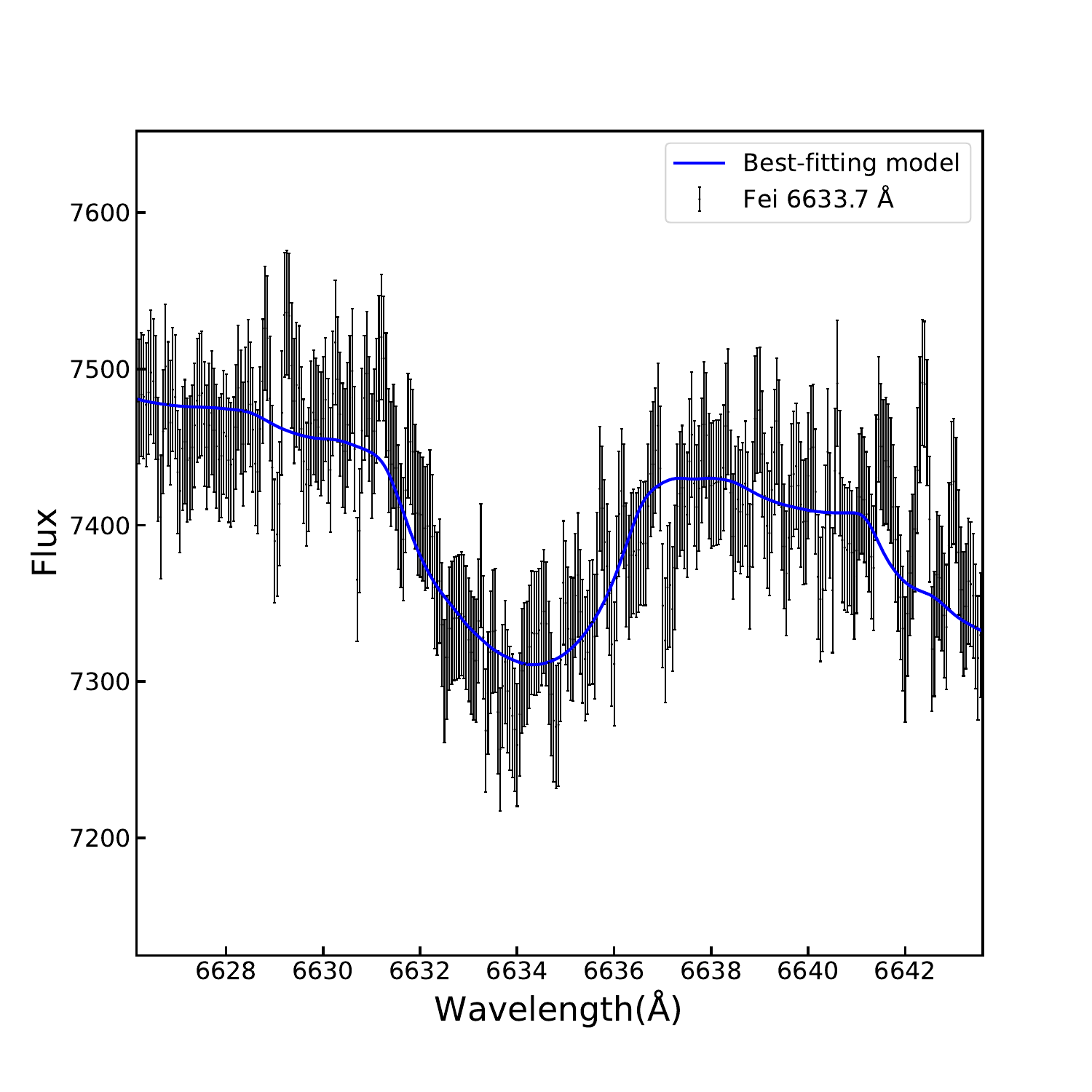 }  & 
    \includegraphics[width=0.31\textwidth]{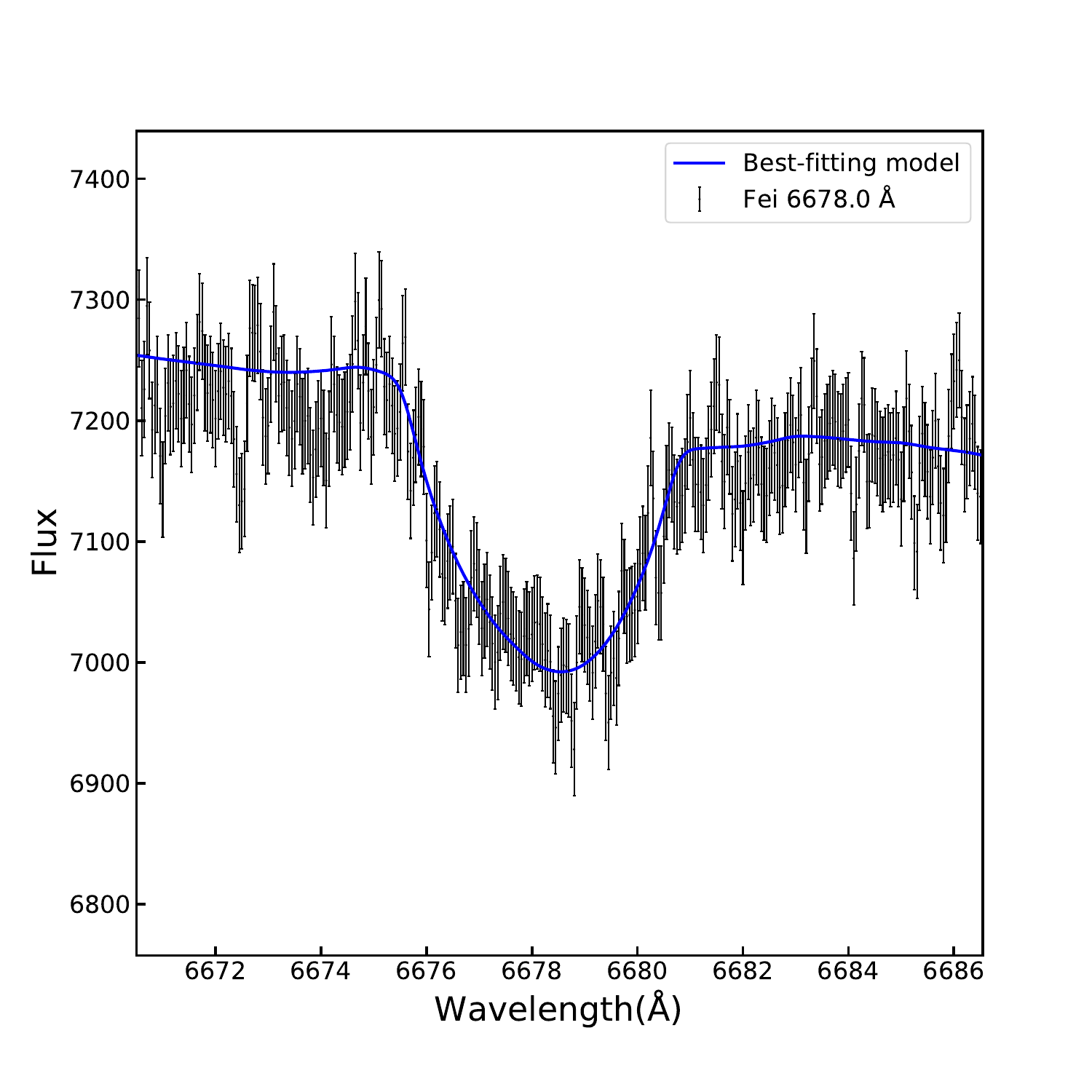}\\
     \includegraphics[width=0.31\textwidth]{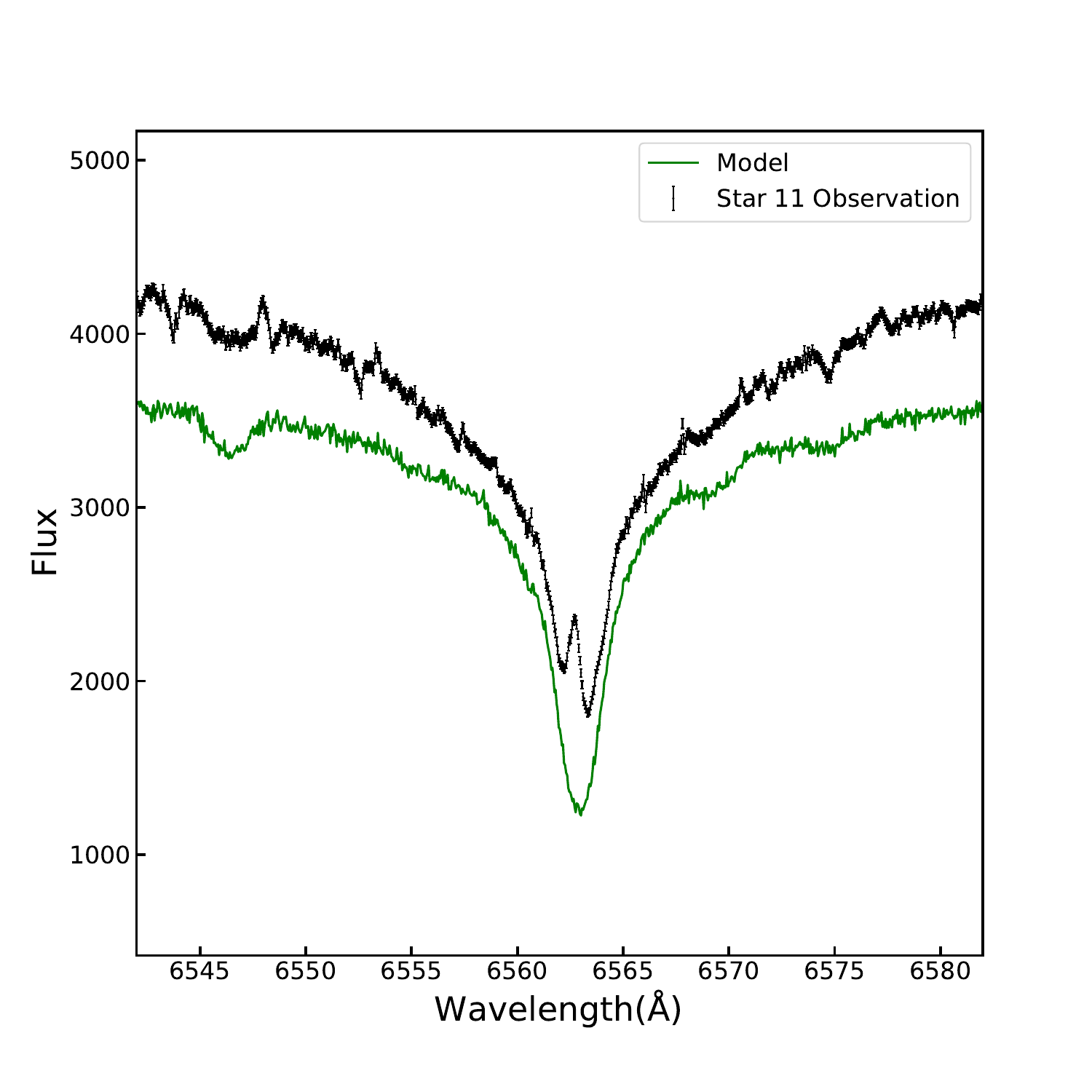}  & 
    \includegraphics[width=0.31\textwidth]{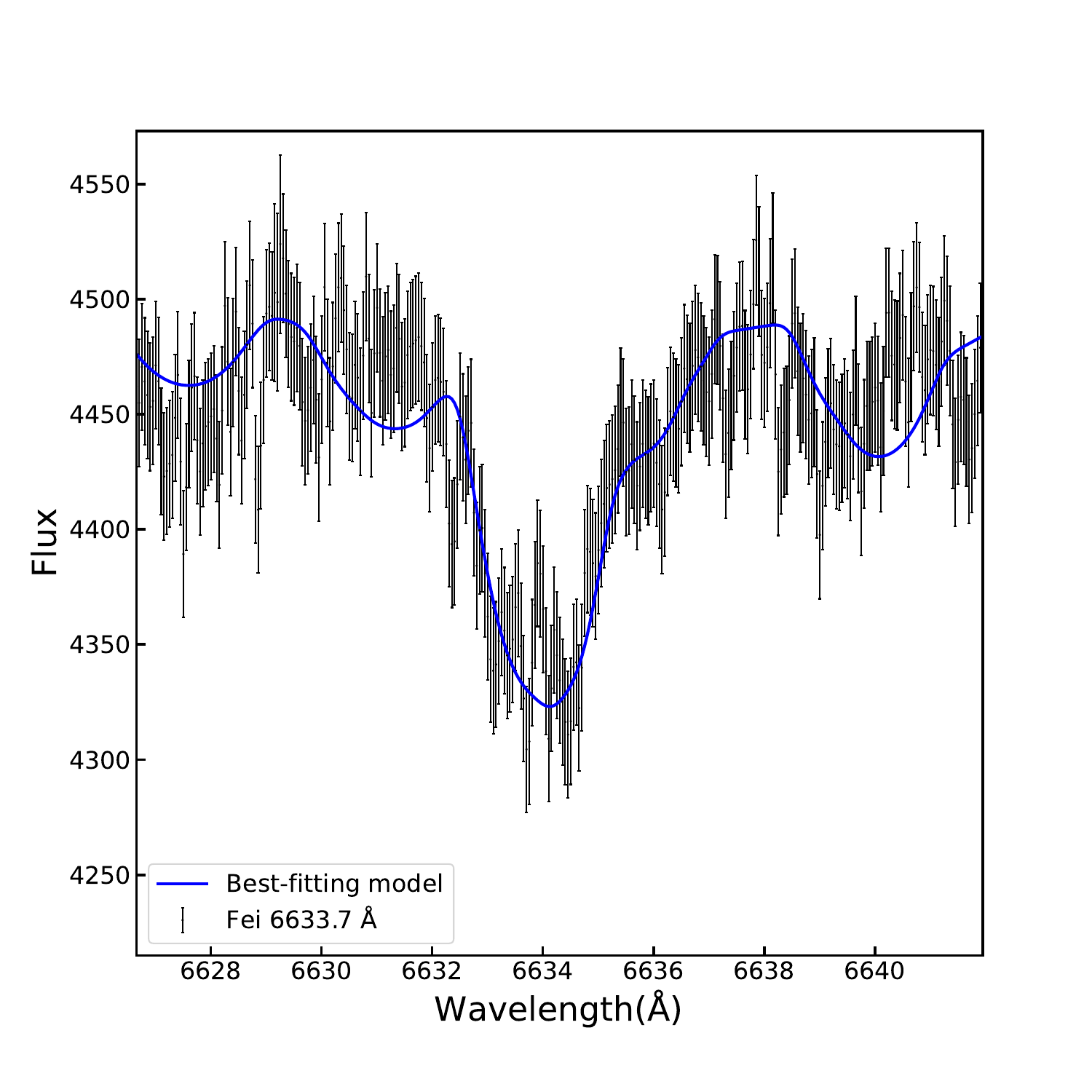 }  & 
    \includegraphics[width=0.31\textwidth]{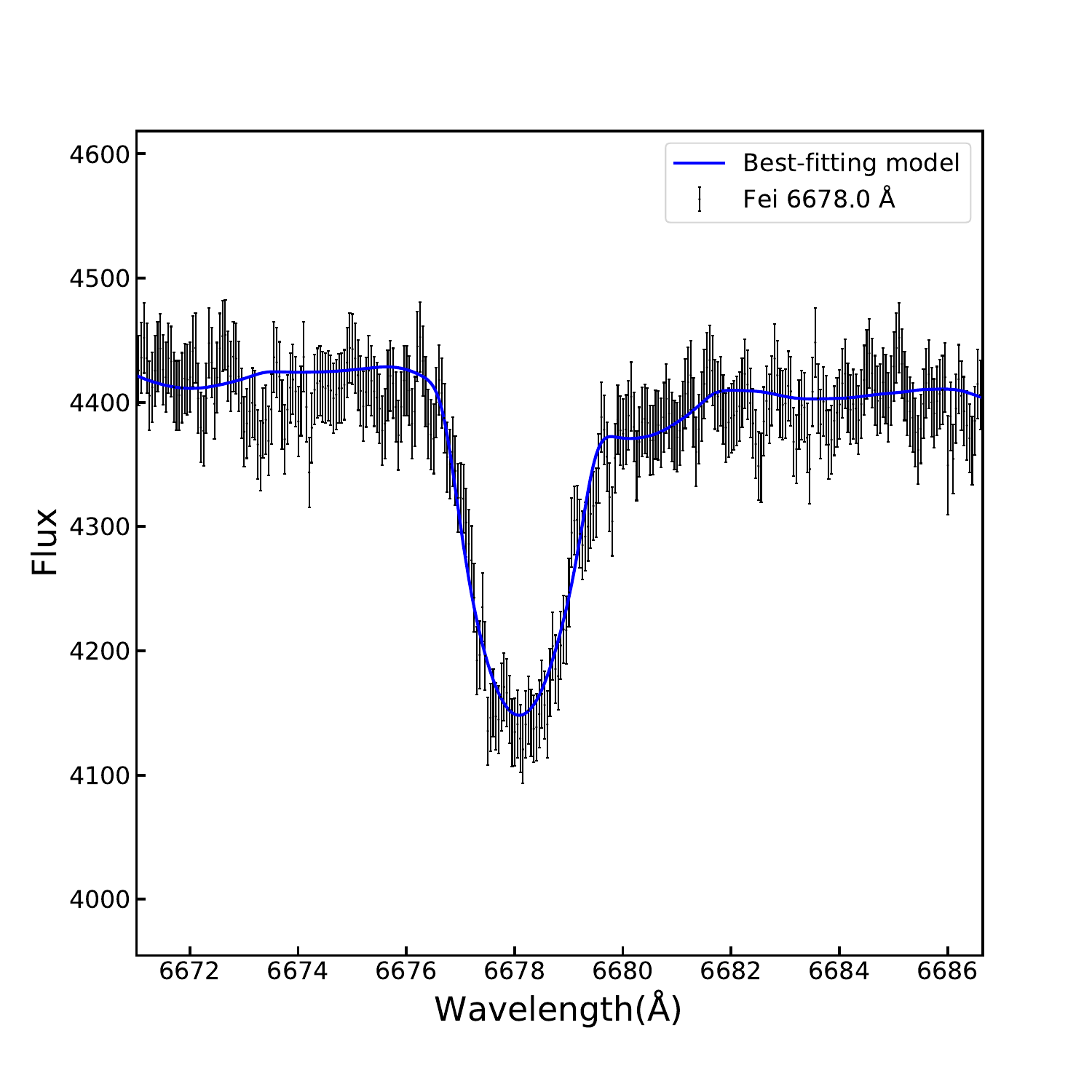}\\
\end{tabular}

	\caption{As Figure \ref{fig:3532 star 2 and 6}, but for Star 8, Star 9, Star 10 and Star 11 listed in Table \ref{tab:3532 results} from the top to the bottom, respectively.}
 \label{fig:3532 star 9}
\end{figure*}

\newpage
\begin{figure*}
\centering
\begin{tabular}{ccc}
    \includegraphics[width=0.31\textwidth]{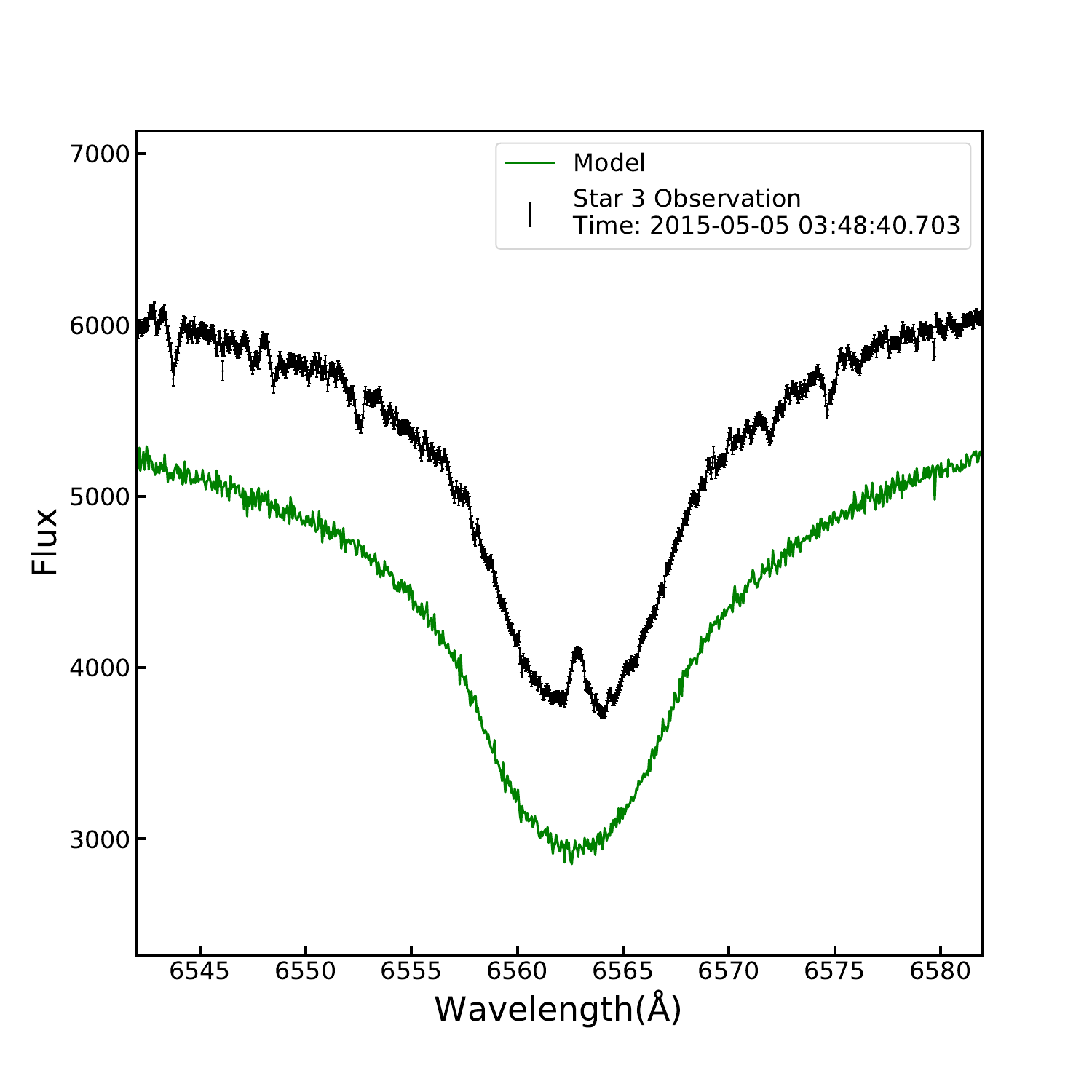}  & 
    \includegraphics[width=0.31\textwidth]{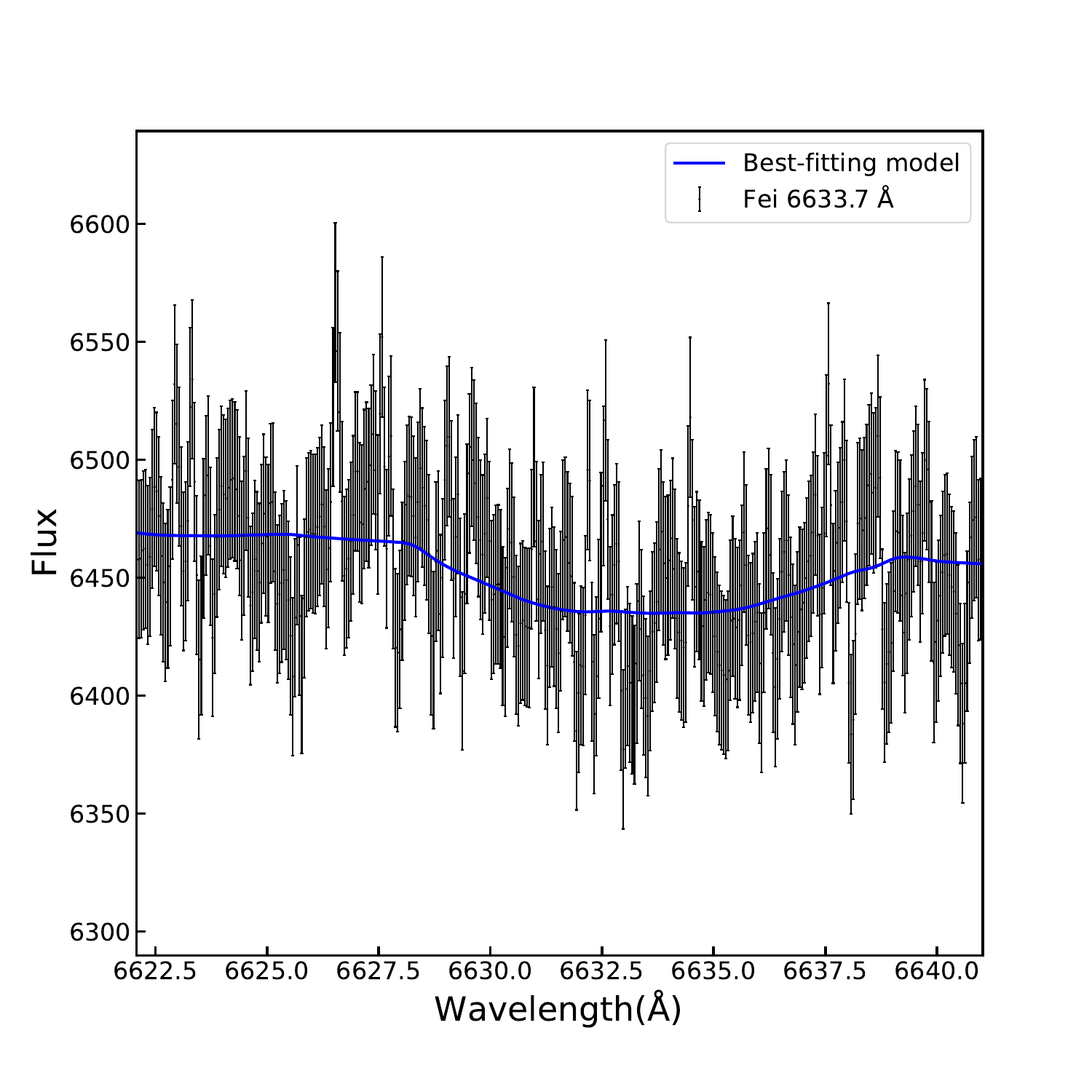}  & 
    \includegraphics[width=0.31\textwidth]{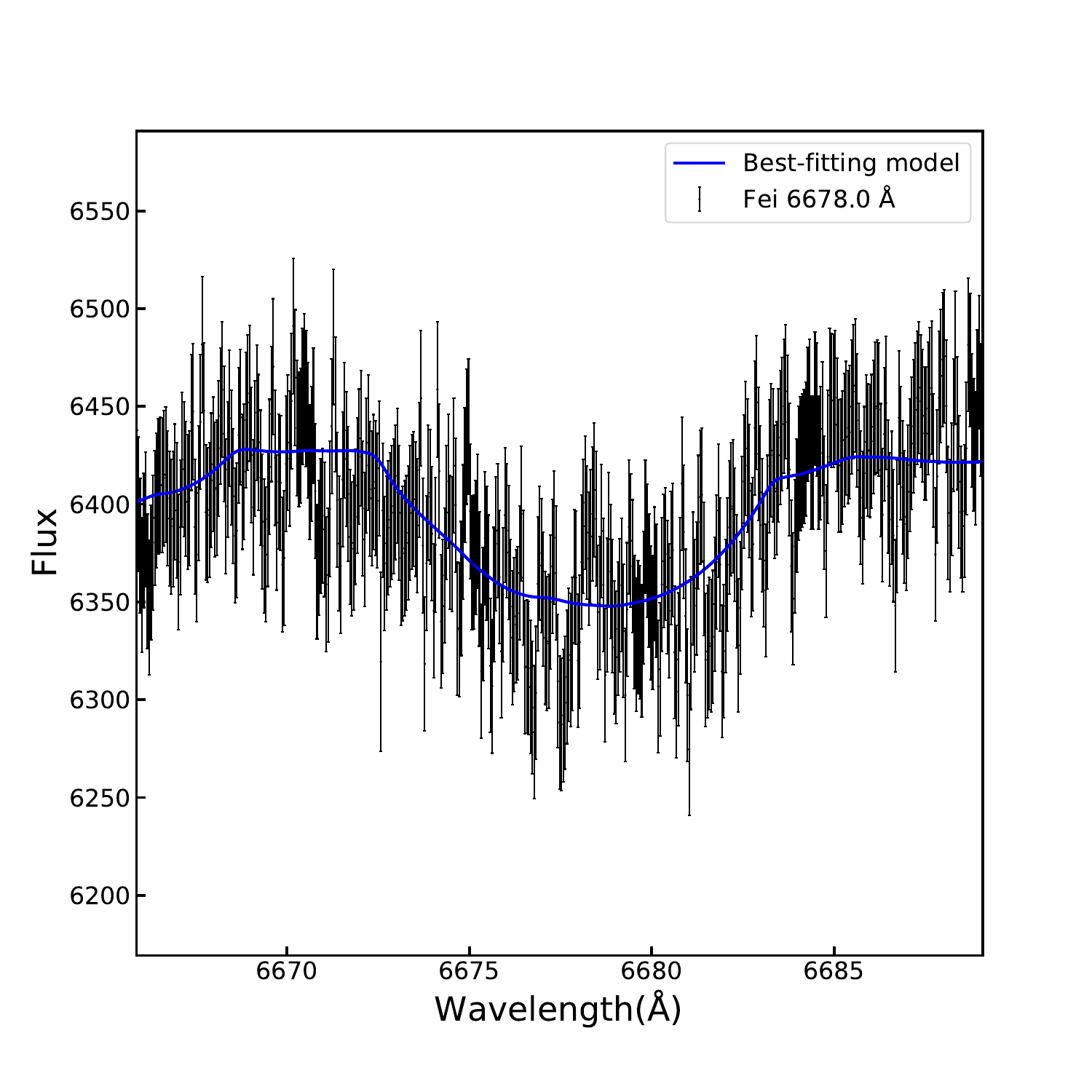}\\
    \includegraphics[width=0.31\textwidth]{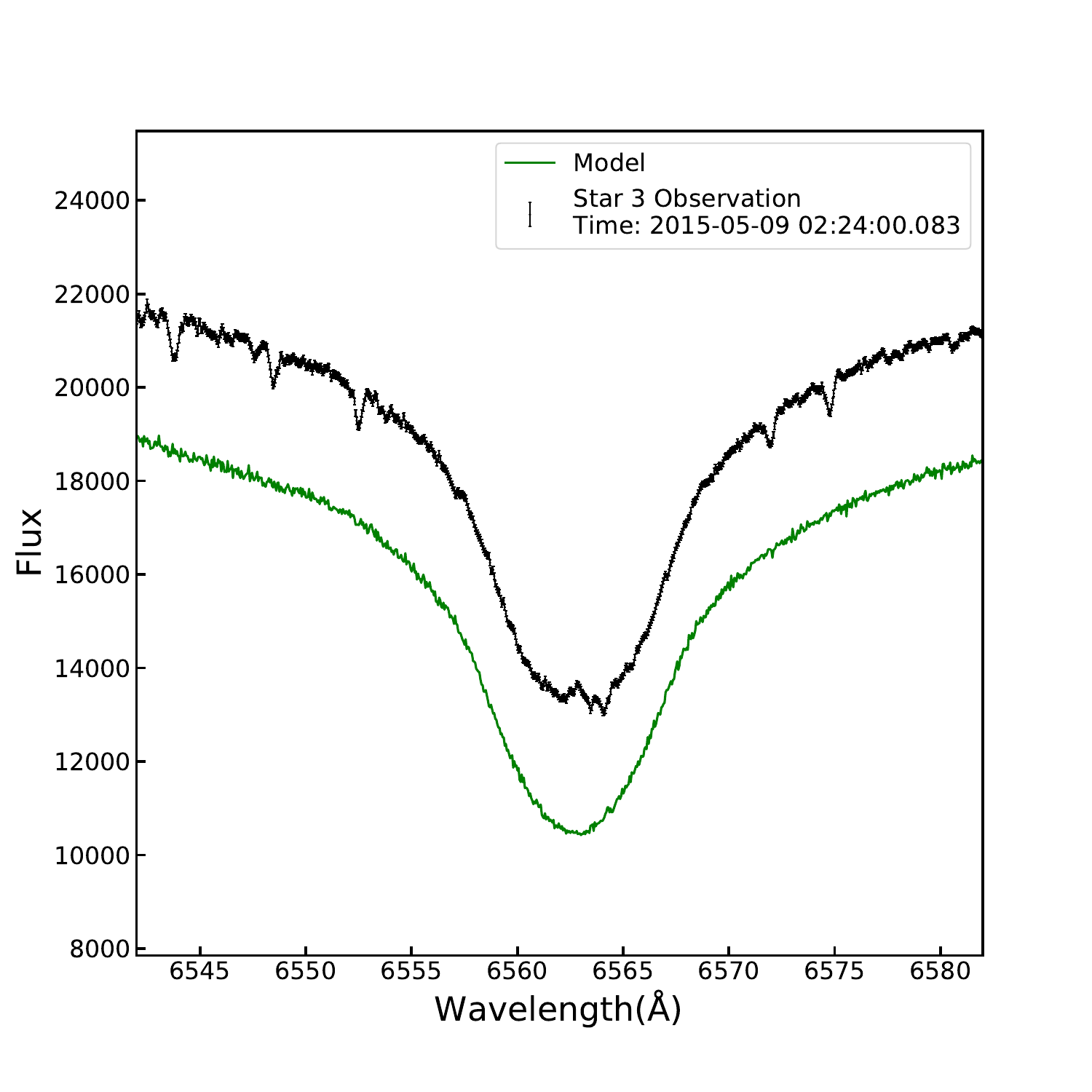}  & 
    \includegraphics[width=0.31\textwidth]{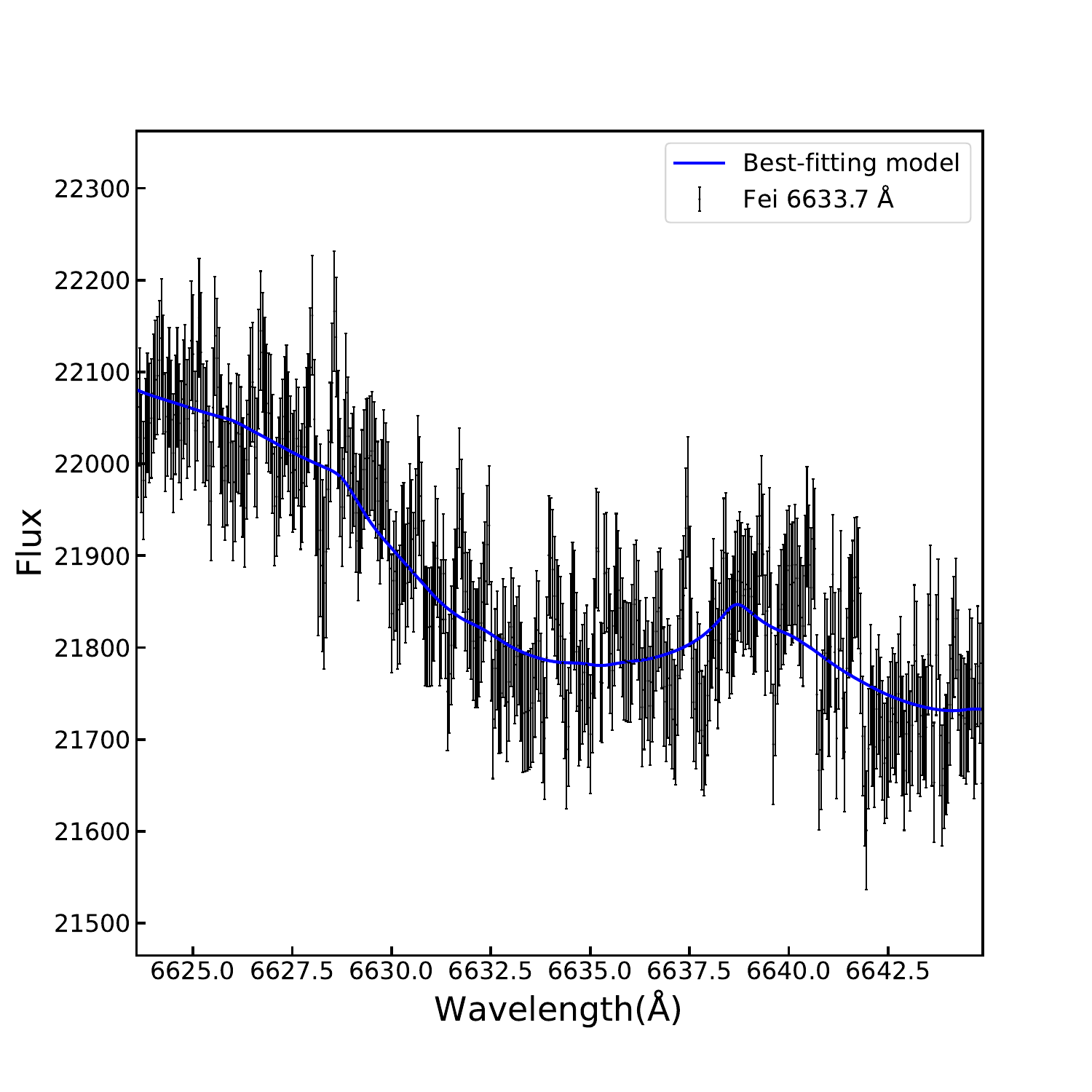 }  & 
    \includegraphics[width=0.31\textwidth]{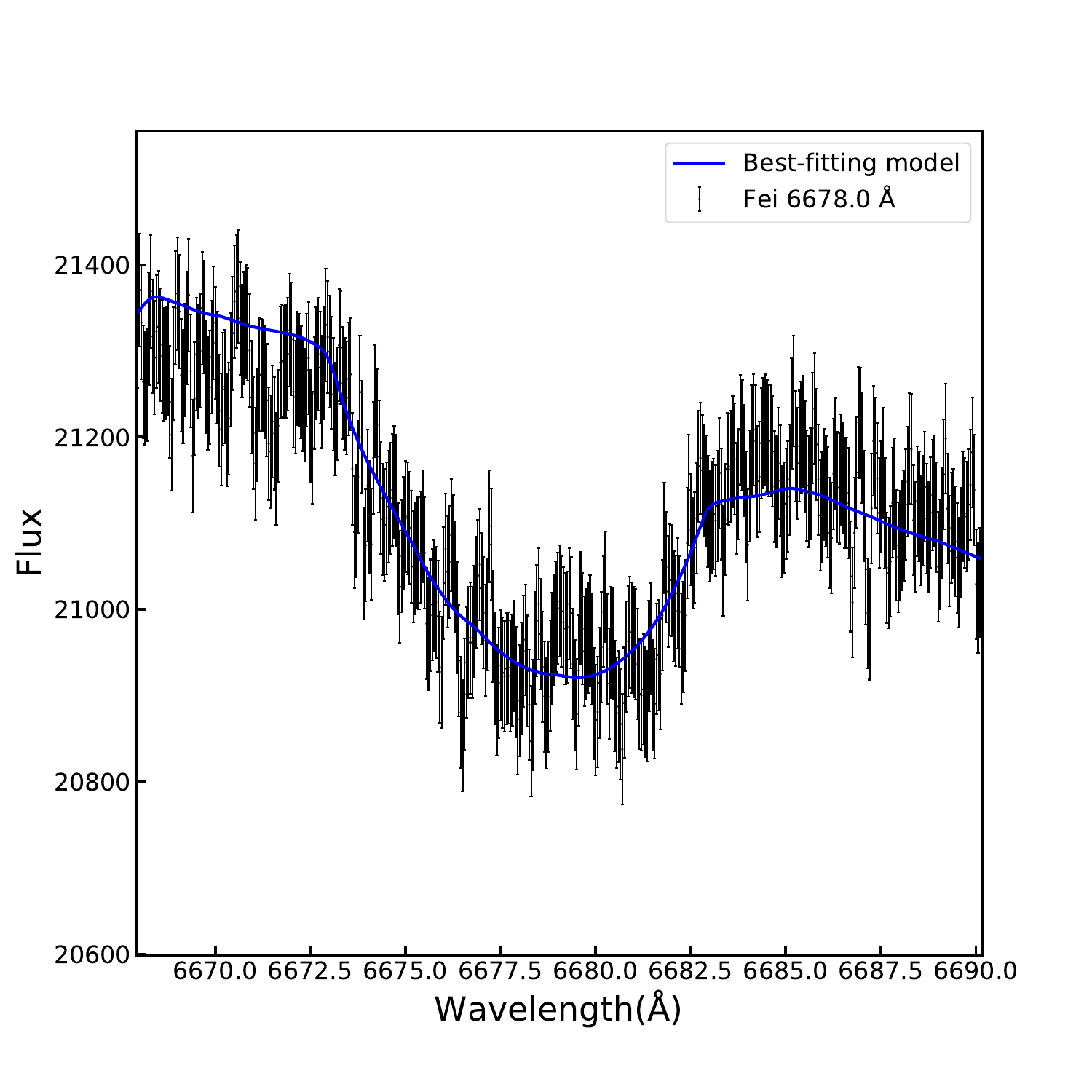}\\
     \includegraphics[width=0.31\textwidth]{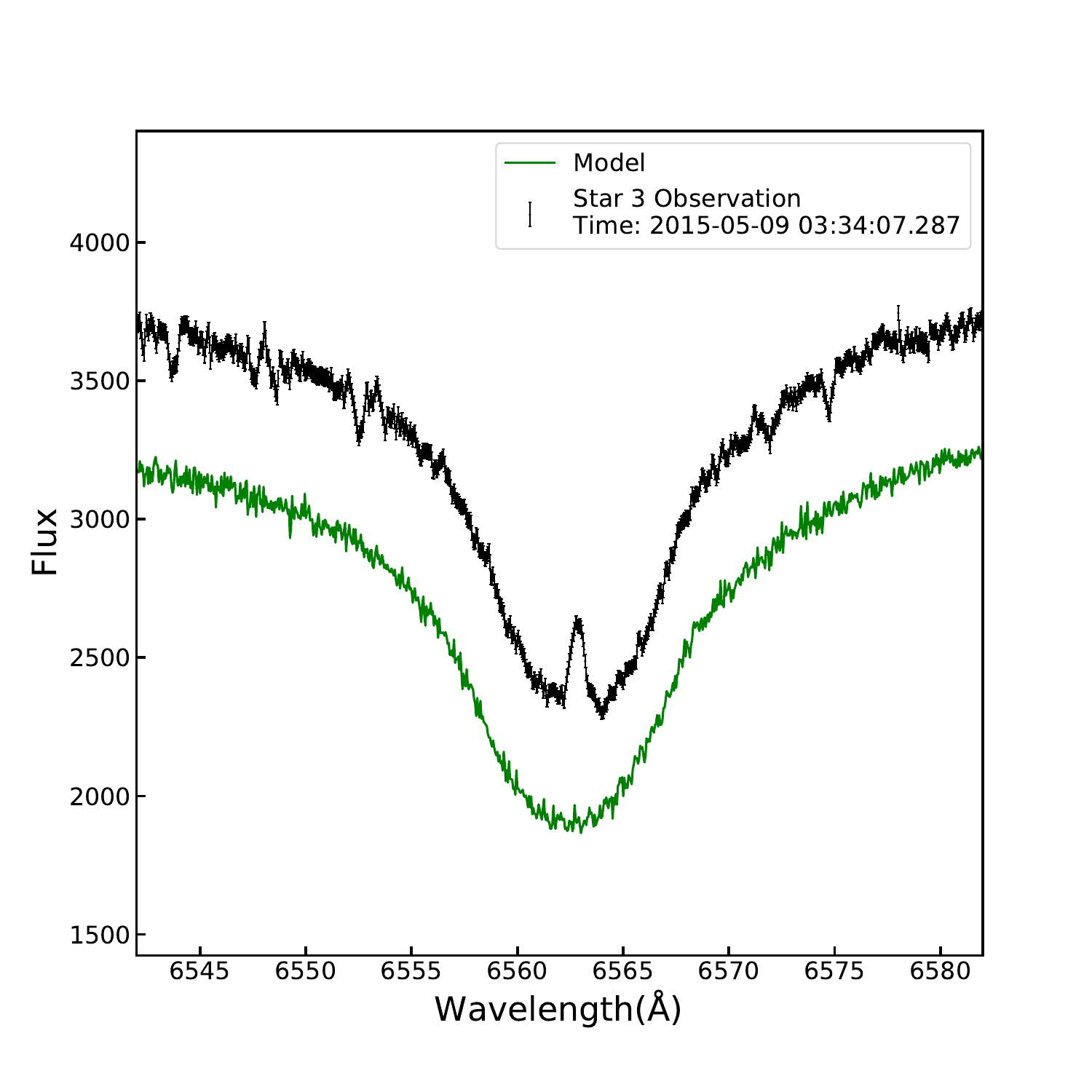}  & 
    \includegraphics[width=0.31\textwidth]{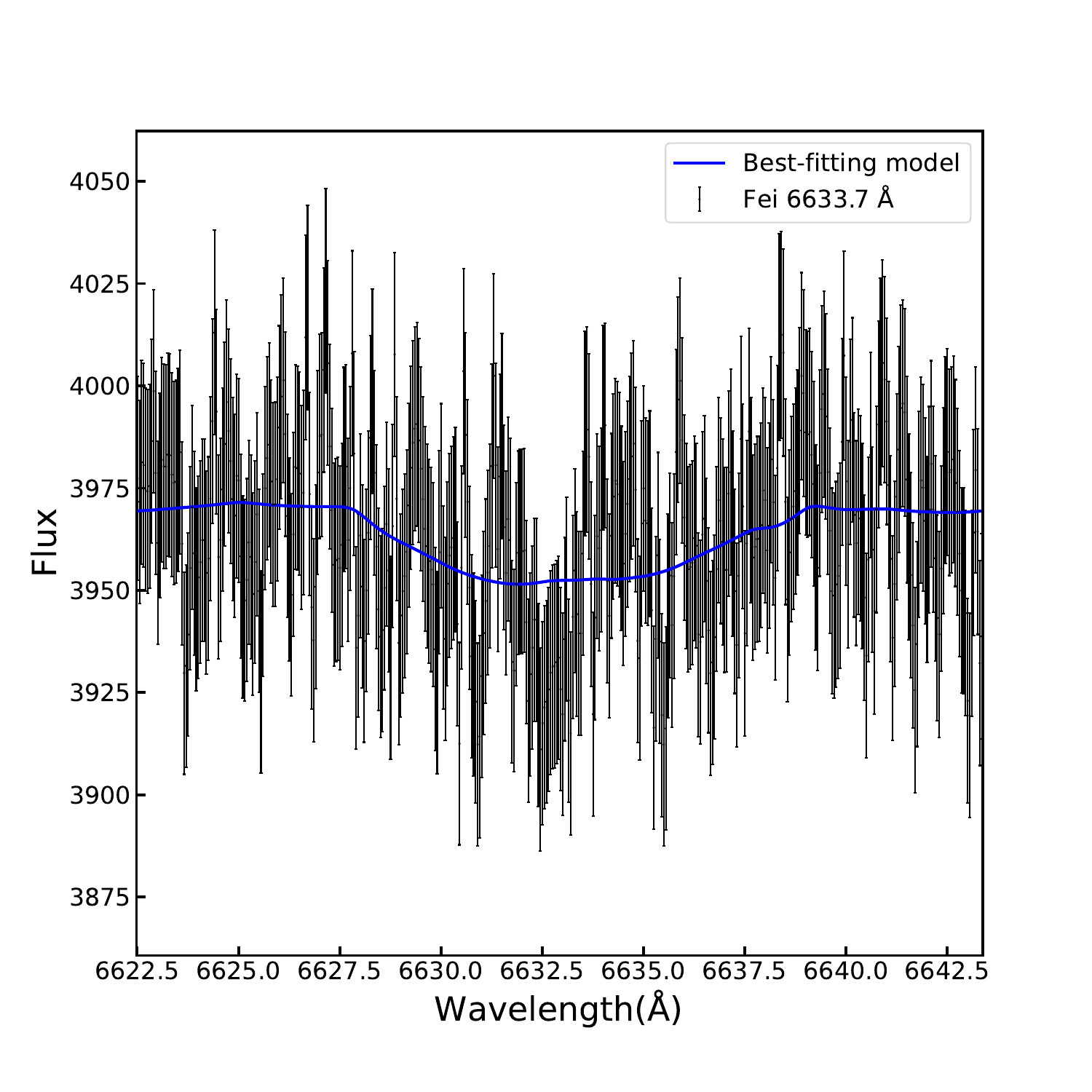 }  & 
    \includegraphics[width=0.31\textwidth]{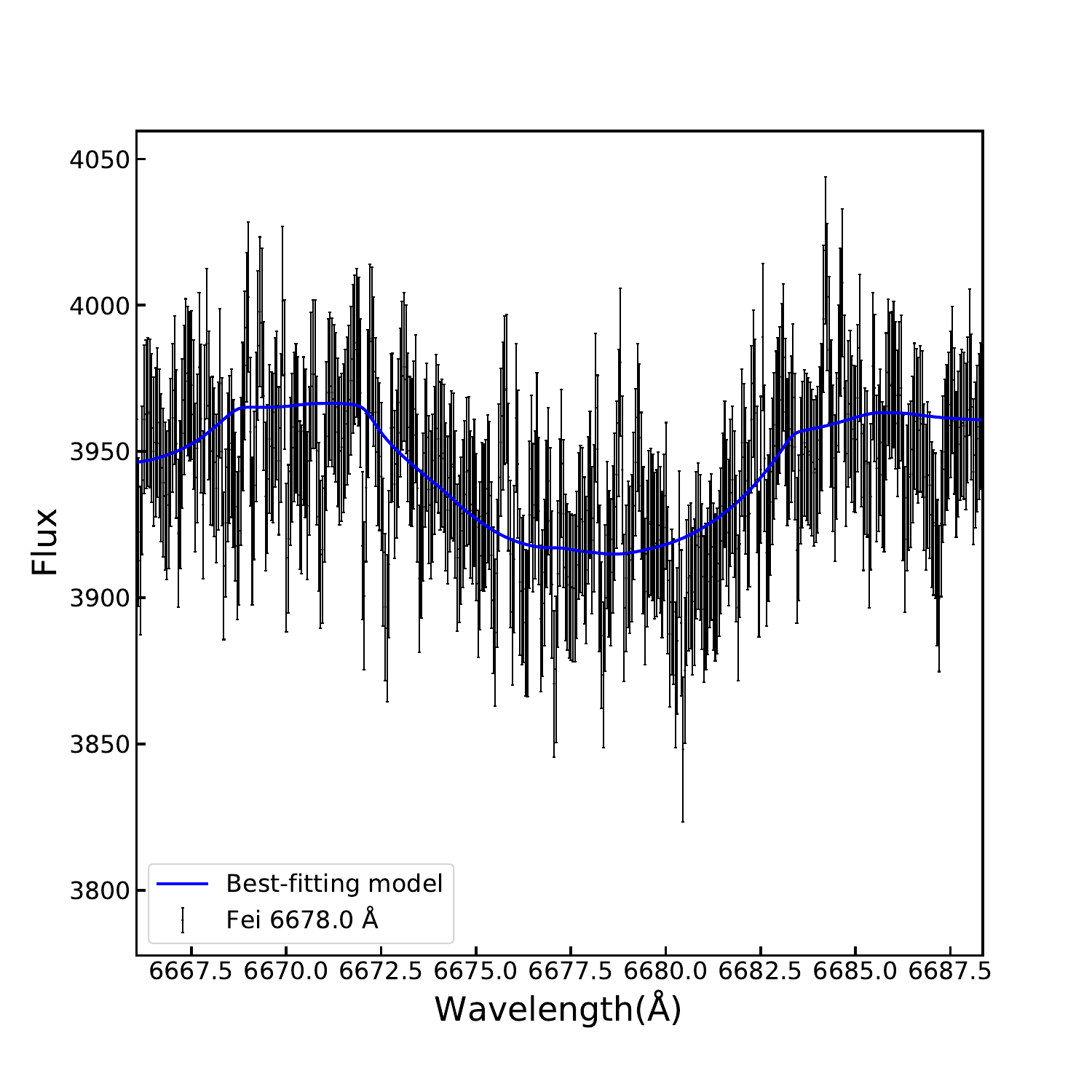}\\  
\end{tabular}

	\caption{As Figure \ref{fig:3532 star 2 and 6}, but for Star 3 listed in Table \ref{tab:3532 results} only. The panels in different lines represents the observed spectra and the corresponding models of different epochs, whose date and time are shown in the legend of the left panels.} 
 \label{fig:3532 star 3}
\end{figure*}

\newpage

\begin{figure}
\centering
\begin{tabular}{ccc}
    \includegraphics[width=0.31\textwidth]{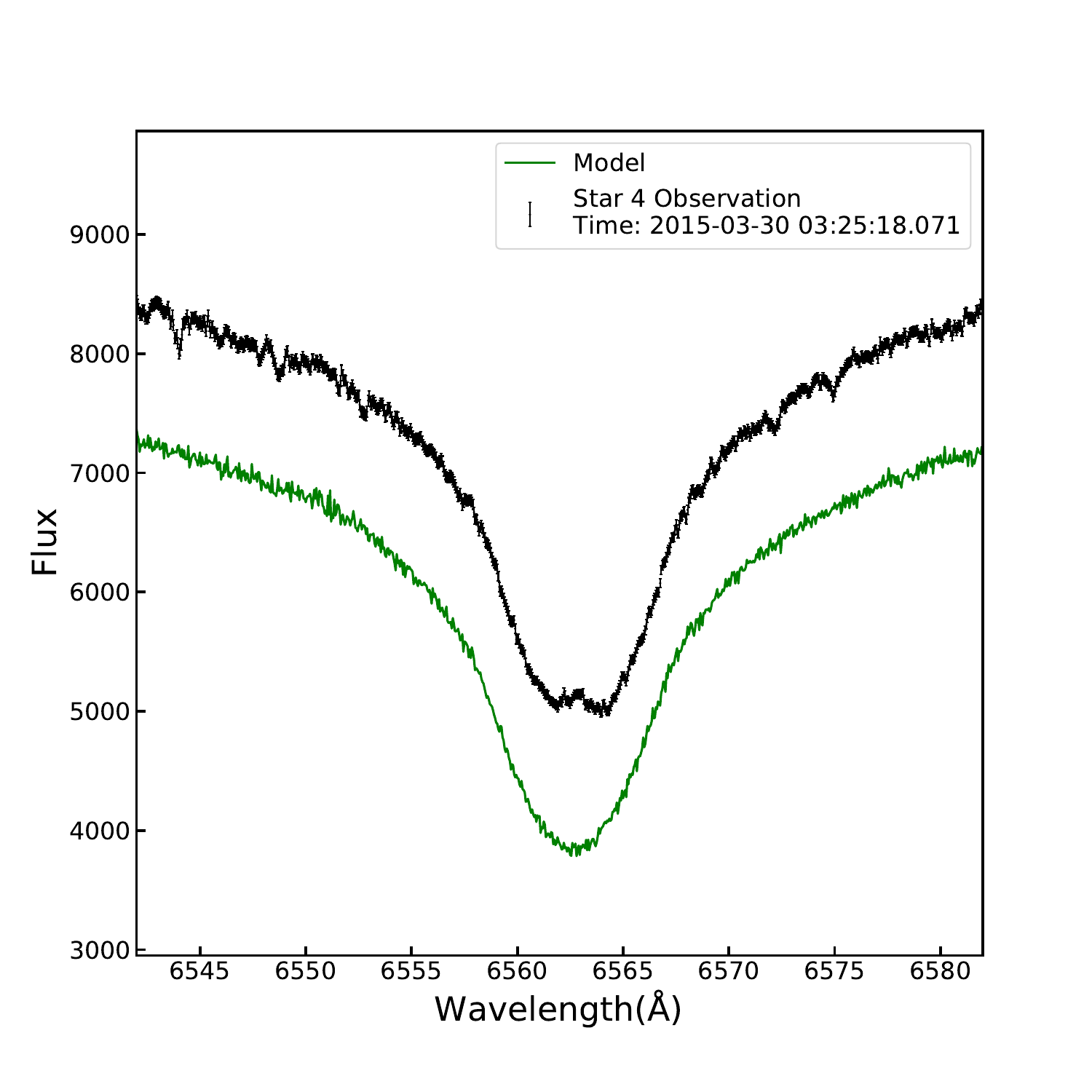}  & 
    \includegraphics[width=0.31\textwidth]{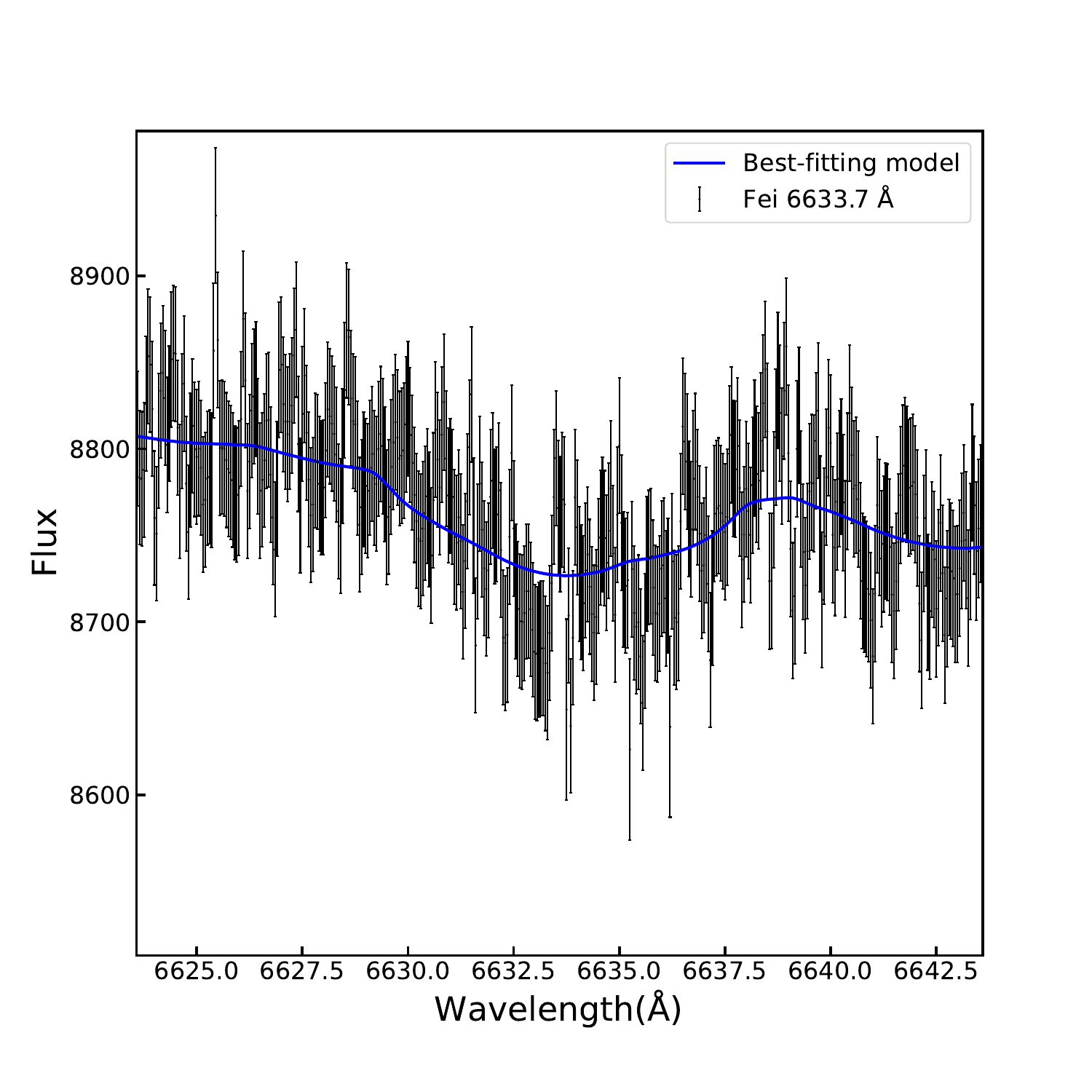}  & 
    \includegraphics[width=0.31\textwidth]{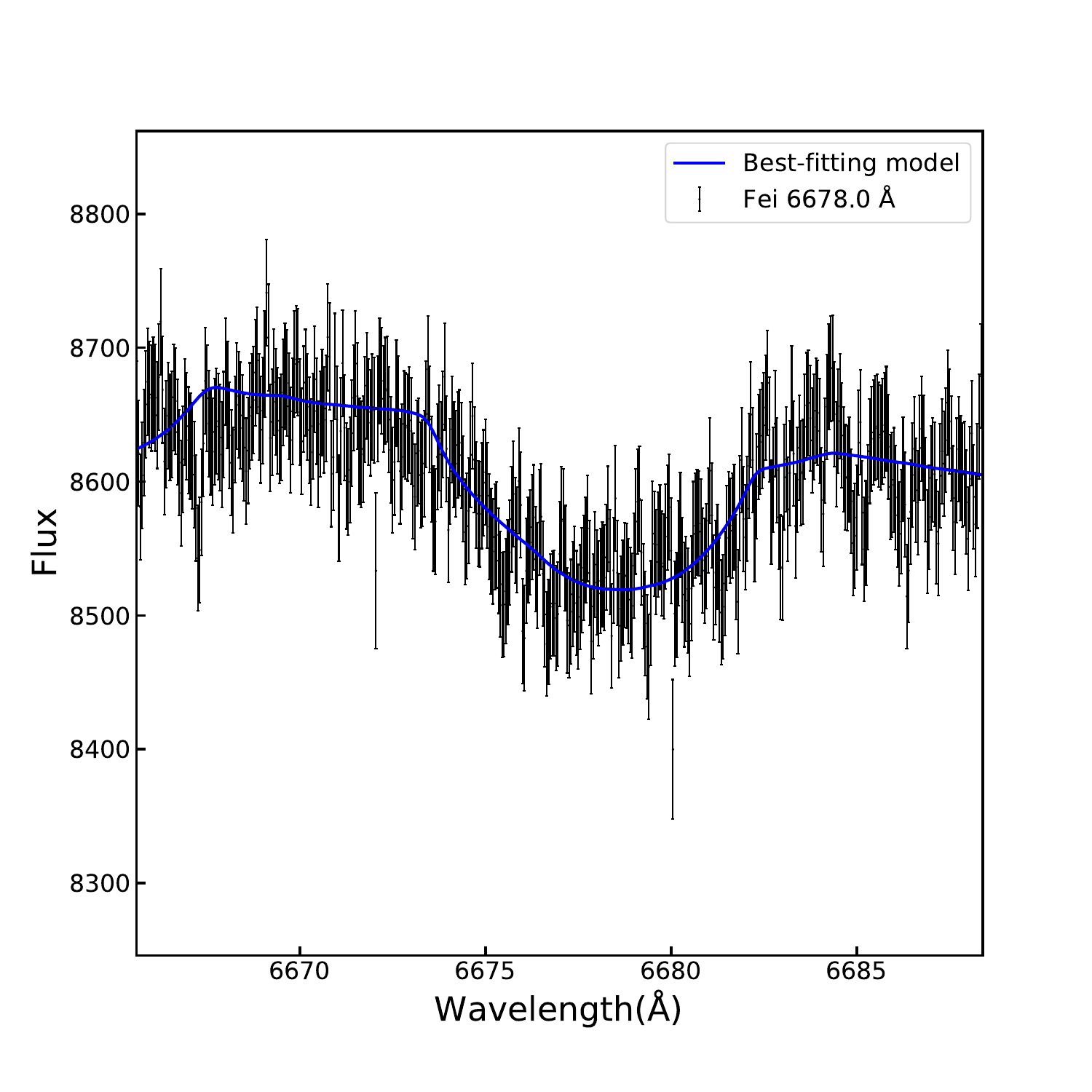}\\
    \includegraphics[width=0.31\textwidth]{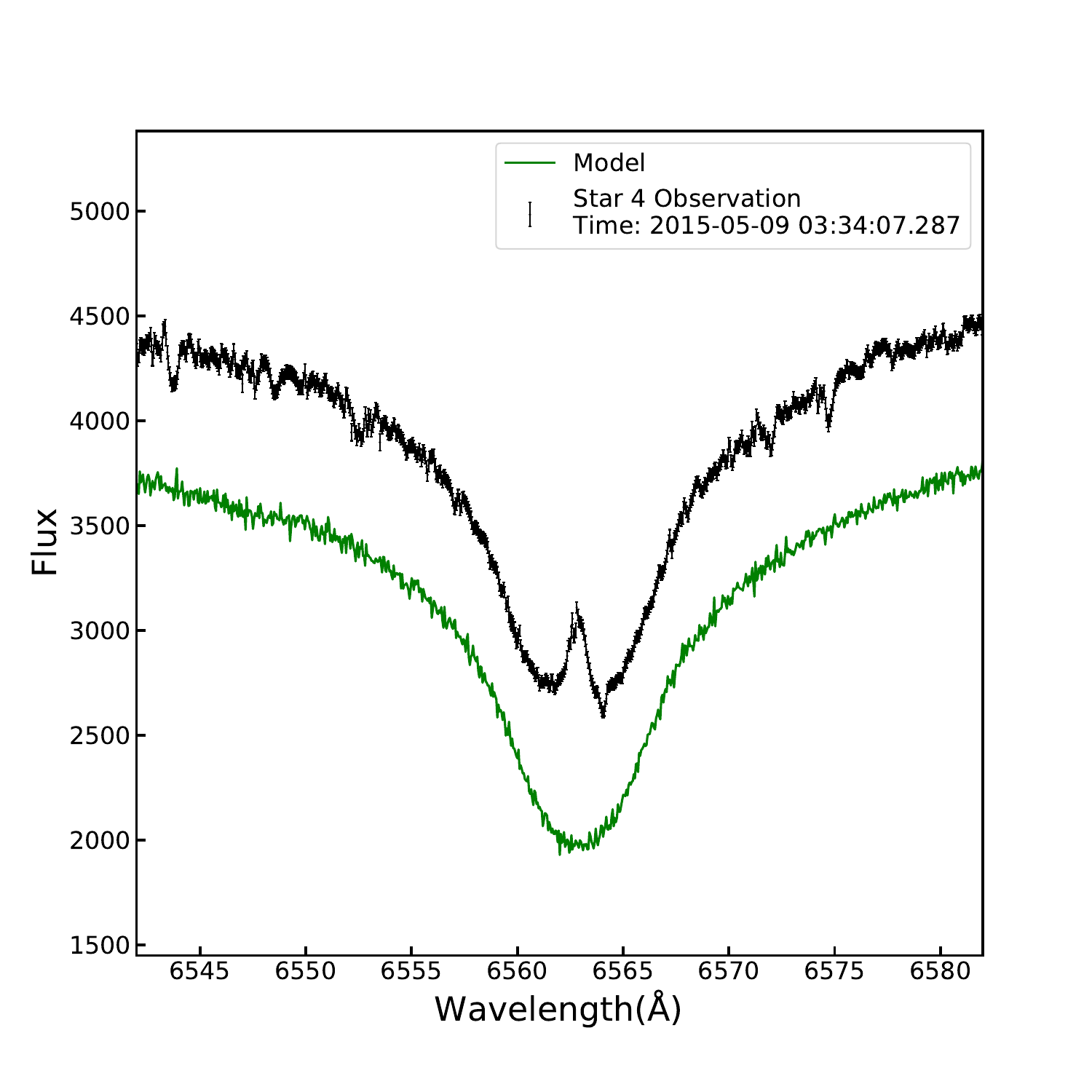}  & 
    \includegraphics[width=0.31\textwidth]{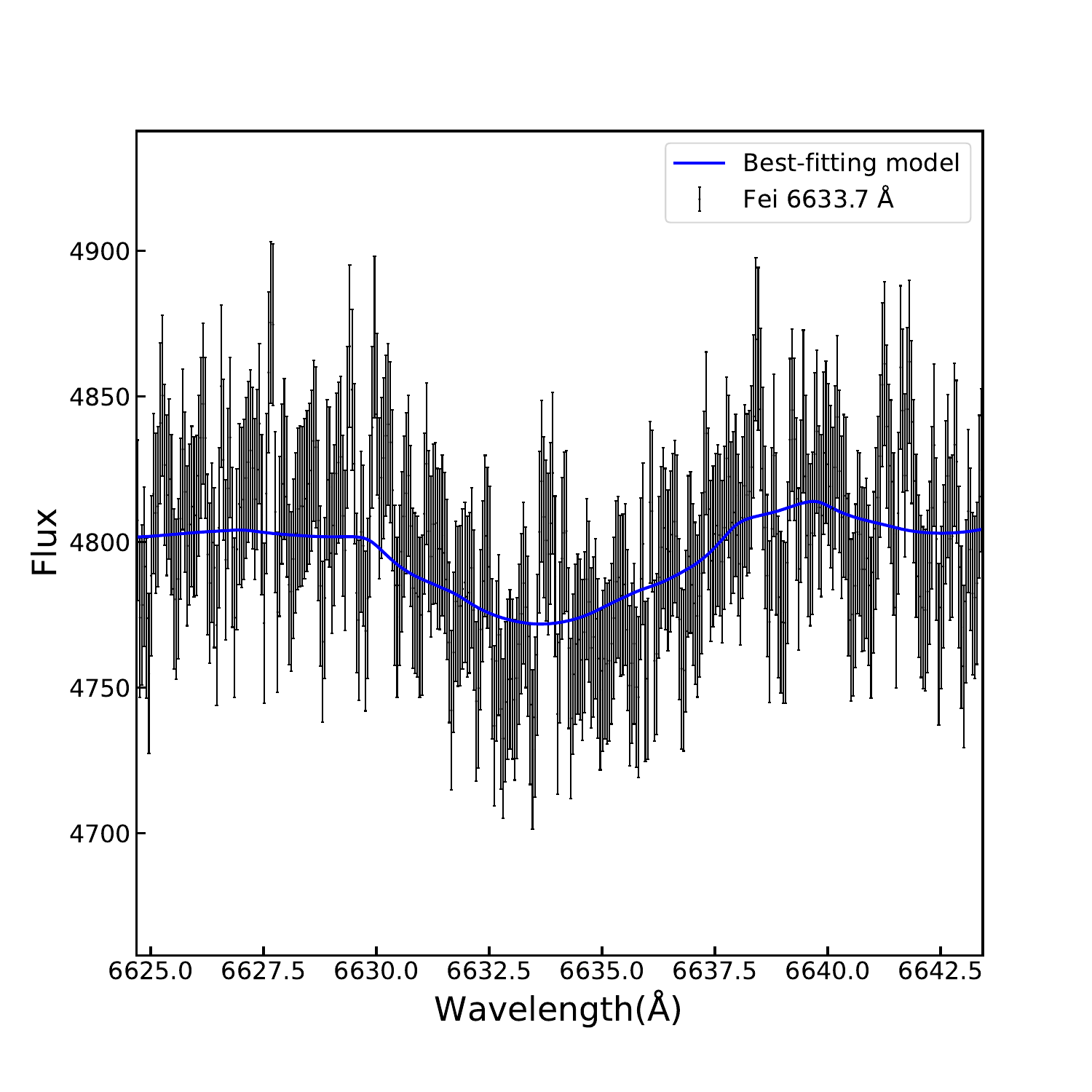}  & 
    \includegraphics[width=0.31\textwidth]{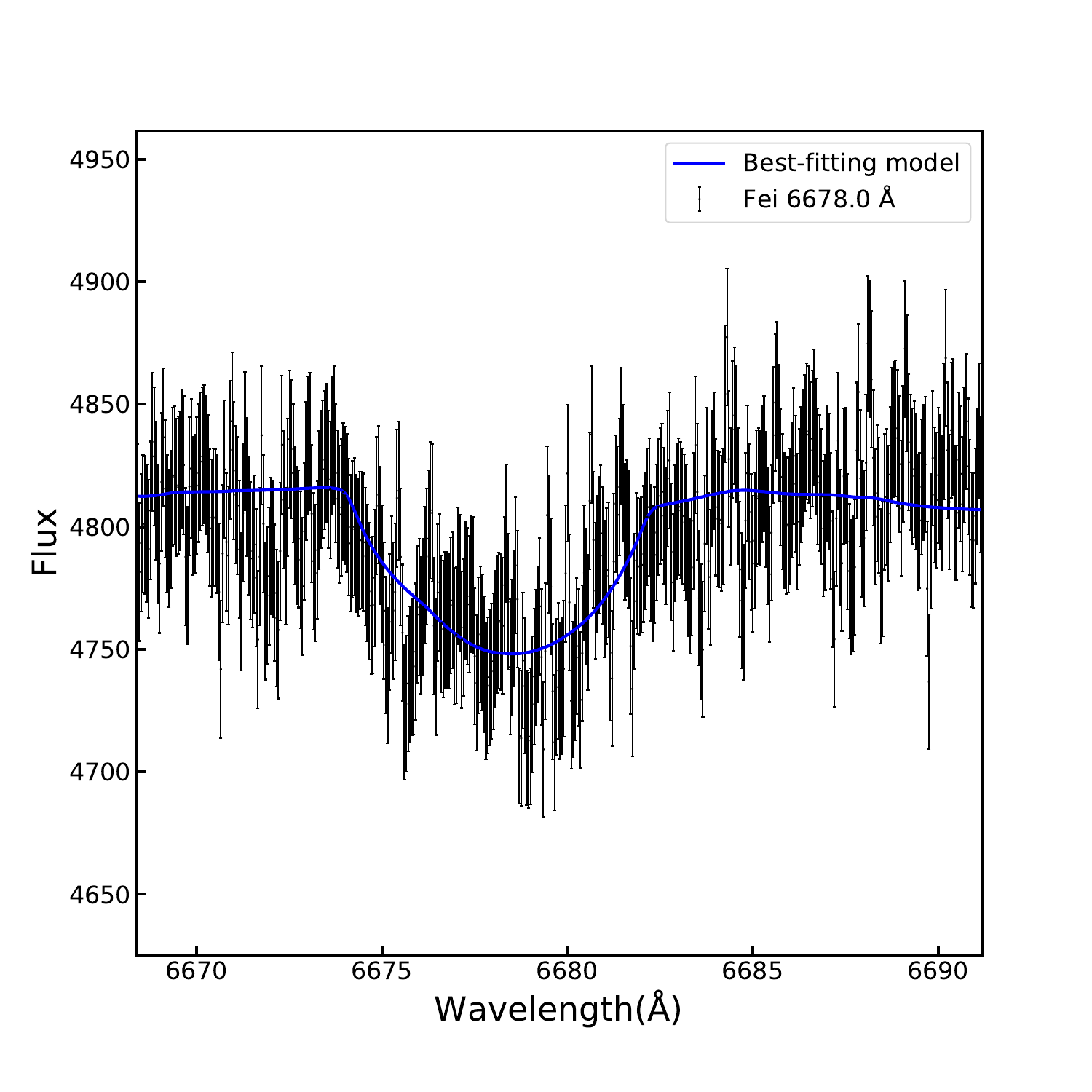}\\     
\end{tabular}

	\caption{As Figure \ref{fig:3532 star 3}，but for Star 4 listed in Table \ref{tab:3532 results}}
 \label{fig:3532 star 4}
\end{figure}

\end{CJK*}
\end{document}